\documentclass{mn2e}

\usepackage[small]{caption}
\usepackage{graphicx}
\usepackage{multirow}
\usepackage{amssymb}
\usepackage{float}
\usepackage{lscape}

\title{Iron line profiles in {\sl Suzaku} spectra of bare Seyfert galaxies}

\author[A. Patrick et al.]{A.R. Patrick$^{1}$, J.N. Reeves$^{1}$, D. Porquet$^{2}$, A.G. Markowitz$^{3}$, A.P. Lobban$^{1}$, Y. Terashima$^{4}$\\
$^{1}$Astrophysics Group, School of Physical Sciences, Keele University, Keele, Staffordshire, ST5 8EH, UK\\
$^{2}$Observatoire astronomique de Strasbourg, Universit�e de Strasbourg, CNRS, UMR 7550, 11 rue de l'Universite, F-67000 Strasbourg,
France\\
$^{3}$Center for Astrophysics and Space Sciences, University of California, San Diego, M.C. 0424, La Jolla, CA, 92093-0424, USA\\
$^{4}$Department of Physics, Ehime University, Matsuyama, Ehime, 790-8577, Japan\\}

\pagerange{\pageref{firstpage}--\pageref{lastpage}} \pubyear{2010}

\begin{document}

\maketitle
\begin{abstract}
We methodically model the broad-band {\sl Suzaku} spectra of a small sample of six 'bare' Seyfert galaxies: Ark 120, Fairall 9, MCG-02-14-009, Mrk 335, NGC 7469 and SWIFT J2127.4+5654. The analysis of bare Seyferts allows a consistent and physical modelling of AGN due to a weak amount of any intrinsic warm absorption, removing the degeneracy between the spectral curvature due to warm absorption and the red-wing of the Fe\,K region. Through effective modelling of the broad-band spectrum and investigating the presence of narrow neutral or ionized emission lines and reflection from distant material, we obtain an accurate and detailed description of the Fe K line region using models such as \textsc{laor}, \textsc{kerrdisk} and \textsc{kerrconv}.

Results suggest that ionized emission lines at 6.7 keV and 6.97 keV (particularly Fe\,{\rm XXVI}) are relatively common and the inclusion of these lines can greatly affect the parameters obtained with relativistic models i.e. spin, emissivity, inner radius of emission and inclination. Moderately broad components are found in all objects, but typically the emission originates from tens of $r_{\rm g}$, rather than within $<6\,r_{\rm g}$ of the black hole. Results obtained with \textsc{kerrdisk} line profiles suggest an average emissivity of q\,$\sim$\,2.3 at intermediate spin values with all objects ruling out the presence of a maximally spinning black hole at the 90\% confidence level. We also present new spin constraints for Mrk 335 and NGC 7469 with intermediate values of $a=0.70^{+0.12}_{-0.01}$ and $a=0.69^{+0.09}_{-0.09}$ respectively.
\end{abstract}


\section{Introduction}
\begin{table*}
\caption{The {\sl Suzaku} Seyfert sample}
\begin{tabular}{l c c c c}
\hline
Object  & RA (J2000) & Dec (J2000) & Redshift & $N_{H}$ (Gal) ($10^{22}$cm$^{-2}$) \\ 
\hline
Ark 120 & 05 16 11.4 & --00 09 00 & 0.033 & 0.0978 \\
Fairall 9 & 01 23 45.8 & --58 48 21 & 0.047 & 0.0316 \\
MCG-02-14-009 & 05 16 21.2 & --10 33 41 & 0.028 & 0.0924 \\
Mrk 335 & 00 06 19.5 & +20 12 10 & 0.026 & 0.0356 \\
NGC 7469 & 23 00 44.4 & +08 36 17 & 0.016 & 0.0445 \\
SWIFT J2127.4+5654 & 21 27 45.0 & +56 56 40 & 0.014 & 0.7650 \\
\hline
\end{tabular}
\label{tab:sample}
\end{table*} 

\begin{table*}
\caption{Summary of observations for the  objects in the sample. $^1$ The observed {2--10\,keV} flux for XIS and EPIC-pn instruments, 15--50\,keV flux for HXD and 20-100\,keV flux for BAT, in units 10$^{-11}$erg\,cm$^{-2}$\,s$^{-1}$ from Model A.}
\begin{tabular}{l l l c c c c c}
\hline
Object & Mission & Instrument & Date & Exposure (s) & Count rate & Flux$^{1}$ & Obs. ID \\
\hline
\multirow{4}{*}{Ark 120} & \multirow{2}{*}{\sl Suzaku} & XIS & \multirow{2}{*}{2007/04/01} & 100864 & $1.896\pm0.003$ & 3.06 & \multirow{2}{*}{702014010} \\
& & HXD & & 89470 & $0.114\pm0.003$ & 3.46 & \\
& {\sl XMM} & EPIC--pn & 2003/08/24 & 78170 & $25.04\pm0.018$ & 3.87 & 014719010 \\
& {\sl Swift} & BAT & -- & 2453000 & $(6.9\pm0.4)\times10^{-4}$ & 4.88 & \\
\hline
\multirow{4}{*}{Fairall 9} & \multirow{2}{*}{\sl Suzaku} & XIS & \multirow{2}{*}{2007/06/07} & 167814 & $1.718\pm0.002$ & 2.53 & \multirow{2}{*}{702043010} \\
& & HXD & & 127310 & $0.089\pm0.002$ & 2.97 & \\
& {\sl XMM} & EPIC--pn & 2000/07/05 & 25830 & $5.825\pm0.015$ & 1.17 & 0101040201 \\
& {\sl Swift} & BAT & -- & 3280000 & $(4.9\pm0.4)\times10^{-4}$ & 4.35 & \\
\hline
\multirow{3}{*}{MCG-02-14-009} & \multirow{2}{*}{\sl Suzaku} & XIS & \multirow{2}{*}{2008/08/28} & 142152 & $0.216\pm0.001$ & 0.42 & \multirow{2}{*}{703060010} \\
& & HXD & & 120028 & $0.017\pm0.002$ & 0.61 & \\
& {\sl XMM} & EPIC--pn & 2009/02/27 & 82790 & $1.946\pm0.005$ & 0.43 & 0550640101 \\
\hline
\multirow{4}{*}{Mrk 335} & \multirow{2}{*}{\sl Suzaku} & XIS & \multirow{2}{*}{2006/08/21} & 151296 & $1.324\pm0.002$ & 1.49 & \multirow{2}{*}{701031010} \\
& & HXD & & 131744 & $0.012\pm0.001$ & 1.31 & \\
& {\sl XMM} & EPIC--pn & 2006/01/03 & 80360 & $16.43\pm0.014$ & 1.82 & 0306870101 \\
& {\sl Swift} & BAT & -- & 3273000 & $(2.5\pm0.3)\times10^{-4}$ & 1.74 & \\
\hline
\multirow{5}{*}{NGC 7469} & \multirow{2}{*}{\sl Suzaku} & XIS & \multirow{2}{*}{2008/06/24} & 112113 & $1.091\pm0.002$ & 2.11 & \multirow{2}{*}{703028010} \\
& & HXD & & 85315 & $0.068\pm0.002$ & 3.22 & \\
& \multirow{2}{*}{\sl XMM} & EPIC--pn & 2004/11/30 & 53014 & $24.09\pm0.021$ & 2.74 & 0207090101 \\
& & EPIC--pn & 2004/12/03 & 54977 & $16.32\pm0.017$ & 2.89 & 0207090201 \\
& {\sl Swift} & BAT & -- & 3286000 & $(6.6\pm0.3)\times10^{-4}$ & 4.82 & \\
\hline
\multirow{3}{*}{SWIFT J2127.4+5654} & \multirow{2}{*}{\sl Suzaku} & XIS & \multirow{2}{*}{2007/12/09} & 91730 & $1.373\pm0.004$ & 3.77 & \multirow{2}{*}{702122010} \\
& & HXD & & 83321 & $0.074\pm0.002$ & 3.33 & \\
& {\sl Swift} & BAT & -- & 3999000 & $(4.2\pm0.3)\times10^{-4}$ & 4.16 & \\
\hline
\end{tabular}
\label{tab:observations}
\end{table*}

Analysis of the Fe\,K line profile can help us determine some of the intrinsic properties of the central black hole in AGN and its accretion disc (e.g. Fabian et al. 1989; Laor 1991). Foremost in the current climate is the determination of black hole (BH) spin, for example recent observations by Miniutti et al. (2007 \& 2009) and Schmoll et al. (2009) have analysed the broad Fe\,K$\alpha$ region in order to constrain the spin of the central BH. A black hole can be characterised simply by its mass and its spin. Many objects have now been classified in terms of their mass into three categories: Galactic BH, Intermediate Mass BH and Supermassive BH. The spin of the black hole determines the nature of the space-time metric in the regions close to it. The spin parameter $a=cJ/GM^2$ (where J=angular momentum and $0<a<0.998$) is used to describe the spin of the BH, where $a$ is limited to a maximum value of $a=0.998$ at the Thorne limit. This is due to photon capture in which photons travelling on 'negative' angular momentum orbits are preferentially captured by the black hole therefore producing an upper bound to the spin parameter $a$ and hence limiting the innermost stable circular orbit to a minimum of $r_{\rm ISCO}=1.235\,r_{\rm g}$ (Thorne 1974). Constraining the spin of SMBHs in AGN and studying the distribution of black hole spin can aid our understanding of the evolution of AGN and the black holes themselves e.g. mergers, relativistic jets and variability (Blandford \& Znajek 1977; Volonteri et al. 2007; King, Pringle \& Hofmann 2008). 

Line emission from the inner regions of the accretion disc can become broadened due to relativistic effects and Doppler motions (Fabian et al. 1989) resulting in an asymmetric profile. Evidence for such broadening was found using the X-ray CCD detectors onboard {\sl ASCA} (Tanaka, Inoue \& Holt 1994) by Mushotzky et al. (1995), Tanaka et al. (1995) and Nandra et al. (1997) typically over the 0.5--10.0\,keV range. Now with X-ray spectra of increasing quality and over wider energy ranges such as that obtained with the XIS and high energy HXD detectors onboard {\sl Suzaku} (0.5--60.0\,keV, Mitsuda et al. 2007), the Fe\,K region of AGN can be examined in detail. High energy X-ray data are important since it allows the reflection component and its strength to be properly fit, assessing its contribution to the continuum and Fe\,K region (e.g. Reeves et al. 2007). Fitting features such as the Compton hump at $\sim30$\,keV allows, for example, the ionization state of the reflecting material to be determined (Ross \& Fabian 2005). With the aim of measuring properties of the accretion disc and the central black hole itself, broad-band data allows us to start making constraints on parameters in these regions based upon the shape of the Fe\,K line profile. 

Previous studies of iron lines have been made using data from {\sl XMM-Newton} over the 2.5--10.0\,keV (Nandra et al. 2007) and 0.6--10.0\,keV energy ranges (Brenneman \& Reynolds 2009), finding complex emission in the Fe\,K band in the majority of Type 1 Seyfert AGN over and above narrow line components originating from distant material. In a sample of 26 objects Nandra et al. (2007) found that narrow 6.4\,keV emission is ubiquitous amongst AGN and broad Fe\,K lines feature in $\sim$40--50\% of AGN and ionized emission due to Fe\,{\rm XXVI} and Fe\,{\rm XXV} is relatively rare amongst AGN. Brenneman \& Reynolds (2009) found that 4/8 AGN were best fit by a model consisting of relativistically blurred reflection from the inner regions of the accretion disc with 2/8 objects suggesting non-zero spin, however noting that the 10.0\,keV cutoff with the EPIC-pn camera limits their findings of the reflection continuum. 

This small sample features predominantly bare Seyferts i.e. with a very weak warm absorber. This is in an attempt to simplify the modelling of the broad-band spectrum and therefore provide a basic understanding of the properties of the Fe\,K region and any accompanying broad red-wing without the need for debate over differing interpretations of the origin of various absorption components in the spectrum (Turner \& Miller 2009). Using data from {\sl Suzaku's} XIS (Koyama et al. 2007) and HXD (Takahashi et al. 2007) detectors spanning 0.5--60.0\,keV and BAT data (from the {\sl Swift} 22 month all sky survey, Tueller et al. 2010) over 20.0--100.0\,keV provides the broad--band spectra necessary for detailed modelling and measurement of the Fe\,K region and the associated Compton reflection hump. This approach ensures that a robust model of the Fe\,K line region (and broad-band spectrum) can be made with the aim of establishing the degree to which narrow ionized emission lines and relativistically broadened components are required. 

\section{Observations \& Data Reduction}
\subsection{Observations}
The objects in this sample are described in Table \ref{tab:sample} and are all the Seyfert 1, radio quiet AGN with low intrinsic absorption available in the public {\sl Suzaku} data archive with spectra from the XIS and HXD detectors. All objects are detected above 15\,keV and are relatively nearby with redshift $z<0.05$. Hard X-ray data from the BAT instrument onboard {\sl Swift} is used in addition to the HXD data in all observations other than MCG-02-14-009, for which no BAT spectrum was publically available (it was however detected by BAT in the 39 month survey with a 14--150\,keV flux of $1.5\times10^{-11}$\,erg\,cm$^{-2}$\,s$^{-1}$, Cusumano et al. 2010). In addition to the {\sl Suzaku} data, observations with {\sl XMM-Newton} are also used for comparison, except for SWIFT J2127.4+5654 which was not publically available. The details of the observations used in this analysis are given in Table \ref{tab:observations}. 

The objects featuring in this sample, however, are all objects with a very low degree of warm absoprtion. Three objects in this analysis have been noted as having some additional absorption in previous observations: MCG-02-14-009, Mrk 335 and NGC 7469. Gallo et al. (2006) found a small neutral absorber ($N_{\rm H}<10^{21}$\,cm$^{-2}$) and a O\,{\rm VII} edge in a re-analysis of a short ($\sim5$\,ks net exposure) observation of MCG-02-14-009 with {\sl XMM--Newton}. However including these components in the {\sl Suzaku} data used here makes no improvement ($N_{\rm H}<3\times10^{20}$\,cm$^{-2}$) and the optical depth of the O\,{\rm VII} edge can only be constrained to $\tau<0.04$. Therefore MCG-02-14-009 can be considered as 'bare' for the purposes of this paper. 

Mrk 335 has also shown evidence for a warm emitter whilst in a low state when observed with {\sl XMM--Newton} (Grupe et al. 2008), however in the {\sl Suzaku} data used here Mrk 335 has a $0.5-2.0$\,keV flux $14\times$ higher than in the low state observation and as such any warm emission features are entirely dominated by the continuum. As a result of this Mrk 335 is also suitable for inclusion within this sample. Previous observations of NGC 7469 have noted some degree of X-ray and UV absorption ($N_{\rm H}\sim10^{20}-10^{21}$\,cm$^{-2}$, Scott et al. 2005; Blustin et al. 2007). The {\sl Suzaku} data used here is consistent with the previous work using data from {\sl XMM--Newton}, requiring an O\,{\rm VII} edge depth of $\tau<0.1$, however there is no effect upon the Fe\,K parameters and NGC 7469 has also been included within this sample. Figure \ref{fig:continuum} shows no significant additional absorption features below 2\,keV within the {\sl Suzaku} data.

\subsection{Data reduction}
All the {\sl Suzaku} data in this paper were reduced using the HEASOFT reduction and analysis package (version 6.8). XIS source spectra were extracted from circular regions of 3.0\arcmin within \textsc{XSELECT} centred upon the source at the on-axis pointing position. Similarly background spectra were extracted from 3.0\arcmin circular regions, taking care not to include the source or the Fe\,55 calibration sources in the corners of the CCD's field of view. Only data from the front-illuminated XIS cameras were used i.e. the XIS\,0 and XIS\,3 cameras -- due to their greater sensitivity at Fe\,K energies, however the observation of Mrk 335 (Obs. ID 701031010) also includes the now non-operational front-illuminated XIS\,2. It should be noted that only data from the XIS\,3 camera was available for the {\sl Suzaku} observation of SWIFT J2127.4+5654 (Obs. ID 702122010) since the XIS\,0 camera was not operating properly.

XIS redistribution matrix files (rmf) were created using the HEASOFT tool \textsc{xisrmfgen} and the ancilliary response files (arf) using \textsc{xissimarfgen}. Data from the 3x3 and 5x5 modes were grouped together and the data from each of the XIS front-illuminated CCDs were co-added in order to increase signal to noise using \textsc{mathpha, addrmf} and \textsc{addarf}. We ignore all XIS data below 0.5\,keV, above 10.0\,keV and between 1.7--1.95\,keV due to uncertainties in the calibration of the detectors around the Si\,K edge.

The HXD/PIN spectrum was extracted from the cleaned HXD/PIN events files and corrected for dead time using the tool \textsc{hxddtcor}. The tuned HXD/PIN background events were used for background subtraction (Fukazawa et al. 2009) using identical good time intervals (GTIs) as per the source events, generated with $10\times$ the actual background count rate to minimize photon noise. A simulated cosmic X-ray background was also produced using \textsc{XSPEC} v 12.5.1n with a spectral form identical to Gruber et al. (1999) and then added to the corrected non X-ray background file to create a single background file. In the analysis of the HXD/PIN spectra we typically consider data between 15.0--60.0\,keV however in some cases such as Mrk 335 the data was also ignored below 20.0\,keV due to thermal noise. Hard X-ray data also obtained from the {\sl Swift} 22-month BAT catalogue was included for all objects (other than MCG-02-14-009) over the energy range 20.0--100.0\,keV.

\section{Analysis \& Results}

\begin{figure*}
\rotatebox{-90}{\includegraphics[width=5cm]{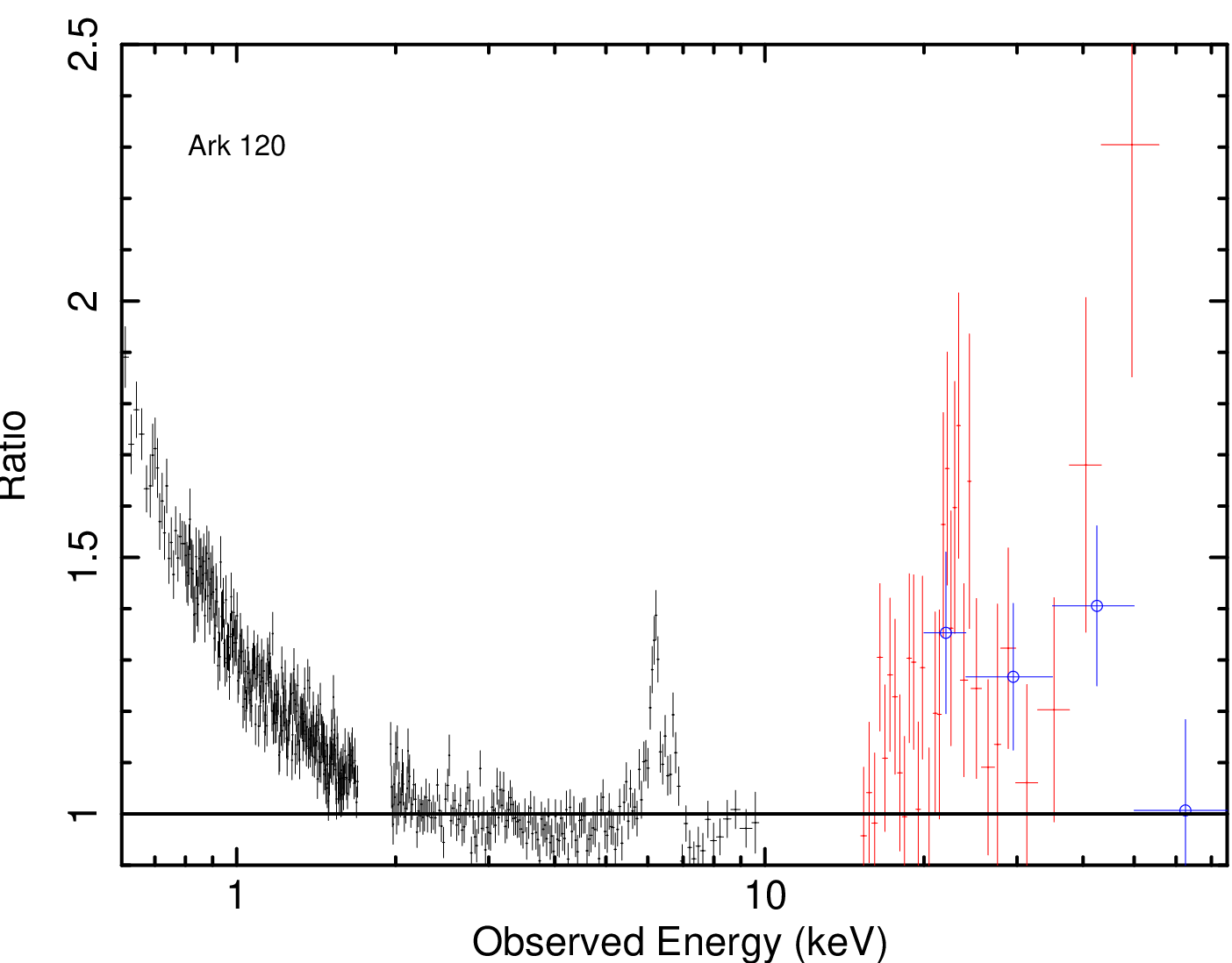}}
\rotatebox{-90}{\includegraphics[width=5cm]{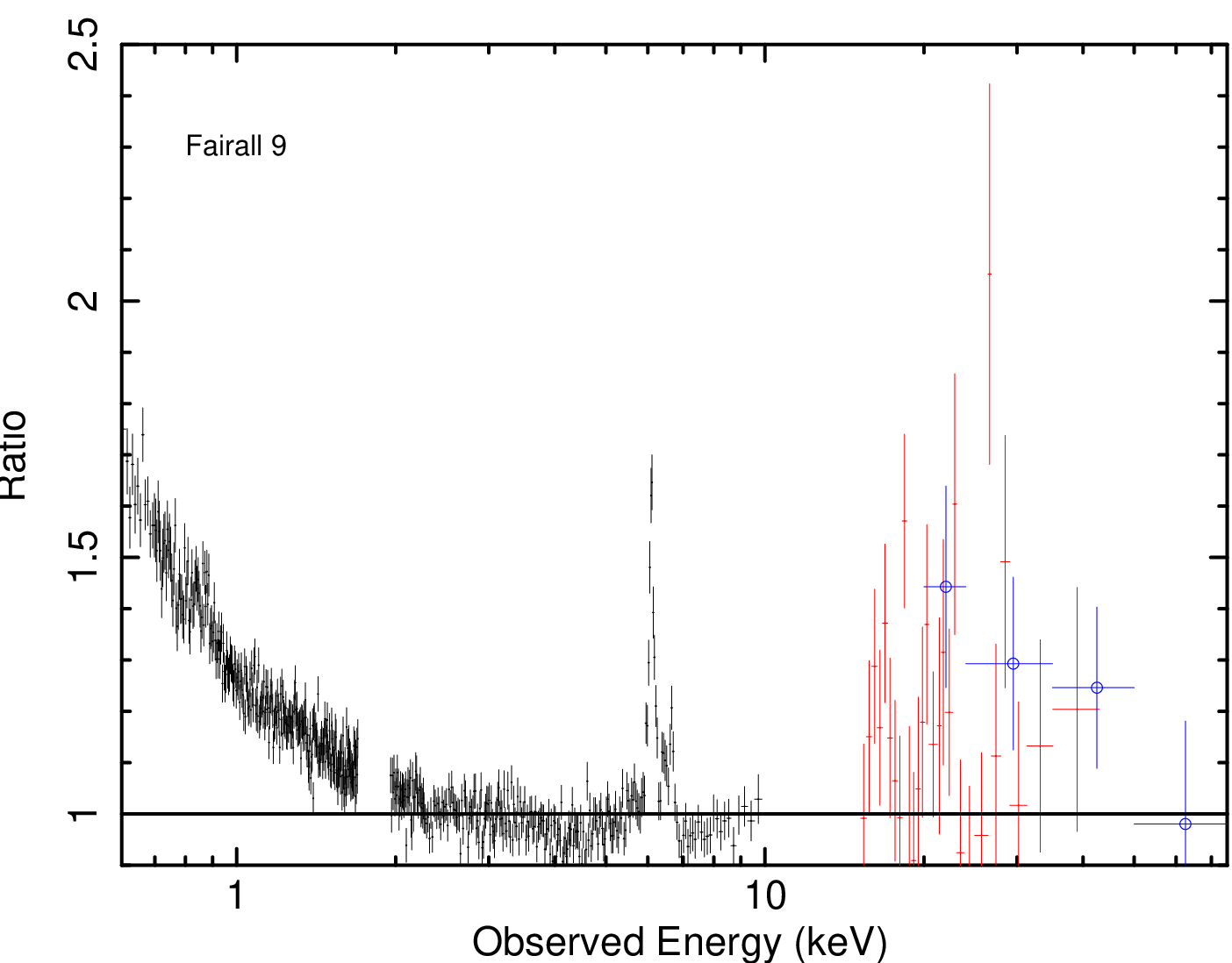}}
\rotatebox{-90}{\includegraphics[width=5cm]{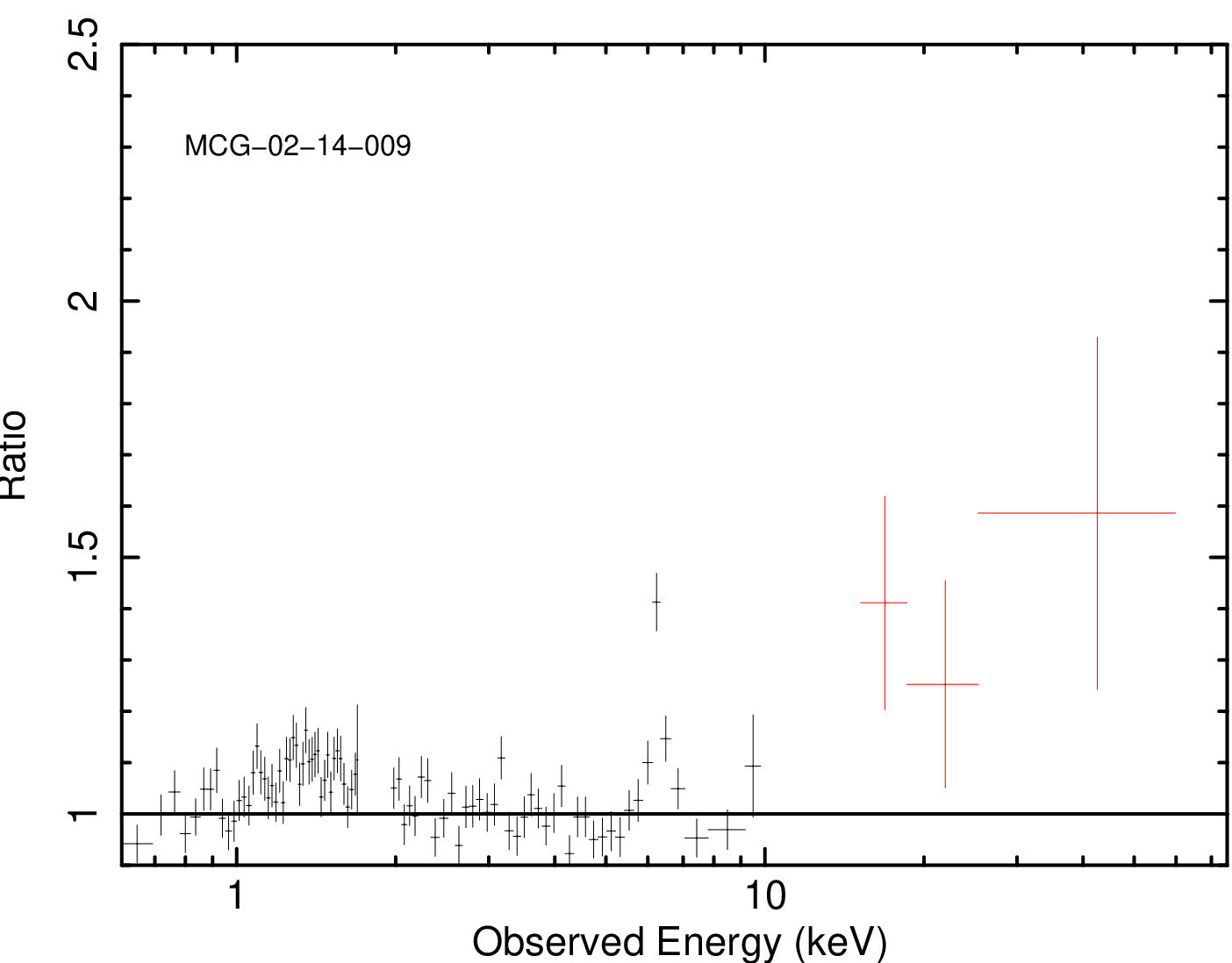}}
\rotatebox{-90}{\includegraphics[width=5cm]{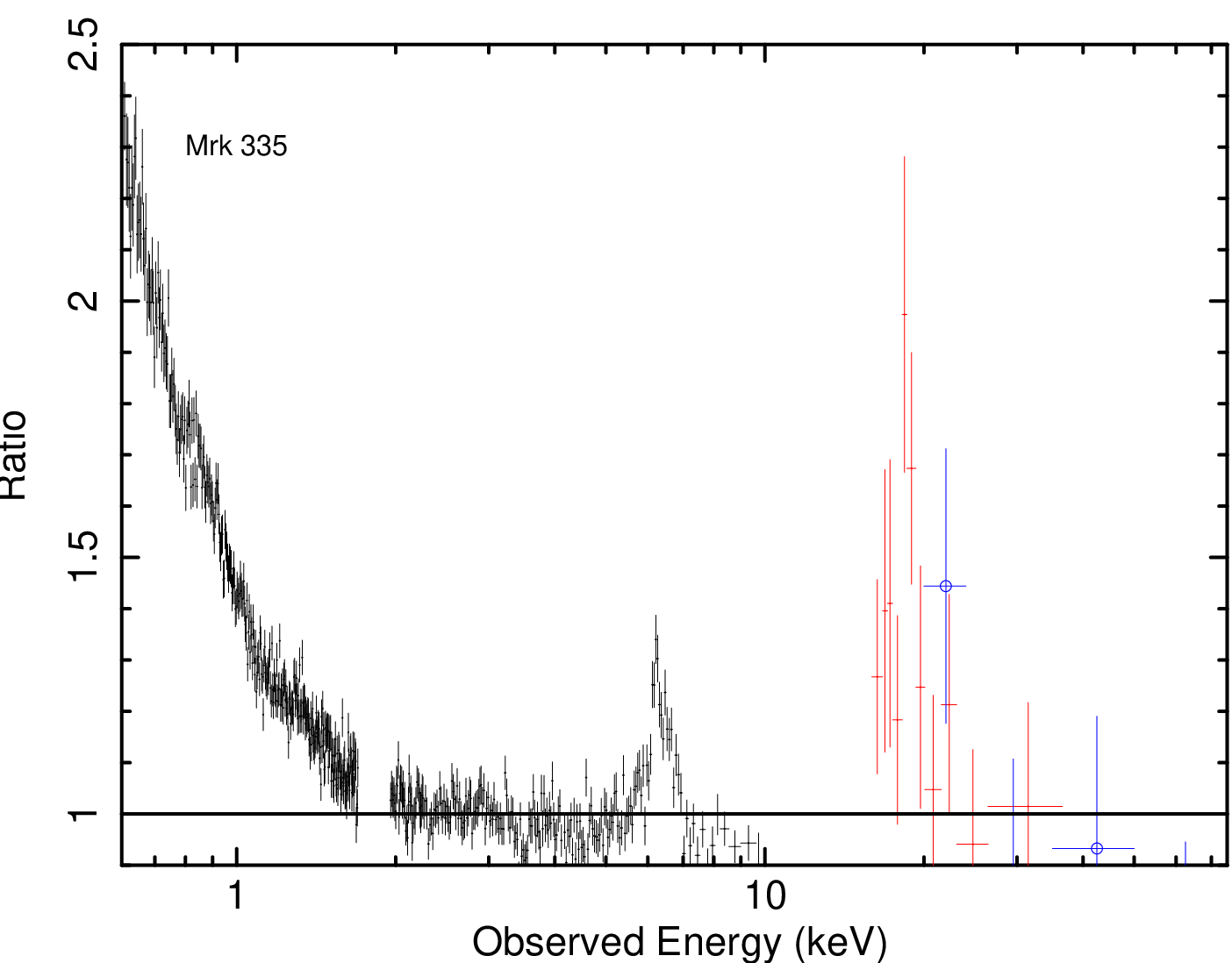}}
\rotatebox{-90}{\includegraphics[width=5cm]{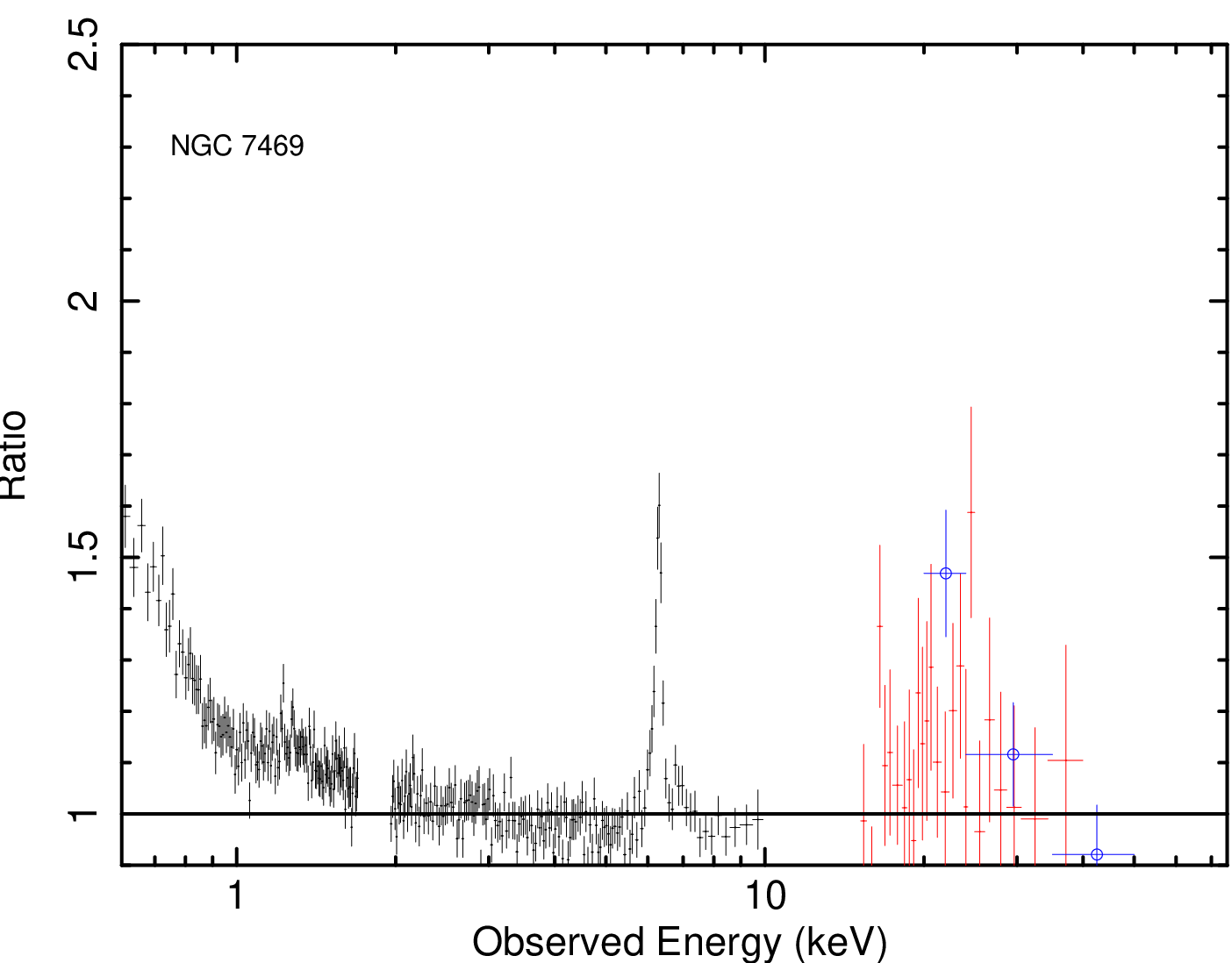}}
\rotatebox{-90}{\includegraphics[width=5cm]{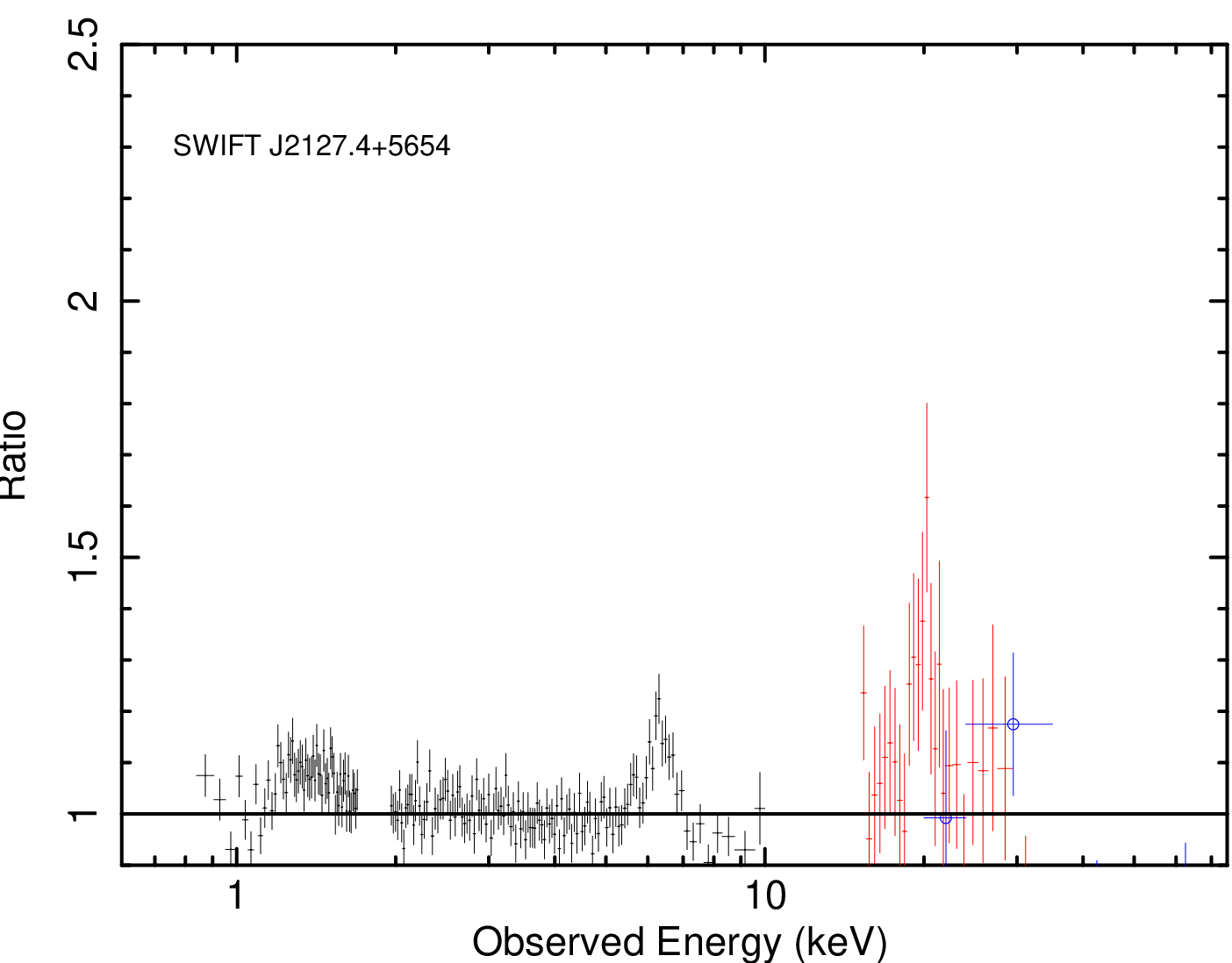}}
\caption{The 0.5--70.0\,keV spectrum and residuals after modelling of the continuum with a simple \textsc{powerlaw} and \textsc{wabs} to account for Galactic absorption, the entire Fe\,K region and any soft excess is left unmodelled. XIS data is in black, HXD in red and BAT data is represented by blue circles.}
\label{fig:soft}
\end{figure*}

\begin{table*}
\caption{Summary of models and components used for each feature.}
\begin{tabular}{l c c c c c }
\hline
Model & Continuum & Distant Reflection & Soft Excess & 6.4\,keV Core & 6.4\,keV Broad \\
\hline
Model A & \textsc{powerlaw} & \textsc{pexrav} & \textsc{compTT} & Gaussian & Gaussian \\
Model B & \textsc{powerlaw} & \textsc{reflionx} & \textsc{compTT} & within \textsc{reflionx} & None \\
Model C & \textsc{powerlaw} & \textsc{reflionx} & \textsc{compTT} & within \textsc{reflionx} & \textsc{laor} \\
Model D & \textsc{powerlaw} & \textsc{reflionx} & \textsc{compTT} & within \textsc{reflionx} & \textsc{kerrdisk} \\
Model E & \textsc{powerlaw} & \textsc{reflionx} & Blurred \textsc{reflionx} & within \textsc{reflionx} & Blurred \textsc{reflionx} \\
Model F & \textsc{powerlaw} & -- & \textsc{compTT} & Gaussian & Blurred \textsc{reflionx} \\
\hline
\end{tabular}
\label{tab:Models}
\end{table*}

Spectral analysis and model fitting is performed with XSPEC v 12.5.1n (Arnaud 1996), all models were modified by Galactic absorption via the \textsc{wabs} multiplicative model (Morrison \& McCammon 1983) using a Galactic column density obtained using the \textsc{nh} {\sl ftool} appropriate for each source giving the weighted average $N_{\rm H}$ value of the (LAB) Survey of Galactic H\,{\rm I} (Kalberla et al. 2005), using abundances from Anders \& Grevesse (1989). In all fits a constant factor was introduced to account for the cross-normalization between the XIS, HXD/PIN and BAT detectors; fixed at 1.16 or 1.18 between XIS and HXD/PIN according to the nominal pointing position\footnote{ http://heasarc.gsfc.nasa.gov/docs/suzaku/analysis/watchout.html} and allowed to vary between XIS and BAT detectors (typically $\sim1.0$, indicating little variability between the {\sl Suzaku} and BAT datasets). Data is fit over the full 0.5--100.0\,keV range, excluding those regions mentioned above. The $\chi^{2}$ minimization technique is used throughout, quoting 90\% errors for one interesting parameter ($\triangle\chi^{2}=2.71$) unless otherwise stated. Where the significance of components is stated according to $\triangle\chi^{2}$, the component has been subtracted from the final model and then refit to ensure the order in which components are added does not effect the quoted statistcal significance. Table \ref{tab:Models} summarises the subsequent models and the components used to model the continuum, soft excess and Fe\,K region.

\subsection{Model A -- Parametrization of spectra}

\begin{figure*}
\rotatebox{-90}{\includegraphics[width=5cm]{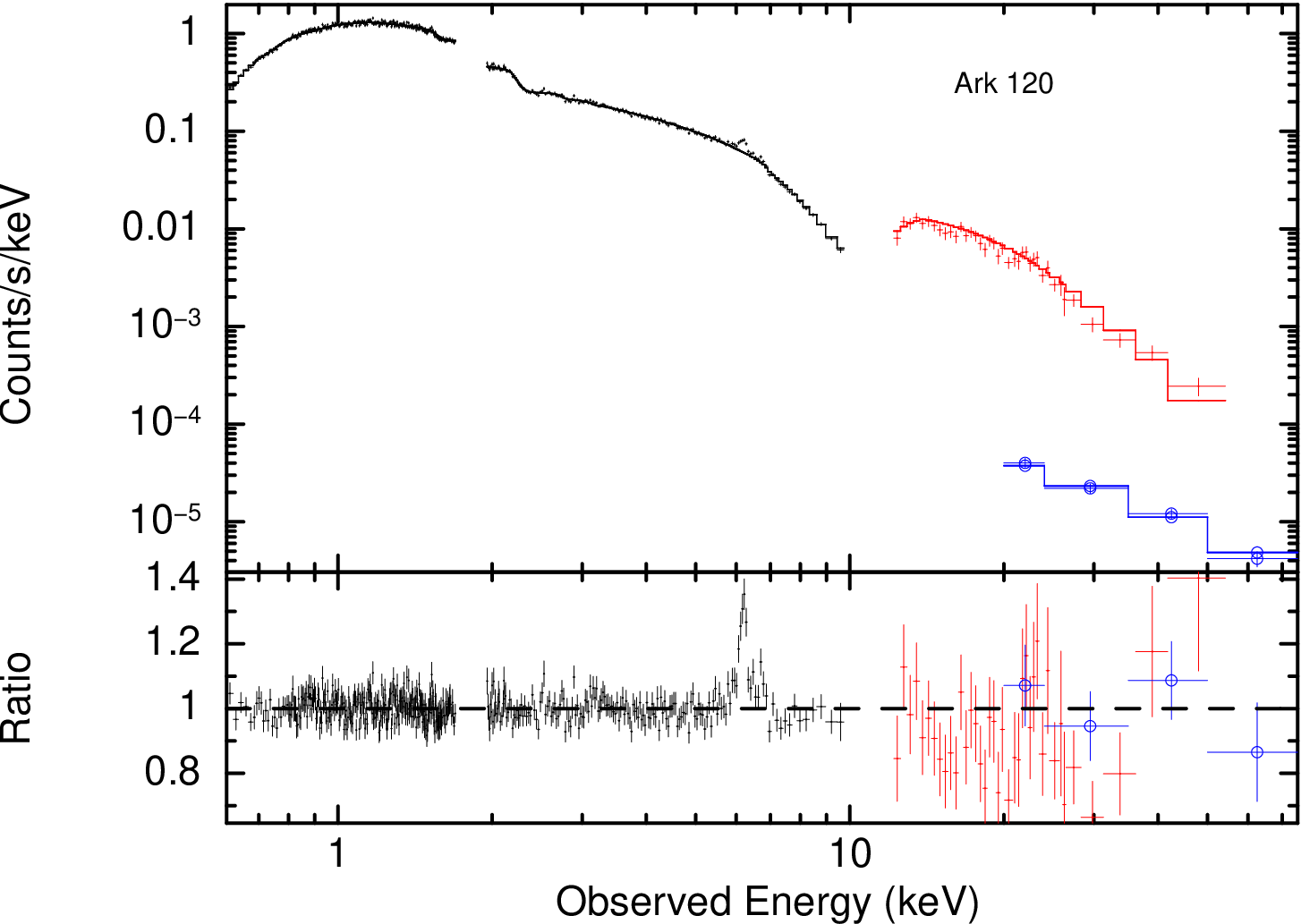}}
\rotatebox{-90}{\includegraphics[width=5cm]{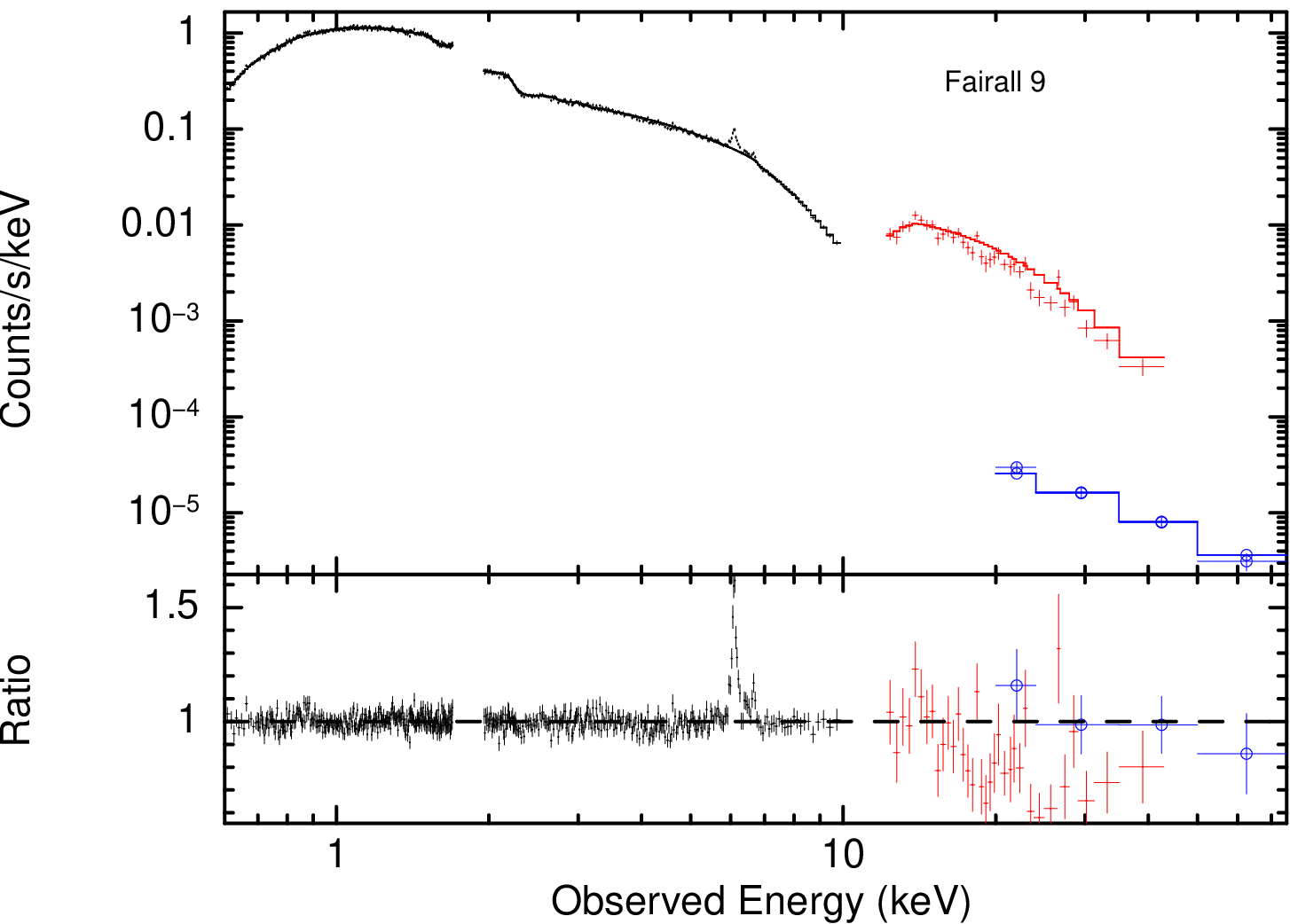}}
\rotatebox{-90}{\includegraphics[width=5cm]{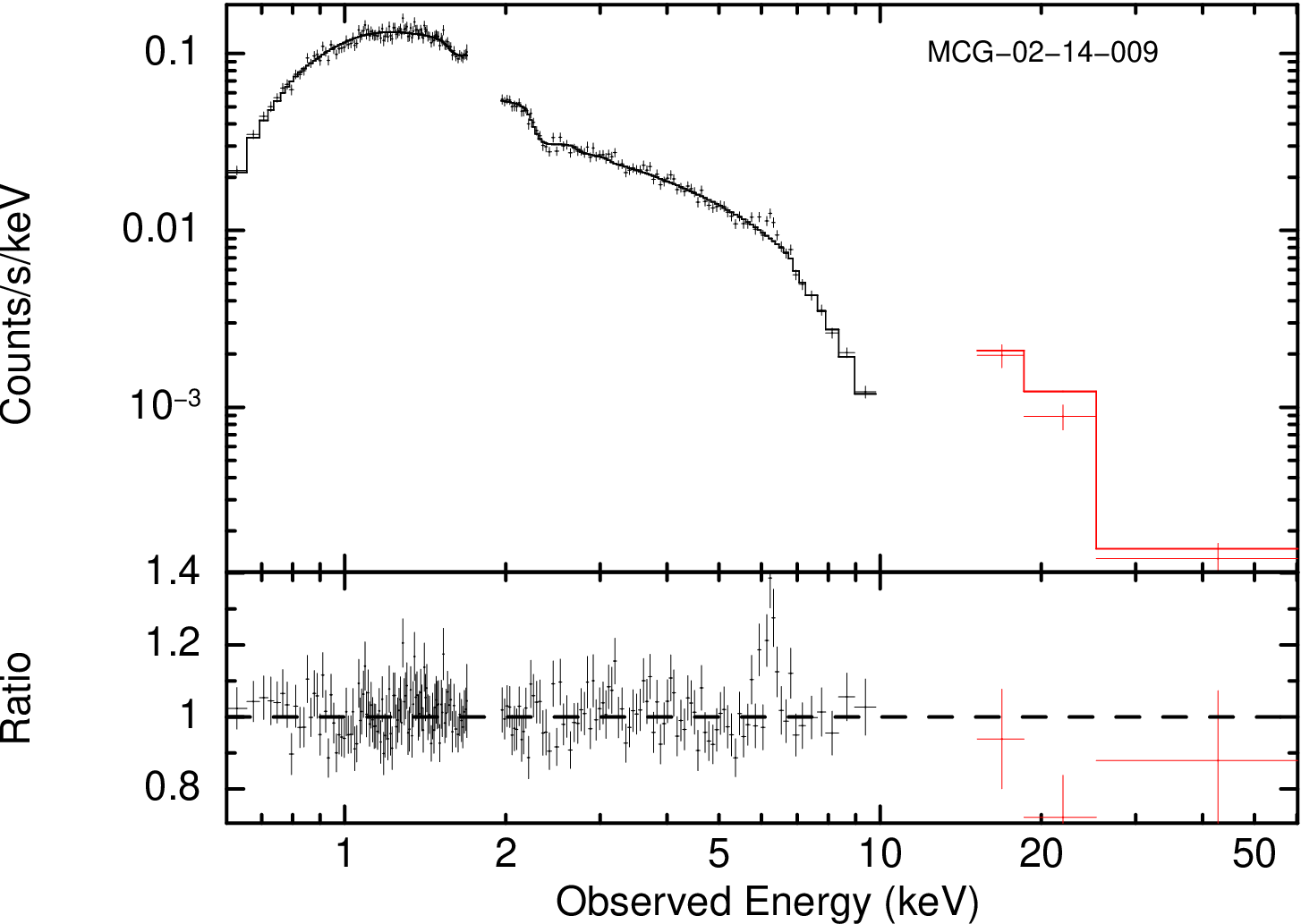}}
\rotatebox{-90}{\includegraphics[width=5cm]{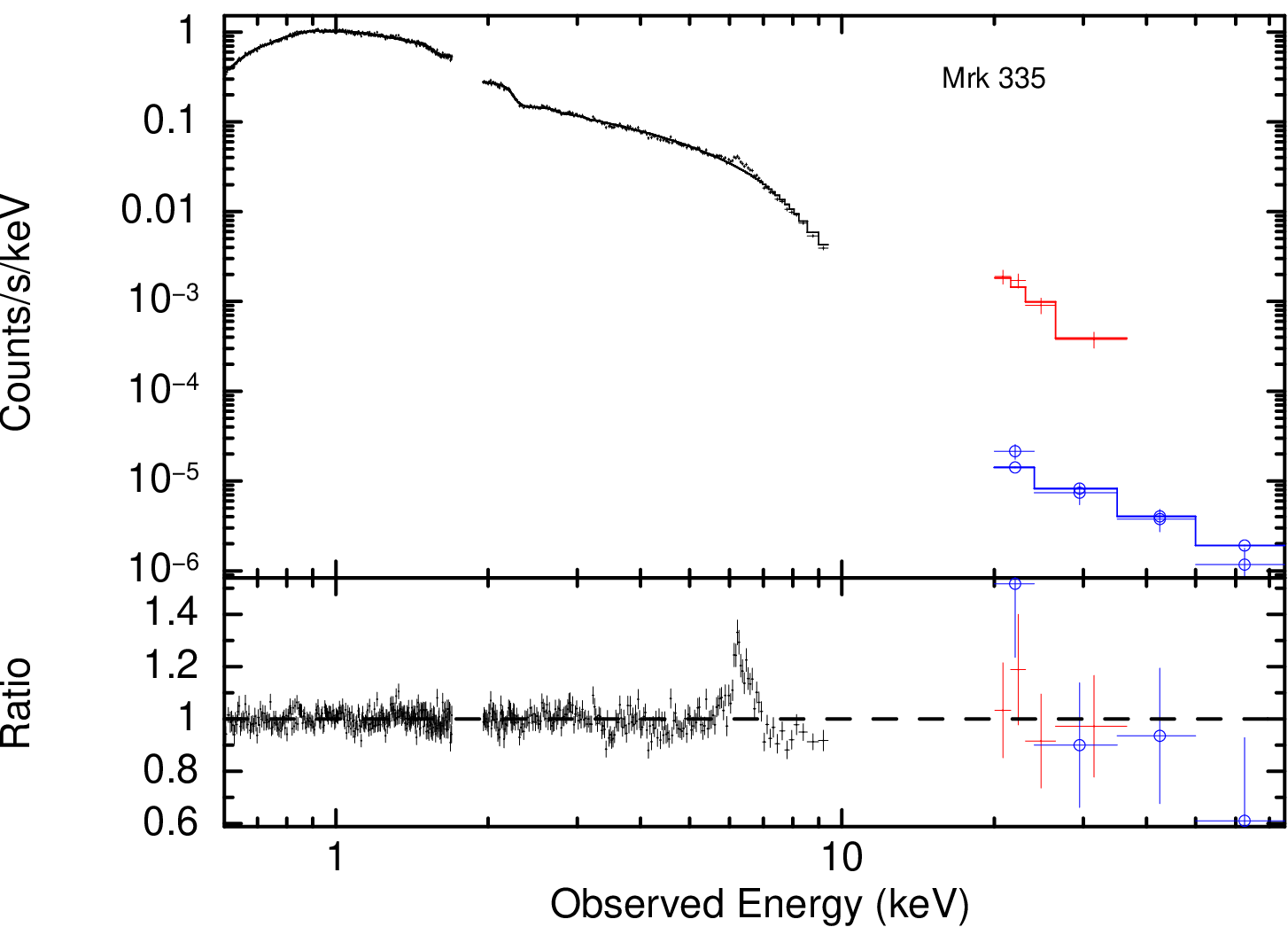}}
\rotatebox{-90}{\includegraphics[width=5cm]{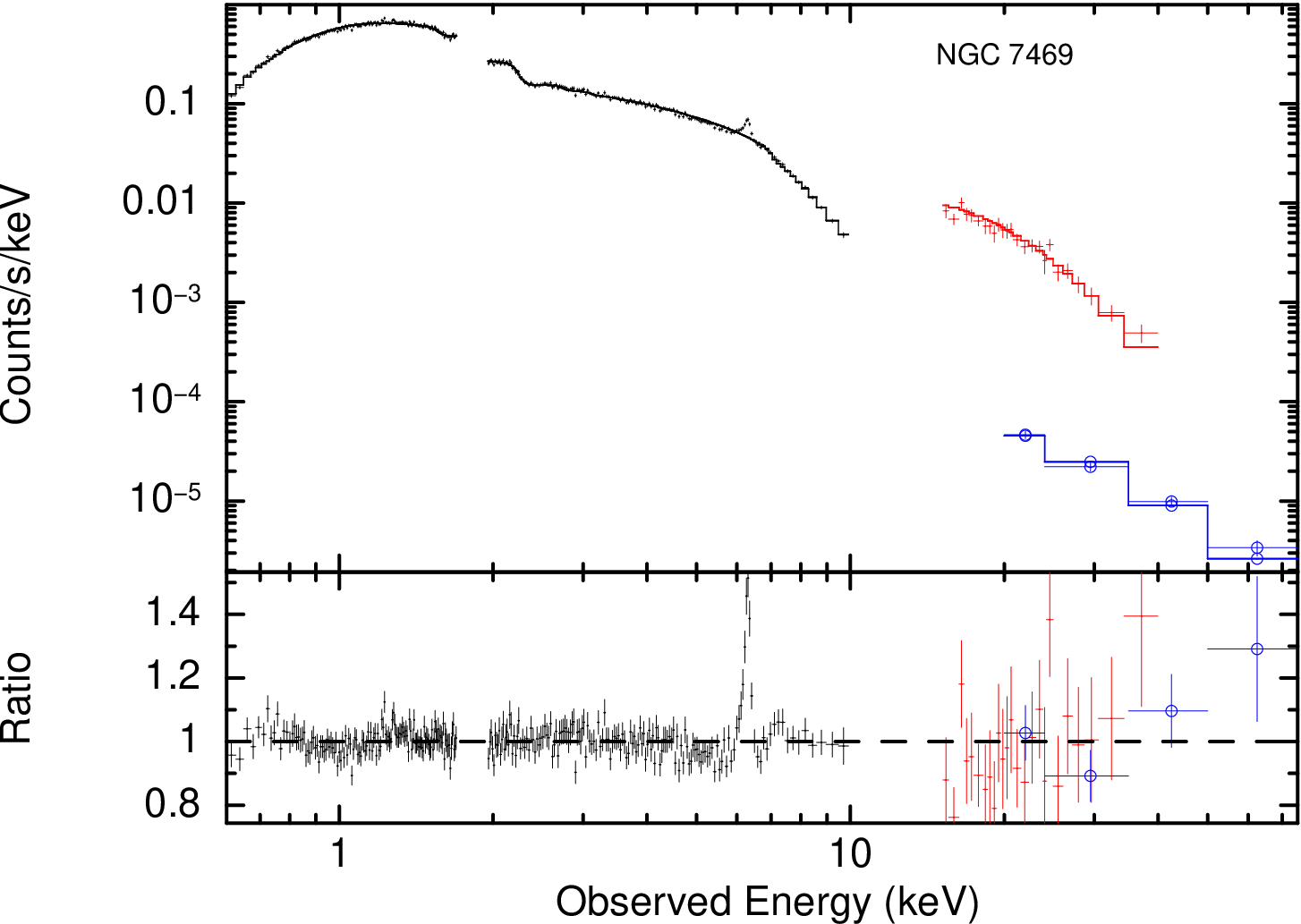}}
\rotatebox{-90}{\includegraphics[width=5cm]{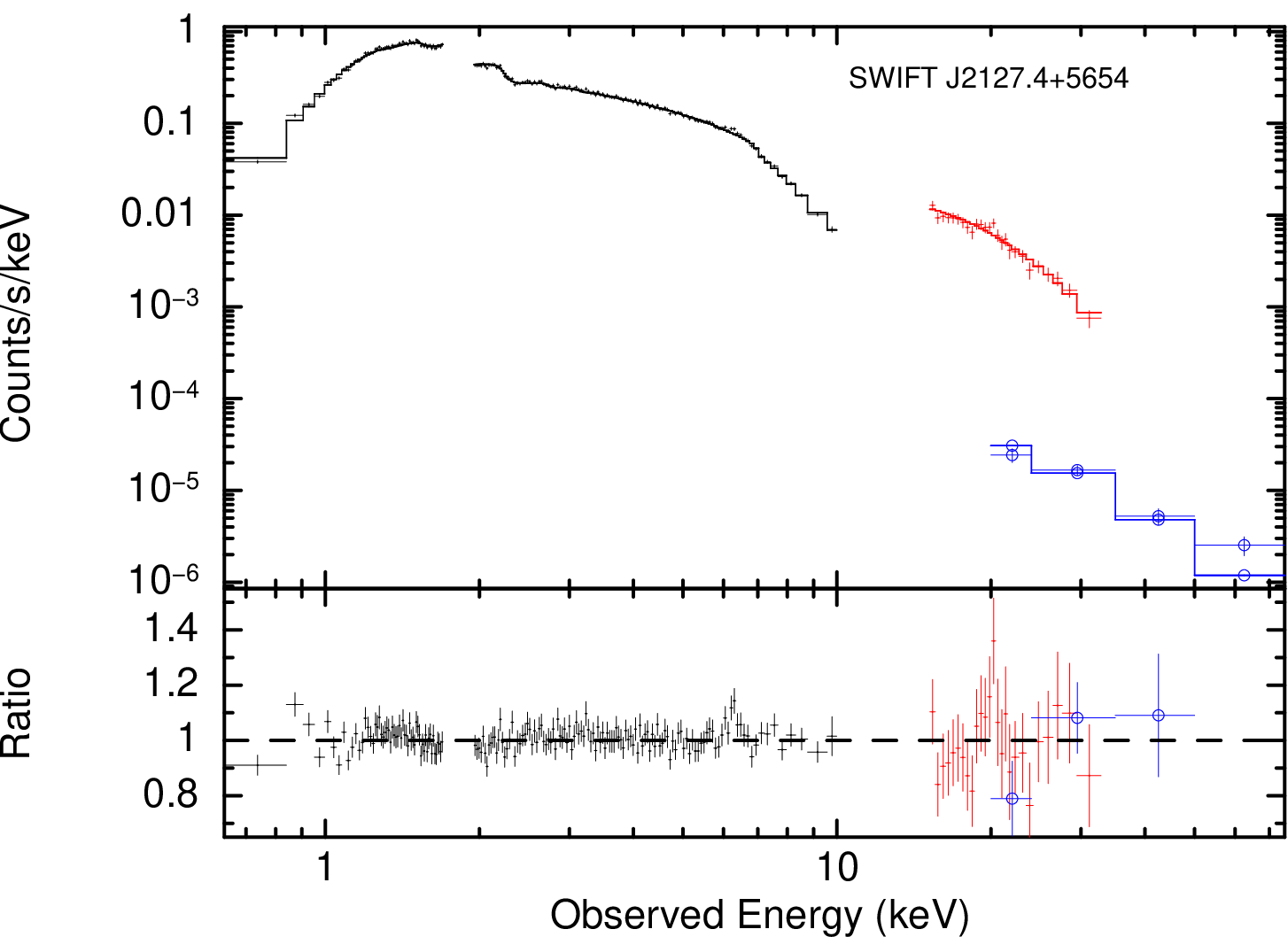}}
\caption{The 0.5--70.0\,keV spectrum and residuals after modelling of the continuum with a \textsc{powerlaw}, neutral reflection from the \textsc{pexrav} model, \textsc{compTT} to model the soft excess and \textsc{wabs} to account for Galactic absorption, the entire Fe\,K region is left unmodelled. XIS data is in black, HXD in red and BAT data is represented by blue circles.}
\label{fig:continuum}
\end{figure*}

\begin{figure*}
\rotatebox{-90}{\includegraphics[width=5cm]{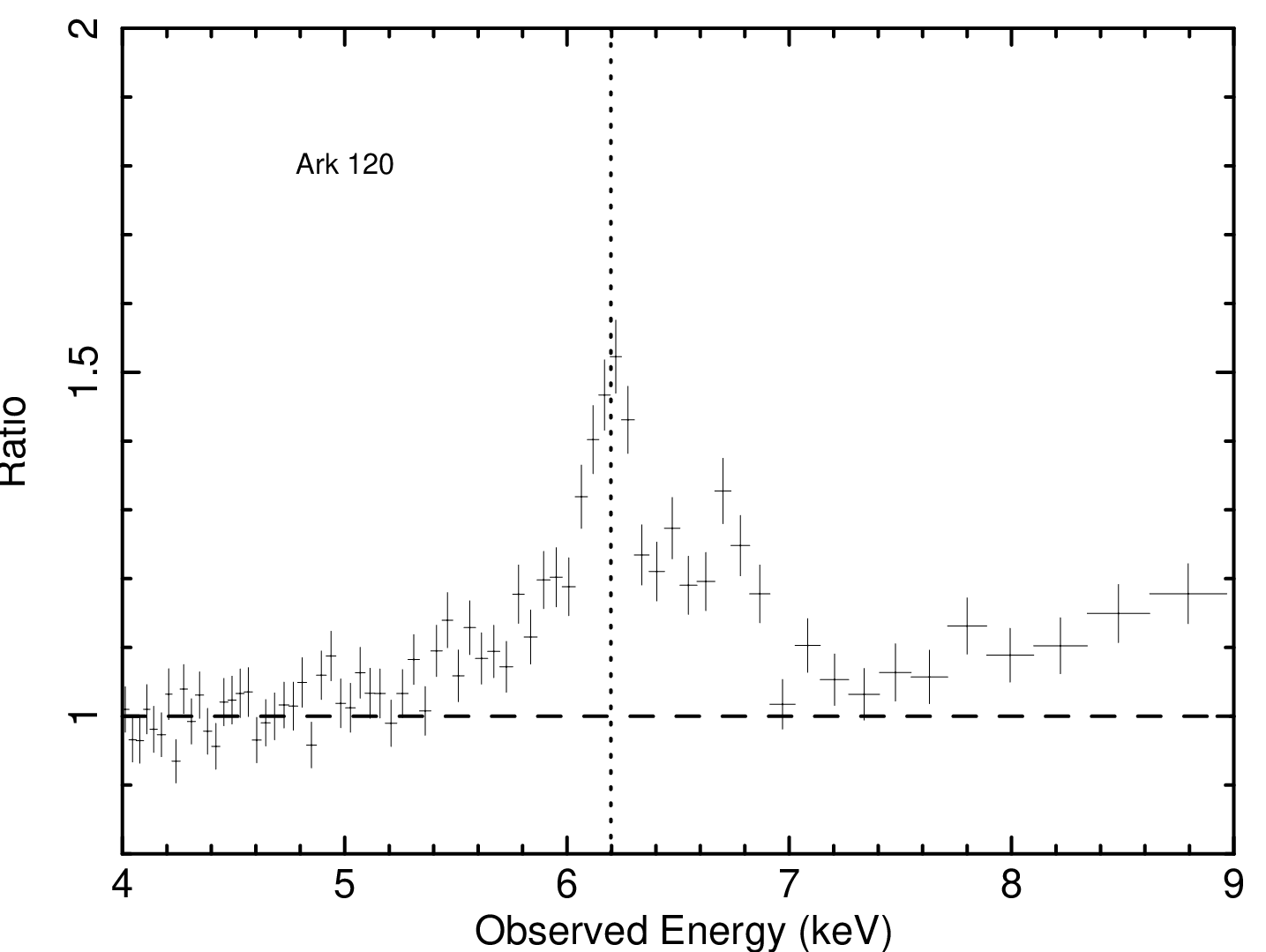}}
\rotatebox{-90}{\includegraphics[width=5cm]{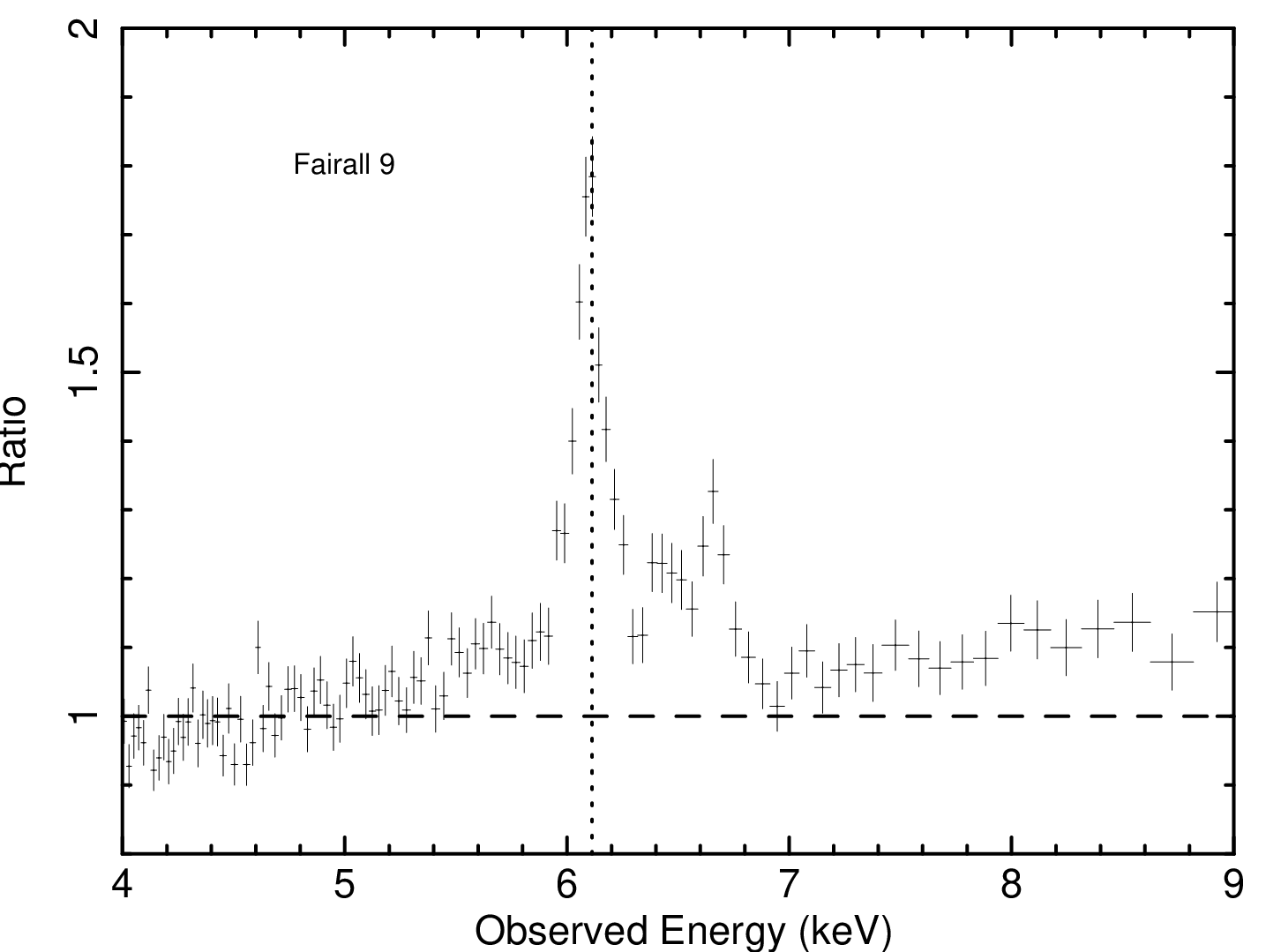}}
\rotatebox{-90}{\includegraphics[width=5cm]{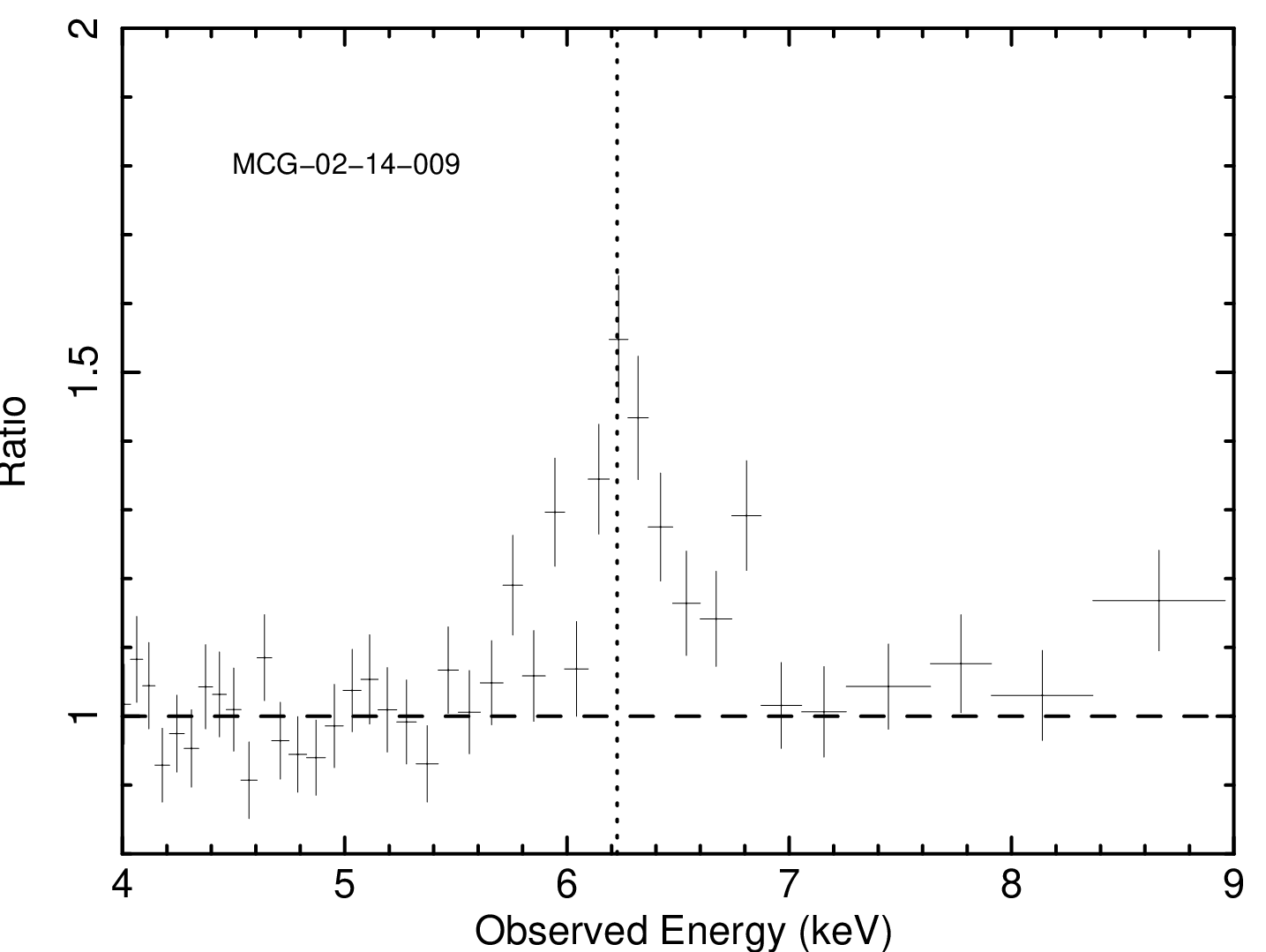}}
\rotatebox{-90}{\includegraphics[width=5cm]{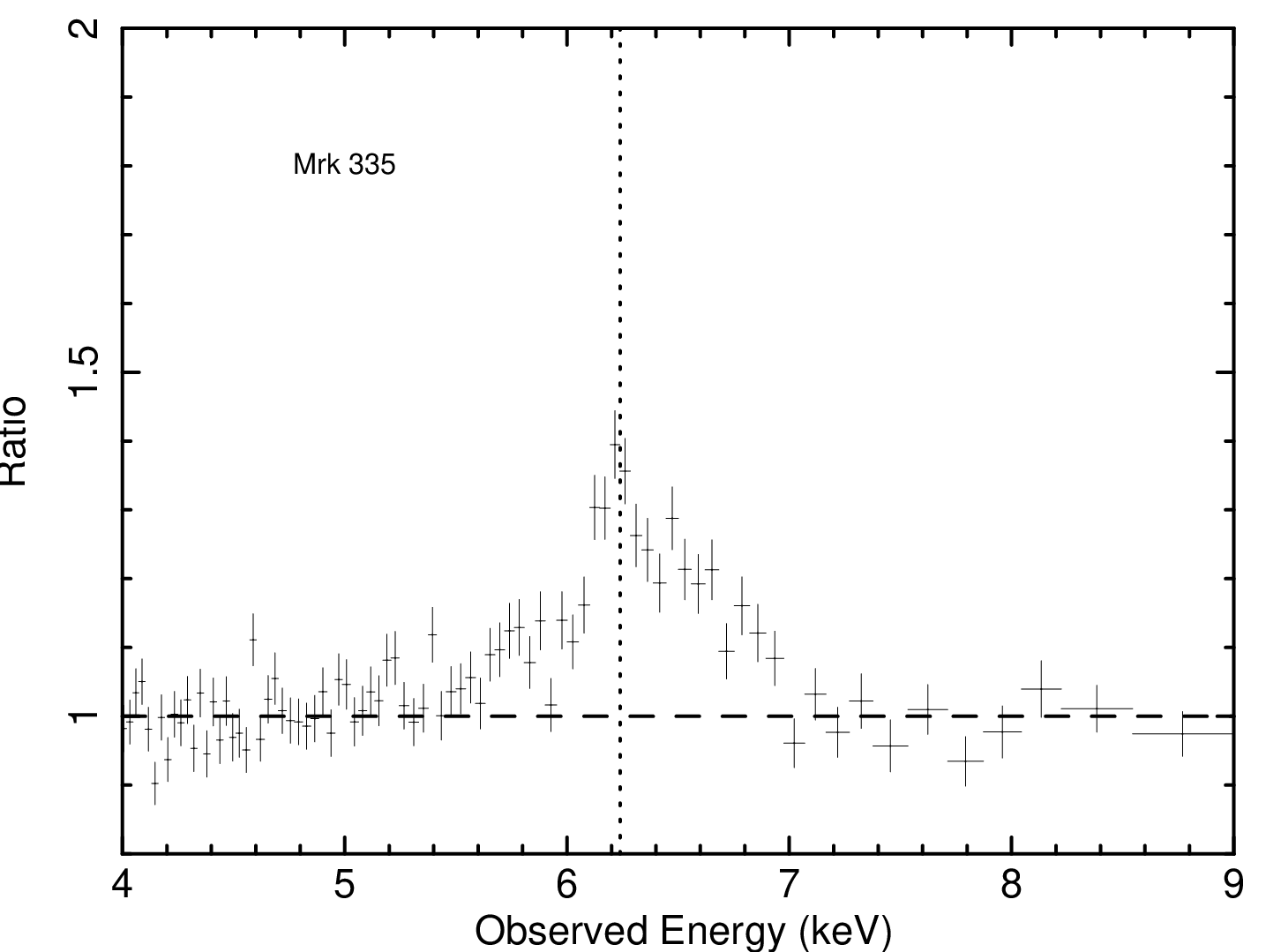}}
\rotatebox{-90}{\includegraphics[width=5cm]{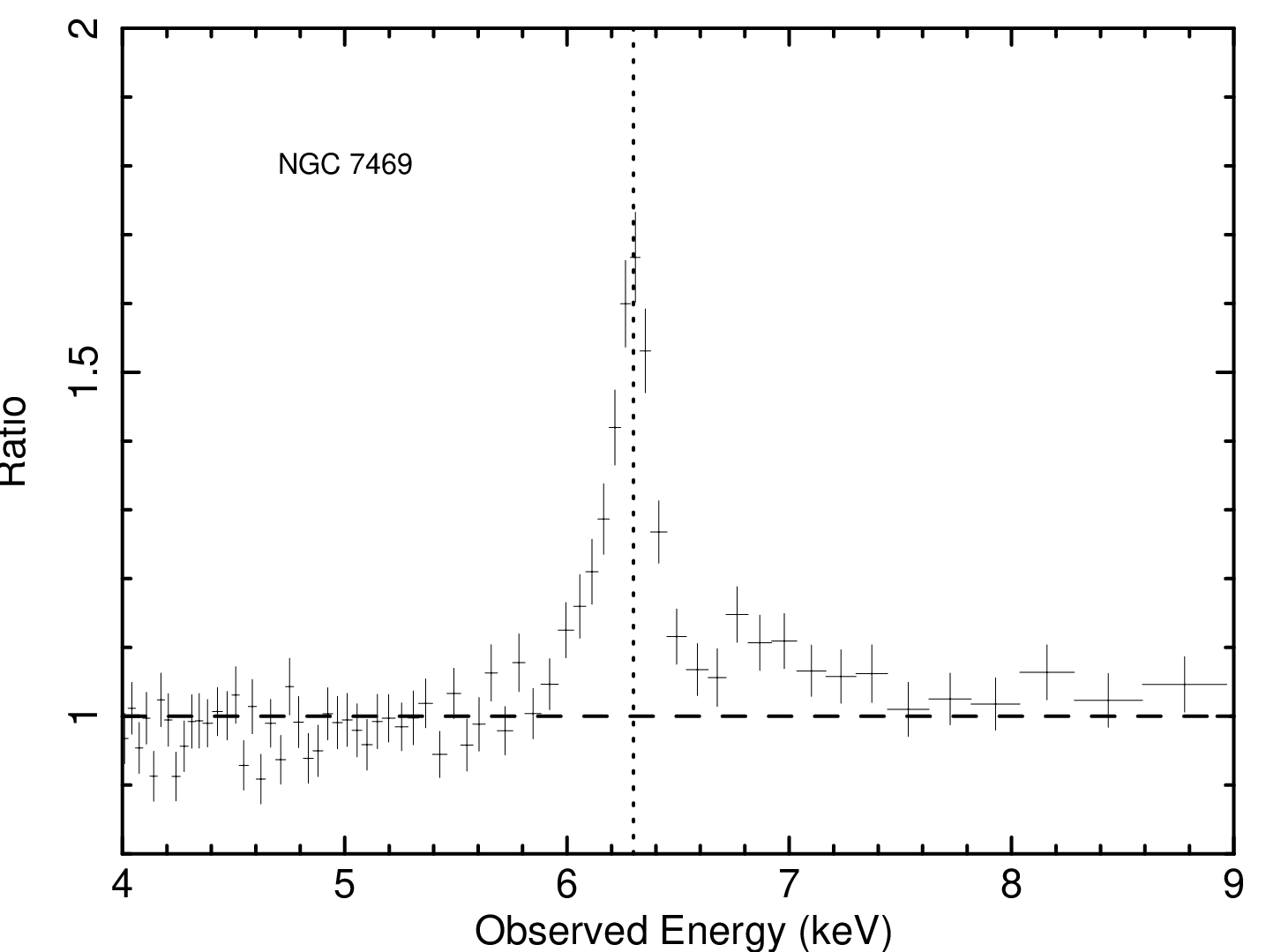}}
\rotatebox{-90}{\includegraphics[width=5cm]{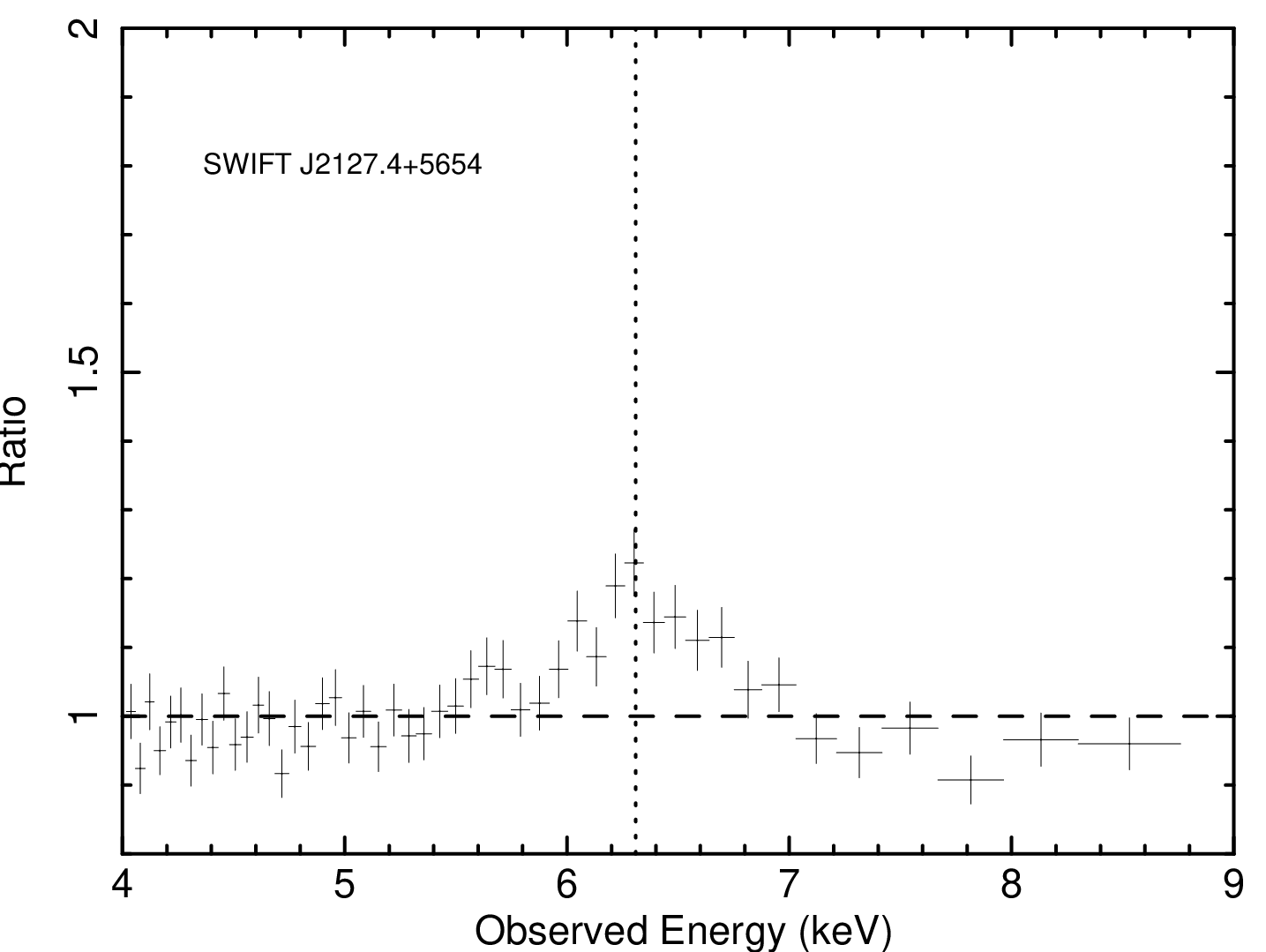}}
\caption{The 4--9\,keV residuals after modelling of the continuum with a powerlaw, \textsc{compTT} to model the soft excess and Galactic photoelectric absorption. The dashed vertical line represents 6.4\,keV in the rest frame.}
\label{fig:po}
\end{figure*}

Model A is intended to provide a simple parametrization of the spectra and give an insight into the presence of basic components of the spectra and the extent to which a possibly relativistically broadened component is required to model the Fe\,K line region. None of the spectra required any significant warm absorber (see Section 2.1), thereby simplifying any present broadening in the region. Additionally, the \textsc{compTT} model representing Comptonization of soft photons in a hot plasma above the disc (Titarchuk 1994) with a soft photon input temperature of 0.02\,keV, is employed to account for the soft excess if present in the spectra (see Figure \ref{fig:soft}). Porquet et al. (2004) also found that in a sample of PG quasars the soft excess was better modelled in this way, rather than by thermal emission from the accretion disc. A second soft \textsc{powerlaw} component instead of \textsc{compTT} also gives a similar parametrization of the soft excess. The narrow 6.4\,keV core due to reflection from distant material is present in all six objects and has been modelled with a narrow Gaussian with width $\sigma_{\rm K{\alpha}}$ free to vary. The narrow component is not resolved in any of the spectra and as such the width is fixed at $\sigma_{\rm K{\alpha}}=0.01$\,keV. Emission resulting from Fe\,K$\beta$ is also accounted for with the line energy fixed at 7.056\,keV, width fixed to that of the narrow K$\alpha$ and flux tied to 13\% of the K$\alpha$ component. 

Consistent with this, neutral distant reflection is accounted for using the \textsc{pexrav} model (Magdziarz \& Zdziarski 1995) applied to the broad--band 0.5--100\,keV spectra. This model requires the input of a photon index $\Gamma$ which is tied to the continuum powerlaw, the normalization of the \textsc{pexrav} component is also tied to that of the powerlaw, abundances are assumed to be Solar (Anders \& Grevesse 1989) and the disc inclination to the observer is fixed at $cos\,i=0.87$ throughout. The reflection fraction $R=\Omega/2\pi$ is left as a free parameter (where $R=1$ denotes reflection from material subtending $2\pi$\,sr). The cut-off energy for the \textsc{pexrav} component is fixed at 1000\,keV, except for NGC 7469 and SWIFT J2127.4+5654 which show indications of a roll-over at high energies, occuring at $E_{\rm c}=119^{+65}_{-31}$\,keV and $E_{\rm c}=49^{+43}_{-14}$\,keV respectively. Such a roll-over was found in SWIFT J2127.4+5654 previously by Malizia et al. (2008) at $E_{\rm c}=33^{+19}_{-10}$\,keV.

Examining the residuals after the application of these components, some objects have more complex features such as excess emission around energies of 6.7\,keV and 6.97\,keV relating to narrow ionized emission from Fe\,{\rm XXV} and Fe\,{\rm XXVI} respectively, again due to distant photoionised gas (see Figure \ref{fig:continuum} and Figure \ref{fig:po}). Similarly to the modelling of the narrow 6.4\,keV core, the energies of these lines are free to vary as is the normalization, however the width remains fixed to $\sigma=0.01$\,keV. A broad Gaussian is also added to the model to account for a red-wing in the spectra with energy, normalization and $\sigma_{\rm Broad}$ as free parameters. In this case the energy is allowed to drop below 6.4\,keV in the rest frame as this model is only intended as a simple parametrization of the Fe\,K region and a test of the significance of the components used to model the region.

\begin{table*}
\caption{Model A components for {\sl Suzaku} XIS, HXD and BAT data from {\sl Swift}. $^{a}$ \textsc{powerlaw} normalization given in units $(10^{-2}\,{\rm ph\,keV^{-1}\,cm^{-2}\,s^{-1}})$. $^{b}$ Flux for \textsc{compTT} quoted over the 0.5-10.0\,keV range in units 10$^{-11}$erg\,cm$^{-2}$\,s$^{-1}$.}
\begin{tabular}{l | c c c c c c}
\hline
& Ark 120 & Fairall 9 & MCG-02-14-009 & Mrk 335 & NGC 7469 & SWIFT J2127.4+5654 \\
\hline
Soft Excess & \checkmark & \checkmark & X & \checkmark & \checkmark & X \\ 
\hline
$\Gamma$ & $1.99^{+0.04}_{-0.04}$ & $1.87^{+0.04}_{-0.04}$ & $1.89^{+0.04}_{-0.04}$ & $2.00^{+0.04}_{-0.02}$ & $1.80^{+0.01}_{-0.01}$ & $2.19^{+0.05}_{-0.05}$ \\ [0.8ex]
Norm$^{a}$ & $1.04^{+0.05}_{-0.06}$ & $0.68^{+0.04}_{-0.05}$ & $0.13^{+0.04}_{-0.04}$ & $0.56^{+0.16}_{-0.02}$ & $0.57^{+0.06}_{-0.07}$ & $1.60^{+0.07}_{-0.06}$ \\
\hline
Plasma Temperature (keV) & $<8.5$ & $<18.6$ & -- & $<8.3$ & $<8.4$ & -- \\ [0.8ex]
Plasma Optical Depth $\tau$ & $<2.4$ & $<2.1$ & -- & $<2.2$ & $0.8^{+0.2}_{-0.5}$& -- \\ [0.8ex]
Flux$^{b}$ & $1.05^{+0.05}_{-0.05}$ & $0.59^{+0.03}_{-0.03}$ & -- & $0.98^{+0.02}_{-0.02}$ & $0.22^{+0.03}_{-0.03}$ & -- \\ [0.8ex]
$\triangle\chi^{2}$ & 250 & 240 & -- & 955 & 194 & -- \\
\hline
Neutral Fe\,K${\alpha}$ Line (keV) & $6.40^{+0.02}_{-0.02}$ & $6.39^{+0.01}_{-0.01}$ & $6.43^{+0.06}_{-0.08}$ & $6.39^{+0.03}_{-0.02}$ & $6.40^{+0.02}_{-0.02}$ & $6.37^{+0.05}_{-0.05}$ \\ [0.8ex]
$\sigma_{\rm Narrow}$ (eV) & $<84$ & $<45$ & $<110$ & $<72$ & $<43$ & $<190$ \\ [0.8ex]
$EW_{\rm Narrow}$ (eV) & $40^{+11}_{-12}$ & $90^{+9}_{-10}$ & $55^{+28}_{-51}$ & $60^{+38}_{-23}$ & $71^{+38}_{-25}$ & $<35$ \\ [0.8ex]
Flux $(10^{-5}\,{\rm ph\,cm^{-2}\,s^{-1}})$ & $1.49^{+0.41}_{-0.46}$ & $2.50^{+0.26}_{-0.27}$ & $0.29^{+0.15}_{-0.27}$ & $1.00^{+0.63}_{-0.38}$ & $1.92^{+1.02}_{-0.69}$ & $<1.49$ \\
\hline
Fe\,{\rm XXV} Line (keV) & -- & $6.74^{+0.04}_{-0.04}$ & -- & $6.69^{+0.06}_{-0.05}$ & -- & -- \\ [0.8ex]
$EW$ (eV) & -- & $17^{+8}_{-9}$ & -- & $33^{+13}_{-7}$ & -- & -- \\ [0.8ex]
Flux $(10^{-5}\,{\rm ph\,cm^{-2}\,s^{-1}})$ & -- & $0.46^{+0.22}_{-0.24}$ & -- & $0.47^{+0.18}_{-0.10}$ -- & -- \\ [0.8ex]
$\triangle\chi^{2}$ & -- & 7 & -- & 14 & -- & -- \\ 
\hline
Fe\,{\rm XXVI} Line (keV) & $6.96^{+0.04}_{-0.04}$ & $6.98^{+0.02}_{-0.02}$ & $6.94^{+0.06}_{-0.09}$ & $6.98^{+0.06}_{-0.14}$ & -- & -- \\ [0.8ex]
$EW$ (eV) & $24^{+11}_{-11}$ & $30^{+8}_{-10}$ & $<51$ & $21^{+13}_{-13}$ & -- & -- \\ [0.8ex]
Flux $(10^{-5}\,{\rm ph\,cm^{-2}\,s^{-1}})$ & $0.70^{+0.32}_{-0.32}$ & $0.70^{+0.19}_{-0.23}$ & $<0.22$ & $0.24^{+0.07}_{-0.06}$ & -- & -- \\ [0.8ex]
$\triangle\chi^{2}$ & 12 & 24 & 3 & 8 & -- & -- \\ 
\hline
Broad Line (keV) & $6.36^{+0.08}_{-0.09}$ & $6.16^{+0.20}_{-0.20}$ & $6.33^{+0.21}_{-0.24}$ & $6.27^{+0.13}_{-0.17}$ & $6.32^{+0.06}_{-0.11}$ & $6.32^{+0.24}_{-0.25}$ \\ [0.8ex]
$\sigma_{\rm Broad}$ (keV) & $0.32^{+0.11}_{-0.09}$ & $0.37^{+0.21}_{-0.13}$ & $0.28^{+0.28}_{-0.13}$ & $0.50^{+0.13}_{-0.11}$ & $0.15^{+0.07}_{-0.03}$ & $0.39^{+0.45}_{-0.26}$ \\ [0.8ex]
$EW_{\rm Broad}$ (eV) & $105^{+26}_{-24}$ & $53^{+27}_{-38}$ & $92^{+59}_{-56}$ & $134^{+42}_{-38}$ & $62^{+28}_{-44}$ & $71^{+53}_{-48}$ \\ [0.8ex]
Flux $(10^{-5}\,{\rm ph\,cm^{-2}\,s^{-1}})$ & $3.74^{+0.93}_{-0.86}$ & $1.64^{+0.83}_{-1.17}$ & $0.49^{+0.31}_{-0.30}$ & $2.28^{+0.72}_{-0.65}$ & $1.55^{+0.70}_{-1.10}$ & $3.43^{+2.56}_{-2.32}$ \\ [0.8ex]
$\triangle\chi^{2}$ & 57 & 28 & 3 & 53 & 24 & 12 \\ 
\hline
$R_{\rm frac}$ & $0.88^{+0.23}_{-0.21}$ & $0.52^{+0.20}_{-0.18}$ & $1.30^{+0.63}_{-0.53}$ & $<0.36$ & $1.46^{+0.37}_{-0.32}$ & $2.5^{+1.1}_{-0.8}$ \\
\hline
BAT const & $1.04^{+0.12}_{-0.11}$ & $0.83^{+0.11}_{-0.10}$ & -- & $1.25^{+0.28 }_{-0.28}$ & $1.24^{+0.16}_{-0.14}$ & $0.78^{+0.13}_{-0.12}$ \\
\hline
NProb & 0.04 & 0.15 & 0.03 & 0.00 & 0.84 & 0.74 \\ [0.8ex]
$\chi^{2}_{\nu}$ & 713.4/648 & 873.4/831 & 606.1/540 & 830.4/720 & 765.7/809 & 839.9/867 \\ 
\hline
\end{tabular}
\label{tab:ModelA}
\end{table*}

\begin{table*}
\caption{Model B components for {\sl Suzaku} XIS, HXD and BAT data from {\sl Swift}. The ionization parameter $\xi$ is given in units erg\,cm\,s$^{-1}$. $^{a}$ \textsc{powerlaw} normalization given in units $(10^{-2}\,{\rm ph\,keV^{-1}\,cm^{-2}\,s^{-1}})$. $^{b}$ \textsc{reflionx} normalisation given in units $10^{-5}$. }
\begin{tabular}{l c c c c c c}
\hline
& Ark 120 & Fairall 9 & MCG-02-14-009 & Mrk 335 & NGC 7469 & SWIFT J2127.4+5654 \\
\hline
$\Gamma$ & $1.90^{+0.01}_{-0.04}$ & $1.83^{+0.01}_{-0.04}$ & $1.86^{+0.01}_{-0.04}$ & $2.00^{+0.02}_{-0.02}$ & $1.78^{+0.07}_{-0.10}$ & $2.11^{+0.03}_{-0.02}$ \\ [0.8ex]
Norm$^{a}$ & $0.95^{+0.06}_{-0.06}$ & $0.64^{+0.02}_{-0.04}$ & $0.12^{+0.04}_{-0.03}$ & $0.51^{+0.01}_{-0.02}$ & $0.54^{+0.04}_{-0.06}$ & $1.52^{+0.04}_{-0.05}$ \\
\hline
$\xi$ & $<22$ & $<20$ & $<21$ & $27^{+9}_{-4}$ & $<11$ & $<13$ \\ [0.8ex]
Fe/Solar & $1.4^{+0.5}_{-0.3}$ & $1.9^{+0.9}_{-0.3}$ & $0.9^{+0.7}_{-0.3}$ & $2.2^{+1.1}_{-0.4}$ & $1.6^{+0.3}_{-0.3}$ & $0.5^{+0.1}_{-0.1}$ \\ [0.8ex]
Norm $^{b}$ & $0.89^{+0.42}_{-0.51}$ & $1.11^{+0.12}_{-0.62}$ & $0.27^{+0.06}_{-0.11}$ & $0.16^{+0.05}_{-0.05}$ & $1.18^{+0.10}_{-0.20}$ & $2.05^{+0.33}_{-1.00}$ \\
\hline
Fe\,{\rm XXV} Line (keV) & $6.66^{+0.04}_{-0.05}$ & $6.73^{+0.03}_{-0.03}$ & -- & $6.68^{+0.03}_{-0.05}$ & -- & $6.66^{+0.07}_{-0.05}$ \\ [0.8ex]
$EW$ (eV) & $20^{+7}_{-7}$ & $19^{+6}_{-6}$ & -- & $40^{+8}_{-8}$ & -- & $25^{+11}_{-11}$ \\ [0.8ex]
Flux $(10^{-5}\,{\rm ph\,cm^{-2}\,s^{-1}})$ & $0.72^{+0.25}_{-0.25}$ & $0.52^{+0.16}_{-0.16}$ & -- & $0.64^{+0.13}_{-0.13}$ & -- & $0.94^{+0.41}_{-0.41}$ \\ [0.8ex]
$\triangle\chi^{2}$ & 20 & 27 & -- & 67 & -- & 16 \\
\hline
Fe\,{\rm XXVI} Line (keV) & $6.95^{+0.03}_{-0.03}$ & $6.98^{+0.02}_{-0.02}$ & $6.94^{+0.05}_{-0.11}$ & $6.96^{+0.06}_{-0.16}$ & -- & $6.98^{+0.09}_{-0.26}$ \\ [0.8ex]
$EW$ (eV) & $31^{+9}_{-9}$ & $31^{+7}_{-7}$ & $39^{+24}_{-23}$ & $21^{+9}_{-9}$ & -- & $16^{+13}_{-13}$ \\ [0.8ex]
Flux $(10^{-5}\,{\rm ph\,cm^{-2}\,s^{-1}})$ & $0.92^{+0.27}_{-0.27}$ & $0.71^{+0.16}_{--0.16}$ & $0.16^{+0.10}_{-0.10}$ & $0.28^{+0.12}_{-0.12}$ & -- & $0.49^{+0.40}_{-0.40}$ \\ [0.8ex]
$\triangle\chi^{2}$ & 31 & 38 & 7 & 14 & -- & 4 \\
\hline
BAT const & $1.07^{+0.12}_{-0.12}$ & $0.83^{+0.10}_{-0.10}$ & -- & $1.03^{+0.23}_{-0.23}$ & $1.03^{+0.10}_{-0.09}$ & $0.70^{+0.10}_{-0.10}$ \\
\hline
NProb & 0.01 & 0.22 & 0.01 & 0.00 & 0.22 & 0.45 \\ [0.8ex]
$\chi^{2}_{\nu}$ & 741.8/649 & 864.7/834 & 623.6/543 & 842.9/723 & 840.2/812 & 871.6/867 \\
\hline
\end{tabular}
\label{tab:ModelB}
\end{table*}

A good fit is obtained by Model A in all objects, particularly Fairall 9, NGC 7469 and SWIFT J2127.4+5654. No significant soft excess is found in MCG-02-14-009 and SWIFT J2127.4+5654, the latter (if present) is likely to be mostly absorbed due to the relatively high amount of Galactic absorption ($N_{\rm H}\approx7.65\times10^{21}$\,cm$^{-2}$). In SWIFT J2127.4+5654 there is some indication of the presence of an intrinsic neutral absorber, albeit of relatively low column density (as found in Miniutti et al. 2009). Instead modelling the spectrum with an absorbed \textsc{powerlaw} using the \textsc{zphabs} model at the redshift of the source improves the fit by $\triangle\chi^{2}\sim59$ for one addition free parameter with intrinsic column density $N_{\rm H}\approx8.2^{+0.6}_{-0.6}\times10^{20}$\,cm$^{-2}$ and photon index $\Gamma=2.19^{+0.05}_{-0.05}$. 

Statistically significant narrow ionized emission is found in most objects, with SWIFT J2127.4+5654 and MCG-02-14-009 having very little requirement for these features. These lines occur at energies likely originating from Fe\,{\rm XXV} and Fe\,{\rm XXVI} with emission at $\sim6.97$\,keV being particularly evident in the spectra, proving strong in Fairall 9 for which the fit improves by $\triangle\chi^{2}\sim24$ with the introduction of a narrow Gaussian (and therefore two additional free parameters) at $6.98^{+0.02}_{-0.02}$\,keV and to a lesser extent in Ark 120 ($\triangle\chi^{2}\sim12$ at $6.96^{+0.04}_{-0.04}$\,keV). The only objects not showing residuals at $\sim6.7$\,keV or $\sim6.97$\,keV are NGC 7469 and SWIFT J2127.4+5654, also in agreement with Miniutti et al. (2009). Residuals at $\sim6.7$\,keV are only found to be significant in two objects: Fairall 9 and Mrk 335 with improvements of $\triangle\chi^{2}\sim7$ and $\sim14$ respectively. 

With the aim of determining the extent to which emission from further in to the black hole is required, a broad component significantly improves the quality of the fit for most objects, typically $\triangle\chi^{2}>20$ for three additional free parameters and with line widths of the order $\sigma_{\rm Broad}\ga0.3$\,keV. Only in MCG-02-14-009 does the addition of a broad Gaussian not make a particularly significant improvement ($\triangle\chi^{2}\sim3$). This is surprising given evidence to the contrary by Porquet (2006) in which a prominent broad and statistically significant iron line was found in a short {\sl XMM-Newton} observation with an equivalent width of {\sl EW}\,$\sim527^{+277}_{-248}$\,eV whereas only {\sl EW}\,$\sim92^{+59}_{-56}$\,eV is found here. However the {\sl XMM-Newton} observation was only 5\,ks net exposure and as a result the Fe\,K line parameters are poorly constrained, while the {\sl Suzaku} data also allow the broad-band continuum to be better constrained (however we cannot rule out some variability between these two observations). In SWIFT J2127.4+5654, this may be due to the relatively high best-fitting value of the reflection component with $R=2.5^{+1.1}_{-0.8}$ which may reduce the significance of a broad component. Mrk 335 features a relatively broad Gaussian with $\sigma_{\rm Broad}=0.50^{+0.13}_{-0.11}$\,keV and {\sl EW}\,$=134^{+42}_{-38}$\,eV, however this feature is not as strong as the one found by Larsson et al. (2008) with $\sigma=0.45^{+0.10}_{-0.06}$\,keV and {\sl EW}\,$=250^{+40}_{-39}$\,eV. Larsson et al. (2008) also find that the Fe\,{\rm XXVI} emission line does not improve the fit, however an improvement of $\triangle\chi^{2}\sim8$ for three additional free parameters is found here with the introduction of a line at $6.98^{+0.06}_{-0.14}$\,keV.

In general, the broad emission found in these objects is typically too strong to be modelled as purely a Compton shoulder despite the similarity of the line energy of the broad Gaussian with the first order Compton shoulder peak energy of $\sim6.24$\,keV (see Table \ref{tab:ModelA}). However we cannot rule out the possibility of a contribution from a Compton shoulder. Given the relatively high $EW$ of the broad line component (mean $EW\sim86$\,eV) compared to the narrow core, the majority of the emission is likely to arise from a broadened component. Indeed the Compton shoulder is unlikely to contribute more than $\sim20\%$ of the 6.4\,keV core flux (e.g. George \& Fabian 1991; Matt 2002). Note that the Compton shoulder is included in subsequent models through the use of a \textsc{reflionx} reflection component.

\subsection{Model B -- Self consistent distant reflection}
\begin{figure*}
\rotatebox{-90}{\includegraphics[width=5cm]{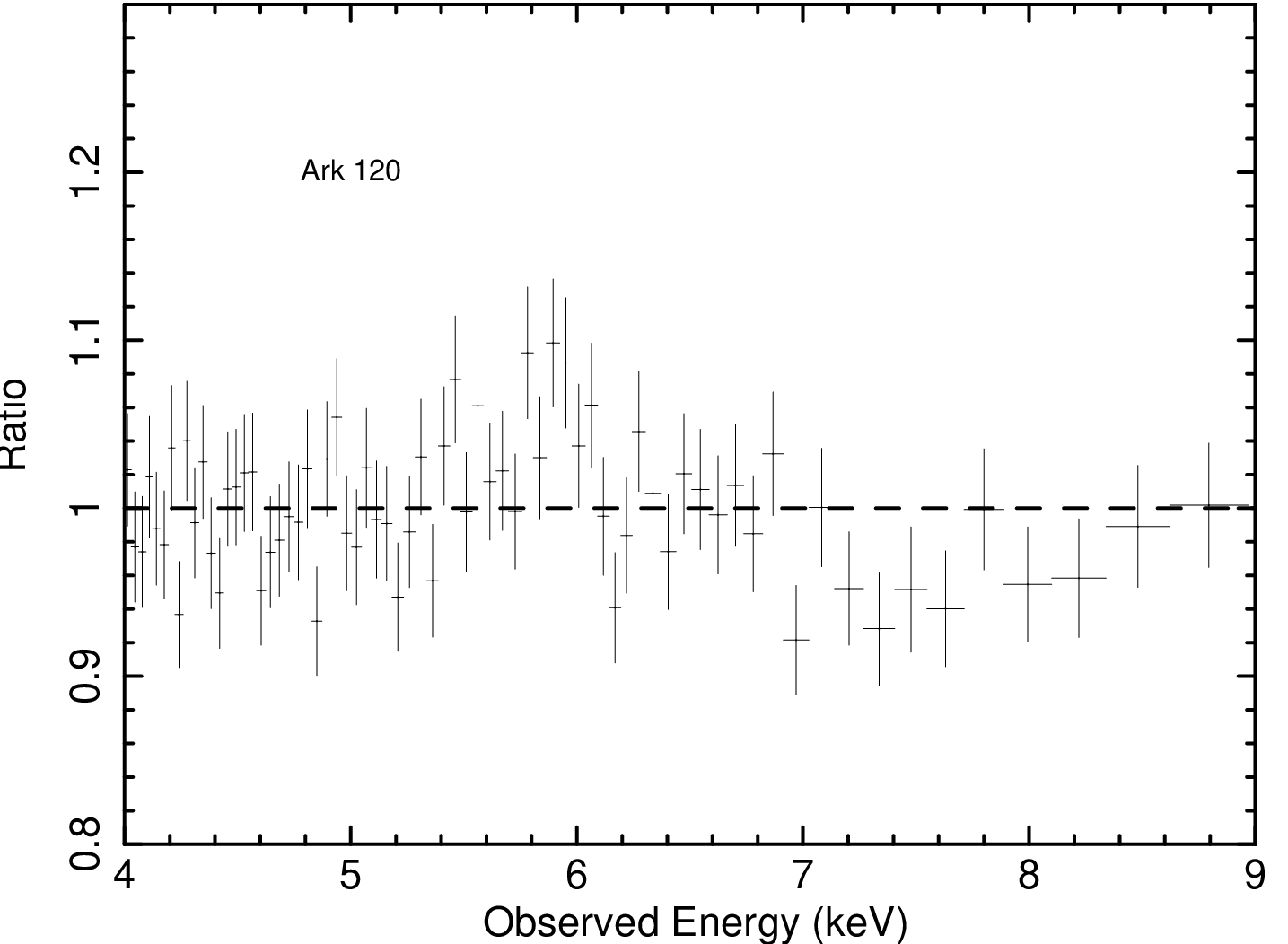}}
\rotatebox{-90}{\includegraphics[width=5cm]{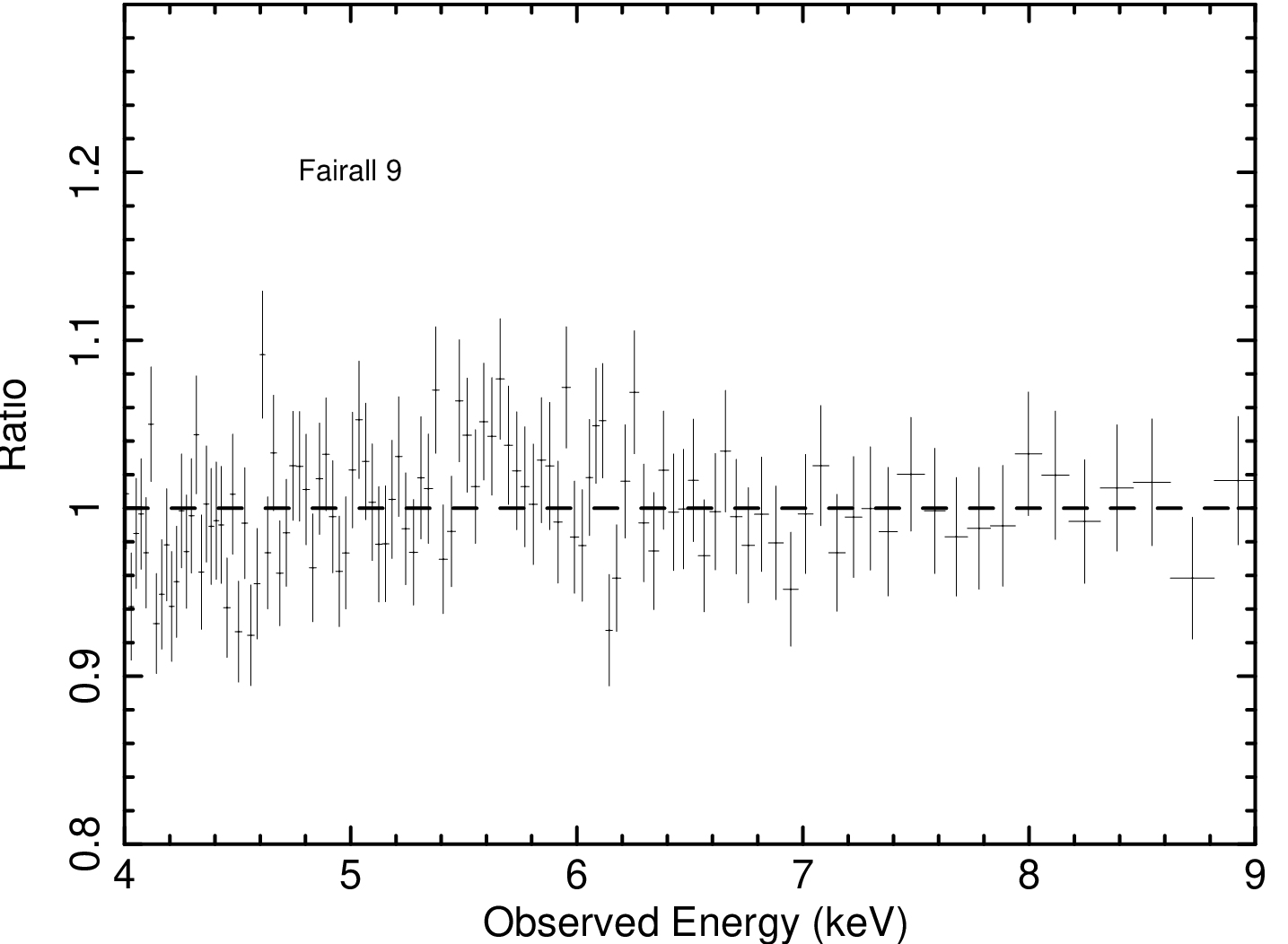}}
\rotatebox{-90}{\includegraphics[width=5cm]{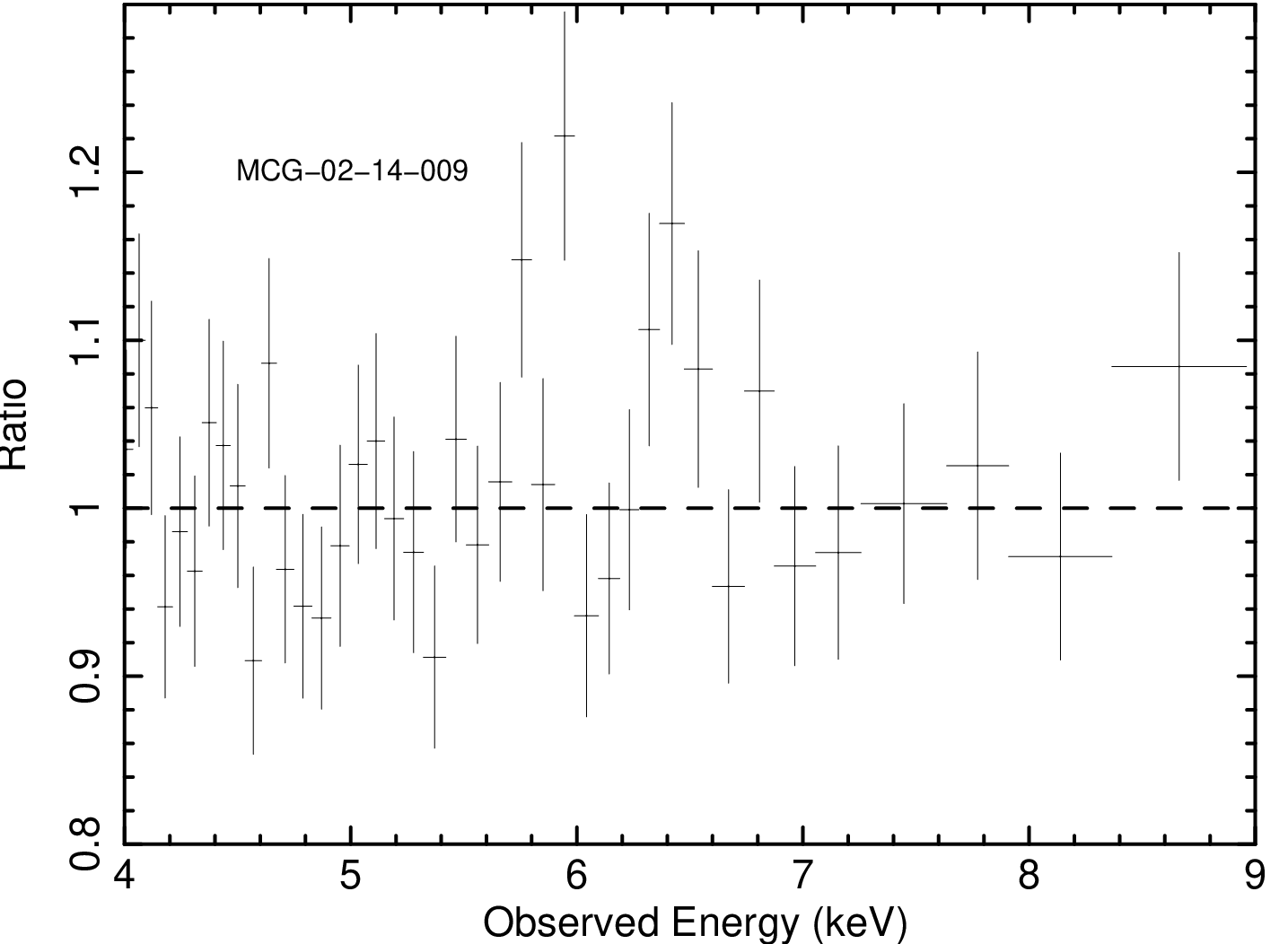}}
\rotatebox{-90}{\includegraphics[width=5cm]{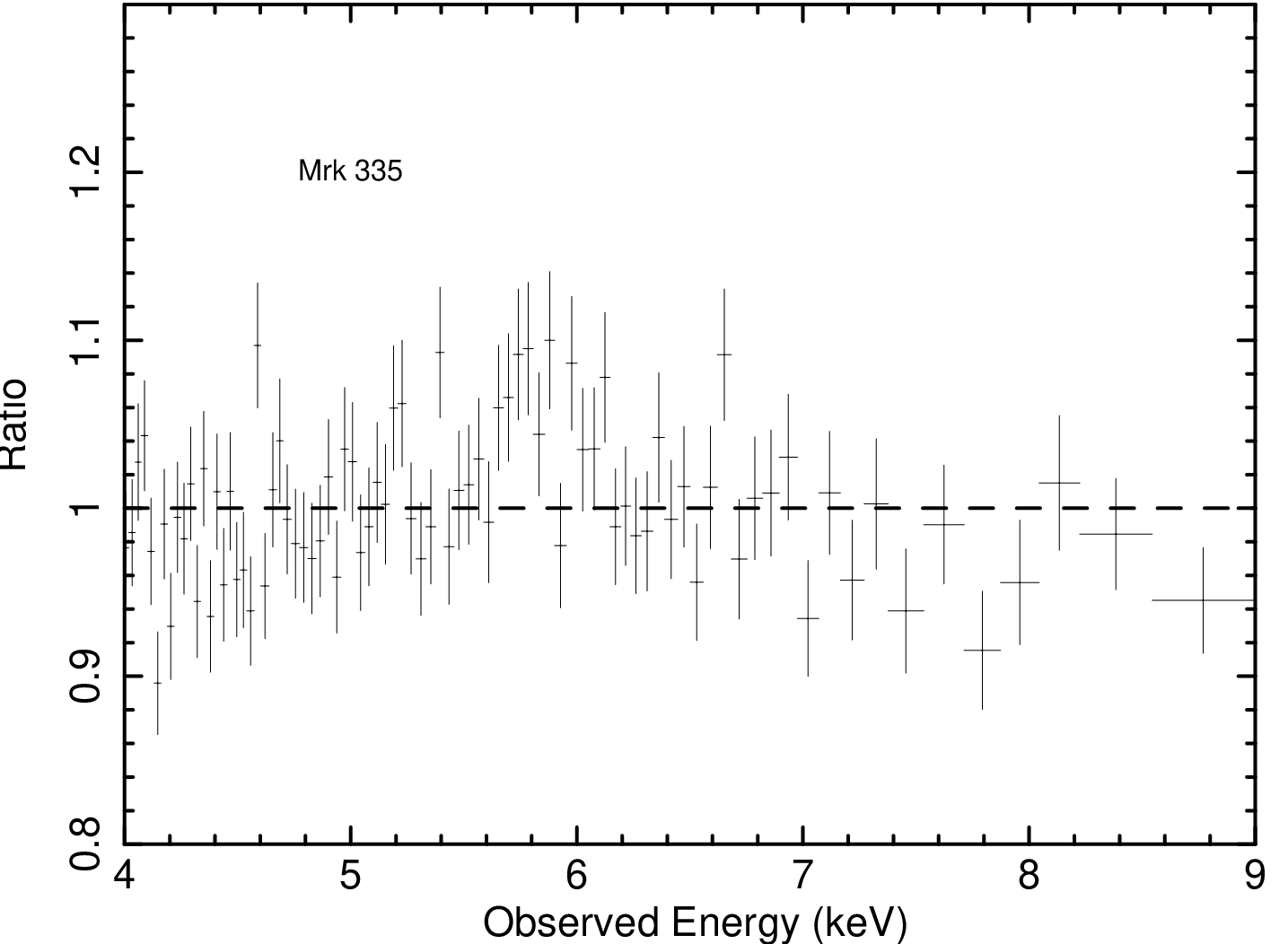}}
\rotatebox{-90}{\includegraphics[width=5cm]{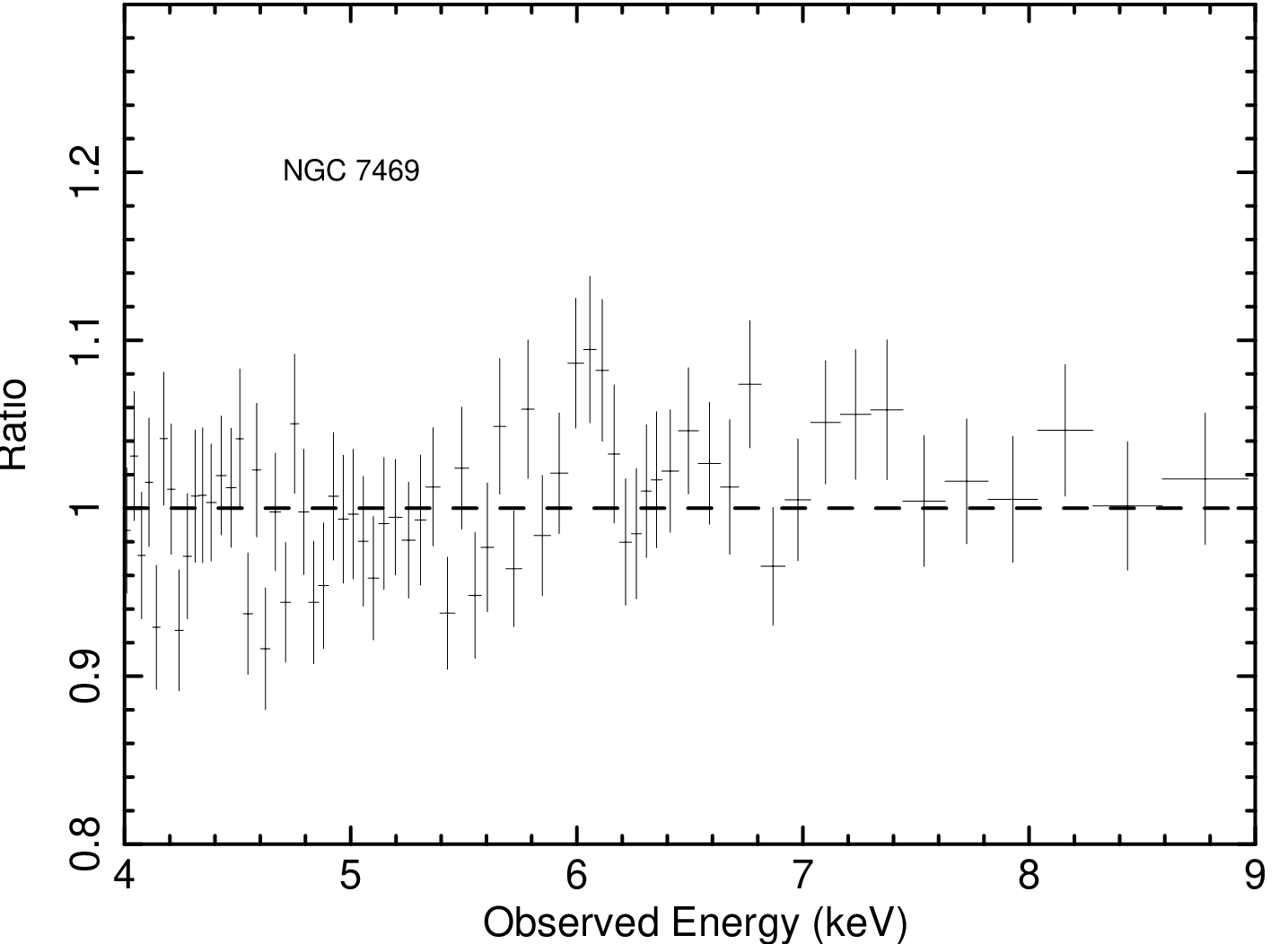}}
\rotatebox{-90}{\includegraphics[width=5cm]{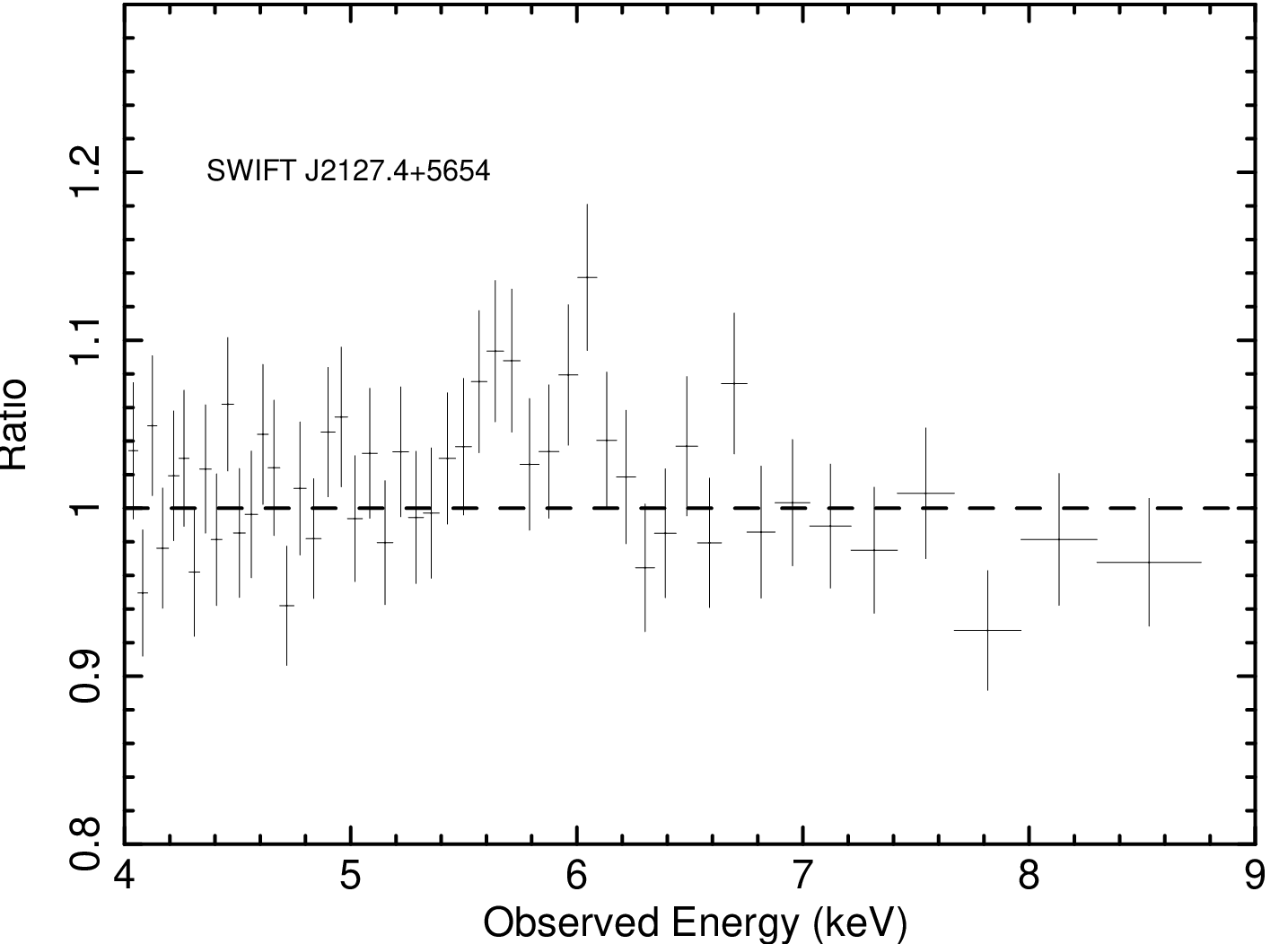}}
\caption{Ratio plots of Model B (i.e. without including a broad Fe\,K$\alpha$) revealing excesses at energies red-ward of 6.4\,keV in some objects. Model B consisting of \textsc{powerlaw} + \textsc{compTT} (where required) + unblurred \textsc{reflionx} + narrow ionized lines as required.}
\label{fig:modelB}
\end{figure*}

In an attempt to build a more self-consistent model the \textsc{pexrav} and narrow 6.4\,keV core are replaced with the \textsc{reflionx} ionized reflection model from Ross \& Fabian (2005) which includes emission lines in addition to the reflection continuum. Model B provides fits of a reasonable quality to all six spectra, however it is not expected to improve upon Model A since it doesn't include a broad Gaussian to parametrize the red-wing in an attempt to retain self-consistency and begin the construction of a more physically motivated model. Model B therefore acts as a 'null hypothesis' model in that all emission and reflection originates from distant matter. In most cases the fit statistic is somewhat worse than in Model A, except for Fairall 9, in which the quality of the fit is actually improved to $\chi^{2}_{\nu}=864.7/834$. The \textsc{compTT} model is still used here to model the soft excess in the four objects requiring it, however the \textsc{compTT} parameters are not quoted in Table \ref{tab:ModelB} (similarly in future tables) since the parameters obtained are consistent with those in Model A. 

The narrow ionized emission lines at 6.7\,keV and 6.97\,keV found in Model A remain in Model B and following the removal of the broad component the significance of these lines is increased. In the case of SWIFT J2127.4+5654, for which there are no statistically significant ionized emission lines in Model A, there are small excesses at these energies, indeed suggesting the possible presence of these lines. Emission lines of high significance in Model A are also similarly significant in Model B e.g. the Fe\,{\rm XXVI} in Fairall 9 is particularly prominent in both A \& B. This is contrary to the analysis of Fairall 9 by Schmoll et al. (2009) in which they did not detect the Fe\,{\rm XXVI} line in the same {\sl Suzaku} data, but they did however detect the He-like Fe\,{\rm XXV} emission line which is also seen here in Models A and B. 

All objects show some indication of at least a small red-wing (see Figure \ref{fig:modelB}), noting that in MCG-02-14-009 this feature is not particularly strong and in all objects the red wing does not extend below $\sim5$\,keV. A smaller than expected excess at energies red-ward of 6.4\,keV is seen in Fairall 9, given the statistical significance of the broad Gaussian in Model A. This could be due in part to a relatively high Fe abundance as measured by the Fe/Solar parameter within \textsc{reflionx} at approximately twice the solar value. Some objects in the sample indicate a super-solar Fe abundance according to the \textsc{reflionx} component (see Table \ref{tab:ModelB}), however the true abundance may in fact be lower than this measured value if part of the Fe\,K$\alpha$ emission additionally arises from Compton-thin matter such as the BLR or NLR. A summary of the results obtained with Model B can be found in Table \ref{tab:ModelB}.

\subsection{Model C -- Laor profile}
The presence of emission from close-in to the black hole has been indicated in the previous two models. In accordance with this, emission from the Fe\,K line region can be modelled with a relativistic component representing line emission from the accretion disc in addition to any appropriate emission from distant material (see Figure \ref{fig:modelB}). Model C advances upon Model B by remodelling the Fe\,K region with the addition of a \textsc{laor} line profile operating under the assumption of a maximally spinning black hole (Laor 1991). The line energy was restricted to 6.4--6.97\,keV in the rest frame, with emissivity, inclination and the inner radius of emission allowed to vary. The outer radius of emission was fixed at 400\,r$_{\rm g}$ throughout. In the cases where the line emission reaches its lower limit it is fixed at 6.4\,keV in the rest frame. Whilst giving a more feasible interpretation of the broad emission in the Fe\,K region it does not provide us with the most physically accurate representation given the assumption of a maximally rotating central black hole with $a=0.998$ within the \textsc{laor} model. Not all of the objects in the sample will feature a maximally rotating black hole, indeed Fairall 9 and SWIFT J2127.4+5654 have previously been found to have intermediate spin values (see Schmoll et al. 2009; Miniutti et al. 2009). The presence of narrow ionized emission lines was reassessed after accounting for the blue-wing of the \textsc{laor} profile. Employment of this model therefore seeks to further parametrize the broad emission in the Fe\,K region, providing suitable and plausible parameters for use in the later \textsc{kerrdisk} models, whilst giving us an indication of the extent to which the spin of the central black hole has an effect upon the observed spectrum. 

\begin{table*}
\caption{Fit parameters from Model C to {\sl Suzaku} XIS, HXD and BAT data from {\sl Swift}. Line energies are quoted in the rest frame. * denotes a frozen parameter, for cases where the emissivity index is fixed at $q=3$ it is unconstrained. The improvement $\triangle\chi^{2}$ in the fit with the introduction of a \textsc{laor} profile is noted in comparison with the purely distant reflection as present in this model. Note that the Fe\,{\rm XXV} emission line is no longer required in Ark 120, Fairall 9 and SWIFT J2127.4+5654. $^{a}$ \textsc{powerlaw} normalization given in units $(10^{-2}\,{\rm ph\,keV^{-1}\,cm^{-2}\,s^{-1}})$. $^{b}$ \textsc{reflionx} normalisation given in units $10^{-5}$.}
\begin{tabular}{l c c c c c c}
\hline
& Ark 120 & Fairall 9 & MCG-02-14-009 & Mrk 335 & NGC 7469 & SWIFT J2127.4+5654 \\
\hline
$\Gamma$ & $1.91^{+0.03}_{-0.03}$ & $1.84^{+0.02}_{-0.03}$ & $1.86^{+0.03}_{-0.03}$ & $2.01^{+0.02}_{-0.02}$ & $1.78^{+0.01}_{-0.01}$ & $2.13^{+0.03}_{-0.03}$ \\ [0.8ex]
Norm$^{a}$ & $0.97^{+0.03}_{-0.05}$ & $0.64^{+0.03}_{-0.02}$ & $0.12^{+0.04}_{-0.03}$ & $0.56^{+0.02}_{-0.02}$ & $0.55^{+0.04}_{-0.06}$ & $1.54^{+0.04}_{-0.05}$ \\
\hline
\multicolumn{7}{c}{Laor Profile} \\
Line E (keV) & 6.4* & $<6.56$ & 6.4* & 6.4* & $<6.44$ & 6.4* \\ [0.8ex]
Eqw (eV) & $85^{+21}_{-17}$ & $79^{+21}_{-22}$ & $119^{+58}_{-52}$ & $126^{+33}_{-35}$ & $60^{+26}_{-22}$ & $140^{+41}_{-41}$ \\ [0.8ex]
q & $3.0^{+1.6}_{-1.0}$ & $3.5^{+2.9}_{-0.9}$ & 3.0* & $3.9^{+1.9}_{-1.0}$ & 3.0* & $2.5^{+0.7}_{-0.6}$ \\ [0.8ex]
R$_{\rm in}$ ($GM/c^{2}$) & $25^{+19}_{-7}$ & $16^{+17}_{-6}$ & $>13$ & $32^{+12}_{-16}$ & $81^{+82}_{-37}$ & $<33$ \\ [0.8ex]
$i^{\circ}$ & $34^{+7}_{-5}$ & $34^{+5}_{-3}$ & $30^{+17}_{-9}$ & $38^{+10}_{-8}$ & $24^{+12}_{-8}$ & $46^{+9}_{-9}$ \\ [0.8ex]
Flux $(10^{-5}\,{\rm ph\,cm^{-2}\,s^{-1}})$ & $3.85^{+0.95}_{-0.77}$ & $1.79^{+0.50}_{-0.50}$ & $0.59^{+0.30}_{-0.26}$ & $2.28^{+0.62}_{-0.64}$ & $2.56^{+1.02}_{-0.84}$ & $5.39^{+1.58}_{-1.58}$ \\ [0.8ex]
$\triangle\chi^{2}$ & 24 & 33 & 14 & 37 & 20 & 34 \\
\hline
$\xi$ & $<19$ & $<21$ & $<15$ & $33^{+18}_{-9}$ & $<13$ & $<15$ \\ [0.8ex]
Fe/Solar & 1.0* & $1.7^{+0.3}_{-0.3}$ & $0.6^{+0.6}_{-0.4}$ & $2.0^{+0.9}_{-0.8}$ & $<1.4$ & $0.4^{+0.1}_{-0.1}$ \\ [0.8ex]
Norm$^{b}$ & $1.06^{+0.15}_{-0.57}$ & $0.97^{+0.23}_{-0.32}$ & $0.23^{+0.07}_{-0.11} $ & $0.11^{+0.06}_{-0.05}$ & $1.07^{+0.08}_{-0.48}$ & $1.96^{+0.24}_{-0.92}$ \\
\hline
Fe\,{\rm XXV} Line (keV) & -- & -- & -- & $6.65^{+0.09}_{-0.07}$ & -- & -- \\ [0.8ex]
$EW$ (eV) & -- & -- & -- & $10^{+8}_{-8}$ & -- & -- \\ [0.8ex]
Flux $(10^{-5}\,{\rm ph\,cm^{-2}\,s^{-1}})$ & -- & -- & -- & $0.17^{+0.13}_{-0.13}$ & -- & -- \\ [0.8ex]
$\triangle\chi^{2}$ & -- & -- & -- & 2 & -- & -- \\
\hline
Fe\,{\rm XXVI} Line (keV) & $6.96^{+0.04}_{-0.04}$ & $6.99^{+0.03}_{-0.03}$ & $6.94^{+0.04}_{-0.06}$ & $7.02^{+0.05}_{-0.06}$ & -- & -- \\ [0.8ex]
$EW$ (eV) & $33^{+12}_{-19}$ & $27^{-9}_{-15}$ & $42^{+23}_{-24}$ & $19^{+9}_{-9}$ & -- & -- \\ [0.8ex]
Flux $(10^{-5}\,{\rm ph\,cm^{-2}\,s^{-1}})$ & $0.98^{+0.34}_{-0.57}$ & $0.63^{+0.21}_{-0.35}$ & $0.18^{+0.10}_{-0.10}$ & $0.25^{+0.12}_{-0.12}$ & -- & -- \\ [0.8ex]
$\triangle\chi^{2}$ & 10 & 13 & 5 & 11 & -- & -- \\
\hline
BAT const & $1.10^{+0.12}_{-0.12}$ & $0.86^{+0.10}_{-0.10}$ & -- & $1.11^{+0.25}_{-0.25}$ & $1.08^{+0.11}_{-0.09}$ & $0.73^{+0.10}_{-0.10}$ \\
\hline
$\chi^{2}_{\nu}$ & 721.6/648 & 857.7/831 & 609.6/540 & 806.3/719 & 820.3/808 & 855.8/867 \\
\hline
\end{tabular}
\label{tab:ModelC}
\end{table*}

The fit to all objects is improved over the purely distant emission in Model B. The 6.97\,keV line is found to be present in all but NGC 7469 and SWIFT J2127.4+5654 whilst the 6.7\,keV line is only found in Mrk 335 (although relatively weak). Model C improves the fit to MCG-02-14-009 the least with only $\triangle\chi^{2}\sim14$ for three additional free parameters whereas the introduction of a \textsc{laor} profile offers a significant improvement for most other objects, particularly Fairall 9, Mrk 335 and SWIFT J2127.4+5654 ($\triangle\chi^{2}\sim33$, 37 \& 34 respectively), see Table \ref{tab:ModelC}.

None of the objects in the sample {\sl require} emission from within 6\,$GM/c^{2}$ indicating that a rotating black hole, whilst possible, is not required to model the spectra. According to the fit parameters obtained with \textsc{laor}, the inner radius of emission lies at tens of $r_{\rm g}$ for all six AGN. In most cases the accretion disc is unlikely to be truncated at these distances from the black hole and it is likely that these values arise from the assumption of a maximally rotating black hole within the \textsc{laor} model. The emissivity indicies also indicate that a high concentration of emission from very close to the black hole is absent in the spectra from these AGN. An emissivity index of $q>5$ would suggest that the emission from the accretion disc is very centrally concentrated (e.g. Miniutti et al. 2003), while for objects such as Mrk 335 and Fairall 9 this is within the error bars, one would also expect the inner radius of emission to be much closer to the black hole if this interpretation were suitable for the spectra. For the cases where the emissivity index cannot be constrained, it is fixed to $q=3.0$ which is consistent with the other objects in the sample. 

Mrk 335 appears to have one of the strongest relativistic lines in the sample with a \textsc{laor} profile equivalent width of $EW=126^{+33}_{-35}$\,eV, and an improvement of $\triangle\chi^{2}\sim37$. However this is not particularly strong in comparison with previous studies of this AGN, for example Longinotti et al. (2007) find an $EW=320^{+170}_{-100}$\,eV at a line energy of $E=6.93^{+0.77}_{-0.27}$\,keV in a 40\,ks {\sl XMM-Newton} observation of Mrk 335 in 2000, also using a \textsc{laor} profile. The width of the iron line emission from the disc and its suggested high ionization state could be due to the lack of a Fe\,{\rm XXVI} line in the Longinotti spectra (possibly due to low S/N), which has been accounted for in this analysis and is noted by O'Neill et al. (2007) in an analysis of the same {\sl XMM-Newton} observation used here. Also in agreement with our results, O'Neill et al. found the equivalent width of the broad line in Mrk 335 to be $EW=115^{+14}_{-14}$\,eV in comparison with $EW=126^{+33}_{-35}$\,eV for the \textsc{laor} profile and $EW=113^{+46}_{-52}$\,eV for the broad Gaussian employed in Model A.

The results obtained here are consistent with a recent analysis of Ark 120 by Nardini et al. (2010, submitted) who initially fit the features in the Fe\,K region with a \textsc{laor} profile. In the Nardini et al. (2010) analysis, both the inclination of the accretion disk and emissivity are frozen at typical values of $i=40^{\circ}$ and $q=3.0$ respectively, finding an inner radius of emission of $r_{\rm in}=13^{+19}_{-7}\,r_{\rm g}$. These values are consistent with those obtained here with Model C, finding $i^{\circ}=34^{+7}_{-5}$, $q=3.0^{+1.6}_{-1.0}$ and $r_{\rm in}=25^{+19}_{-7}\,r_{\rm g}$. Also detected is the Fe\,{\rm XXVI} narrow ionized emission line, again consistent with Nardini et al. (2010).

As in Model A, there are no narrow ionized emission lines found in SWIFT J2127.4+5654 corresponding to Fe\,{\rm XXV} and Fe\,{\rm XXVI}. This implies that the excesses modelled as narrow components in Model B may instead be due to relativistically broadened Fe\,K${\alpha}$ emission. When modelled as a single \textsc{laor} profile all significant excesses in the Fe K band are removed with the profile centroid rest frame energy at 6.4\,keV and the blue-wing peaking at $\sim$\,6.7\,keV in the rest frame. Consequently the observed excess at 6.7\,keV observed in Model A and more significantly in Model B is more convincingly described as a blue-wing. Additionally, the observed intermediate feature in MCG-02-14-009 is modelled well by the blue-wing of the \textsc{laor} profile which occurs at $\sim6.58$\,keV, prompting the removal of the previously employed narrow Gaussian.

\subsection{Model D -- Kerrdisk profile}
Replacing the \textsc{laor} model with a \textsc{kerrdisk} (Brenneman \& Reynolds 2006) line profile further allows a more physically motivated fit to the data. The \textsc{kerrdisk} model allows the spin parameter to be varied between $0<a<0.998$ with the aim of determining the extent to which we can rule in or out a non-rotating or maximally spinning black hole. Similarly to Model C, the outer radius of the disc is fixed at 400\,r$_{\rm ms}$, while assuming a disc of uniform emissivity. Throughout the fits using Model D the inner radius of emission is assumed to extend down to the innermost stable circular orbit (r$_{\rm ISCO}$). The line energy is confined to 6.4--6.97\,keV in the rest frame, being frozen at 6.4\,keV if it reaches its lower limit. 

\begin{table*}
\caption{Fit parameters from Model D to {\sl Suzaku} XIS, HXD and BAT data from {\sl Swift}. Line energies are quoted in the rest frame. * denotes a frozen parameter. $^{a}$ \textsc{powerlaw} normalization given in units $(10^{-2}\,{\rm ph\,keV^{-1}\,cm^{-2}\,s^{-1}})$. $^{b}$ \textsc{reflionx} normalisation given in units $10^{-5}$.}
\begin{tabular}{l c c c c c c}
\hline
& Ark 120 & Fairall 9 & MCG-02-14-009 & Mrk 335 & NGC 7469 & SWIFT J2127.4+5654 \\
\hline
$\Gamma$ & $1.90^{+0.03}_{-0.03}$ & $1.84^{+0.02}_{-0.03}$ & $1.86^{+0.01}_{-0.01}$ & $2.02^{+0.03}_{-0.02}$ & $1.78^{+0.01}_{-0.01}$ & $2.13^{+0.04}_{-0.03}$ \\ [0.8ex]
Norm$^{a}$ & $0.97^{+0.04}_{-0.05}$ & $0.64^{+0.01}_{-0.04}$ & $0.13^{+0.04}_{-0.04}$ & $0.56^{+0.01}_{-0.01}$ & $0.55^{+0.04}_{-0.06}$ & $1.55^{+0.05}_{-0.05}$ \\
\hline
\multicolumn{7}{c}{Kerrdisk Profile} \\
Line E (keV) & 6.4* & 6.4* & $6.47^{+0.03}_{-0.03}$ & 6.4* & 6.4* & 6.4* \\ [0.8ex]
Eqw (eV) & $95^{+32}_{-26}$ & $63^{+36}_{-19}$ & $142^{+47}_{-46}$ & $146^{+39}_{-39}$ & $91^{+9}_{-8}$ & $178^{+82}_{-69}$ \\ [0.8ex]
$q$ & $2.3^{+0.4}_{-0.3}$ & $2.7^{+0.7}_{-0.4}$ & $2.0^{+0.4}_{-0.4}$ & $2.6^{+0.5}_{-0.3}$ & $1.7^{+0.4}_{-0.6}$ & $2.6^{+1.0}_{-0.4}$ \\ [0.8ex]
$a$ & $<0.94$ & $0.44^{+0.04}_{-0.11}$ & $<0.88$ & $0.70^{+0.12}_{-0.01}$ & $0.69^{+0.09}_{-0.09}$ & $0.70^{+0.10}_{-0.14}$ \\ [0.8ex]
$i^{\circ}$ & $33^{+2}_{-17}$ & $38^{+8}_{-5}$ & $24^{+10}_{-9}$ & $38^{+2}_{-2}$ & $23^{+15}_{-7}$ & $43^{+5}_{-10}$ \\ [0.8ex]
Flux $(10^{-5}\,{\rm ph\,cm^{-2}\,s^{-1}})$ & $3.62^{+1.22}_{-0.99}$ & $1.78^{+1.02}_{-0.54}$ & $6.64^{+2.20}_{-2.15}$ & $2.52^{+0.67}_{-0.67}$ & $2.23^{+0.22}_{-0.20}$ & $6.35^{+2.91}_{-2.46}$ \\ [0.8ex]
$\triangle\chi^{2}$ (kerrdisk) & 23 & 30 & 12 & 40 & 25 & 37 \\ [0.8ex]
$\triangle\chi^{2}$ (zero spin) & 1 & 4 & 0 & 4 & 10 & 7 \\
\hline
$\xi$ & $<16$ & $<17$ & $<16$ & $41^{+7}_{-16}$ & $<11$ & $<13$ \\ [0.8ex]
Fe/Solar & 1.0* & $1.6^{+0.3}_{-0.4}$ & $0.8^{+0.6}_{-0.4}$ & $1.9^{+0.9}_{-0.6}$ & $0.8^{+0.4}_{-0.2}$ & $0.3^{+0.2}_{-0.2}$ \\ [0.8ex]
Norm$^{b}$ & $0.97^{+0.50}_{-0.64}$ & $1.09^{+0.09}_{-0.29}$ & $0.25^{+0.07}_{-0.16}$ & $0.07^{+0.07}_{-0.01}$ & $0.69^{+0.22}_{-0.15}$ & $2.12^{+0.63}_{-0.83}$ \\
\hline
Fe\,{\rm XXV} Line (keV) & -- & -- & -- & $6.67^{+0.06}_{-0.05}$ & -- & -- \\ [0.8ex]
$EW$ (eV) & -- & -- & -- & $16^{+8}_{-8}$ & -- & -- \\ [0.8ex]
Flux $(10^{-5}\,{\rm ph\,cm^{-2}\,s^{-1}})$ & -- & -- & -- & $0.26^{+0.13}_{-0.13}$ & -- & -- \\ [0.8ex]
$\triangle\chi^{2}$ & -- & -- & -- & 4 & -- & -- \\ 
\hline 
Fe\,{\rm XXVI} Line (keV) & $6.99^{+0.04}_{-0.06}$ & $6.98^{+0.04}_{-0.03}$ & $6.94^{+0.08}_{-0.09}$ & $7.02^{+0.04}_{_0.05}$ & -- & -- \\ [0.8ex]
$EW$ (eV) & $22^{+23}_{-15}$ & $30^{+8}_{-10}$ & $44^{+26}_{-38}$ & $22^{+10}_{-10}$ & -- & -- \\ [0.8ex]
Flux $(10^{-5}\,{\rm ph\,cm^{-2}\,s^{-1}})$ & $0.65^{+0.68}_{-0.44}$ & $0.70^{+0.19}_{-0.24}$ & $0.18^{+0.11}_{-0.16}$ & $0.28^{+0.12}_{0.12}$ & -- & -- \\ [0.8ex]
$\triangle\chi^{2}$ & 7 & 12 & 9 & 14 & -- & -- \\
\hline
BAT const & $1.05^{+0.12}_{-0.12}$ & $0.85^{+0.10}_{-0.10}$ & -- & $1.05^{+0.23}_{-0.23}$ & $1.06^{+0.11}_{-0.09}$ & $0.67^{+0.09}_{-0.09}$ \\
\hline
$\chi^{2}_{\nu}$ & 724.4/648 & 861.2/832 & 611.6/538 & 803.0/719 & 815.5/808 & 852.3/867 \\
\hline
\end{tabular}
\label{tab:ModelD}
\end{table*}

\begin{figure*}
\rotatebox{-90}{\includegraphics[width=0.33\textwidth]{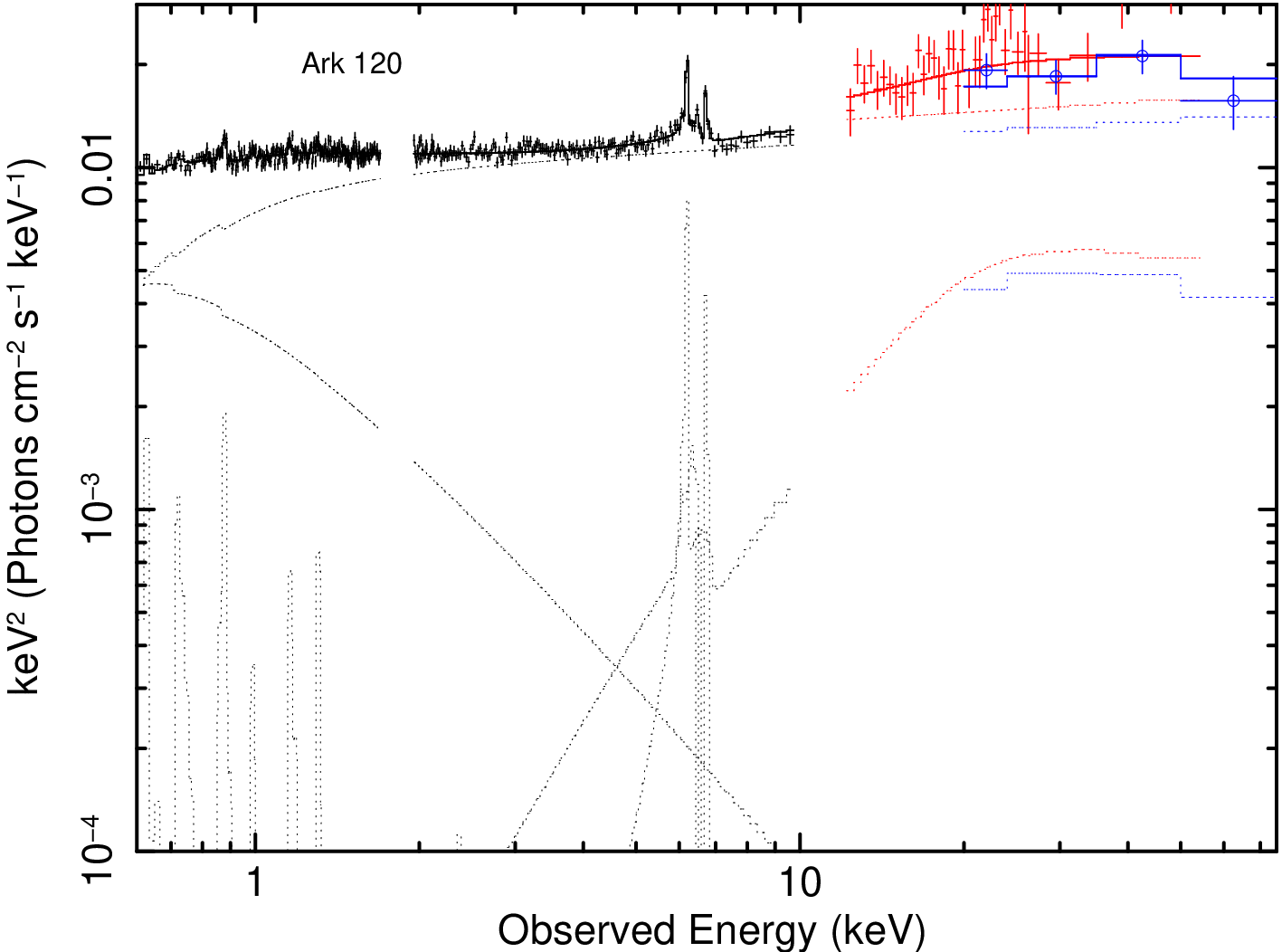}}
\rotatebox{-90}{\includegraphics[width=0.33\textwidth]{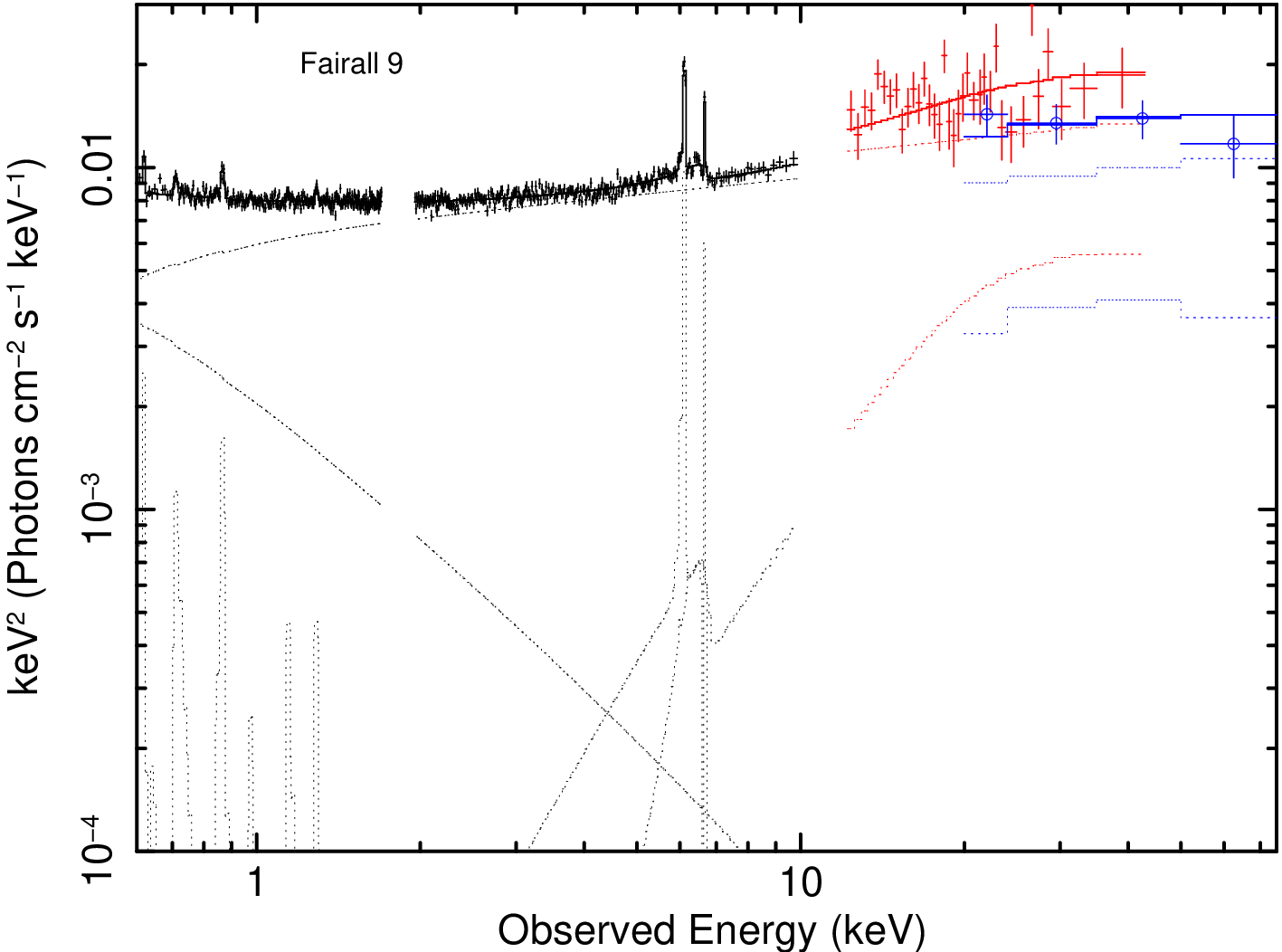}}
\rotatebox{-90}{\includegraphics[width=0.33\textwidth]{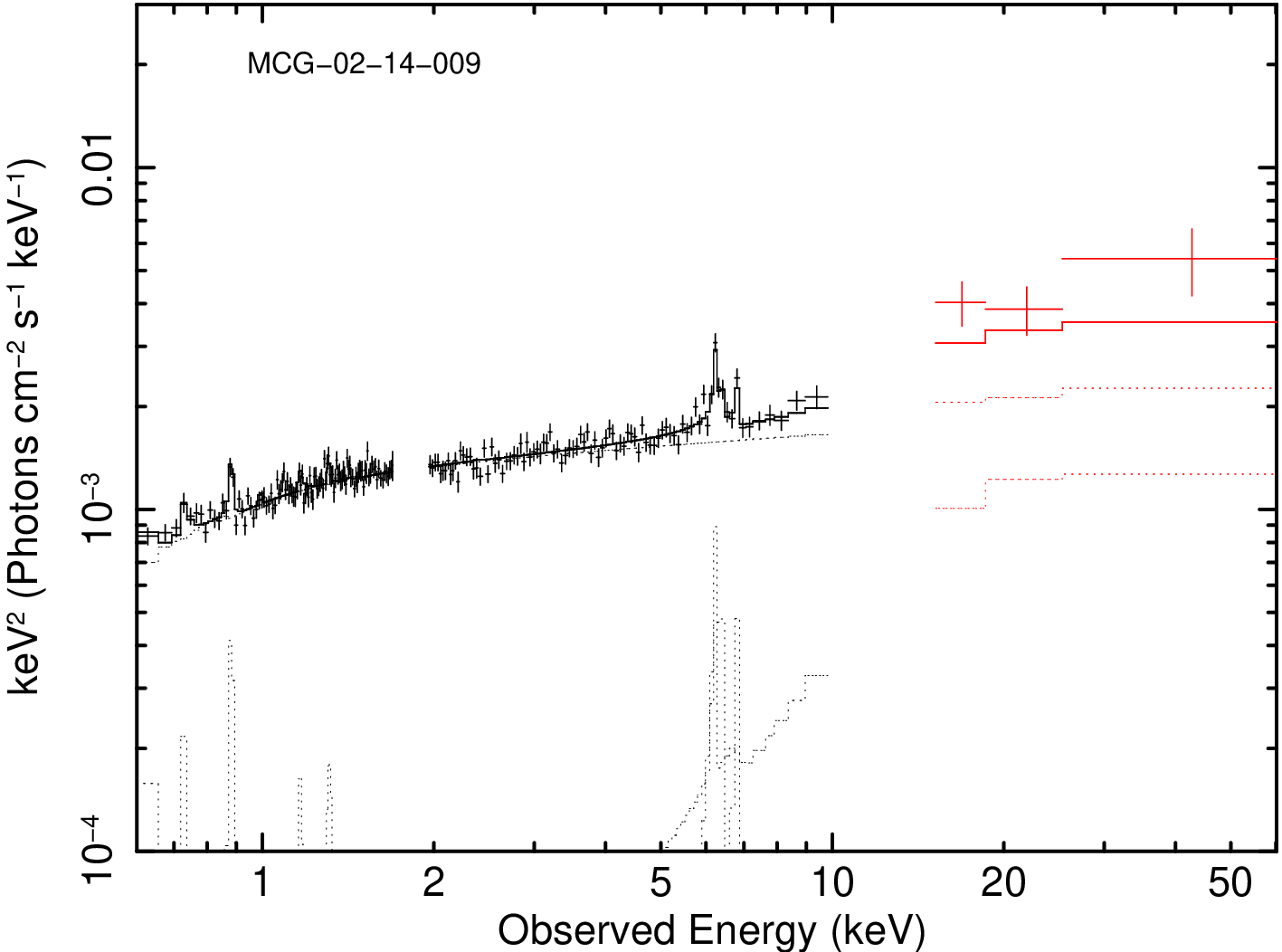}}
\rotatebox{-90}{\includegraphics[width=0.33\textwidth]{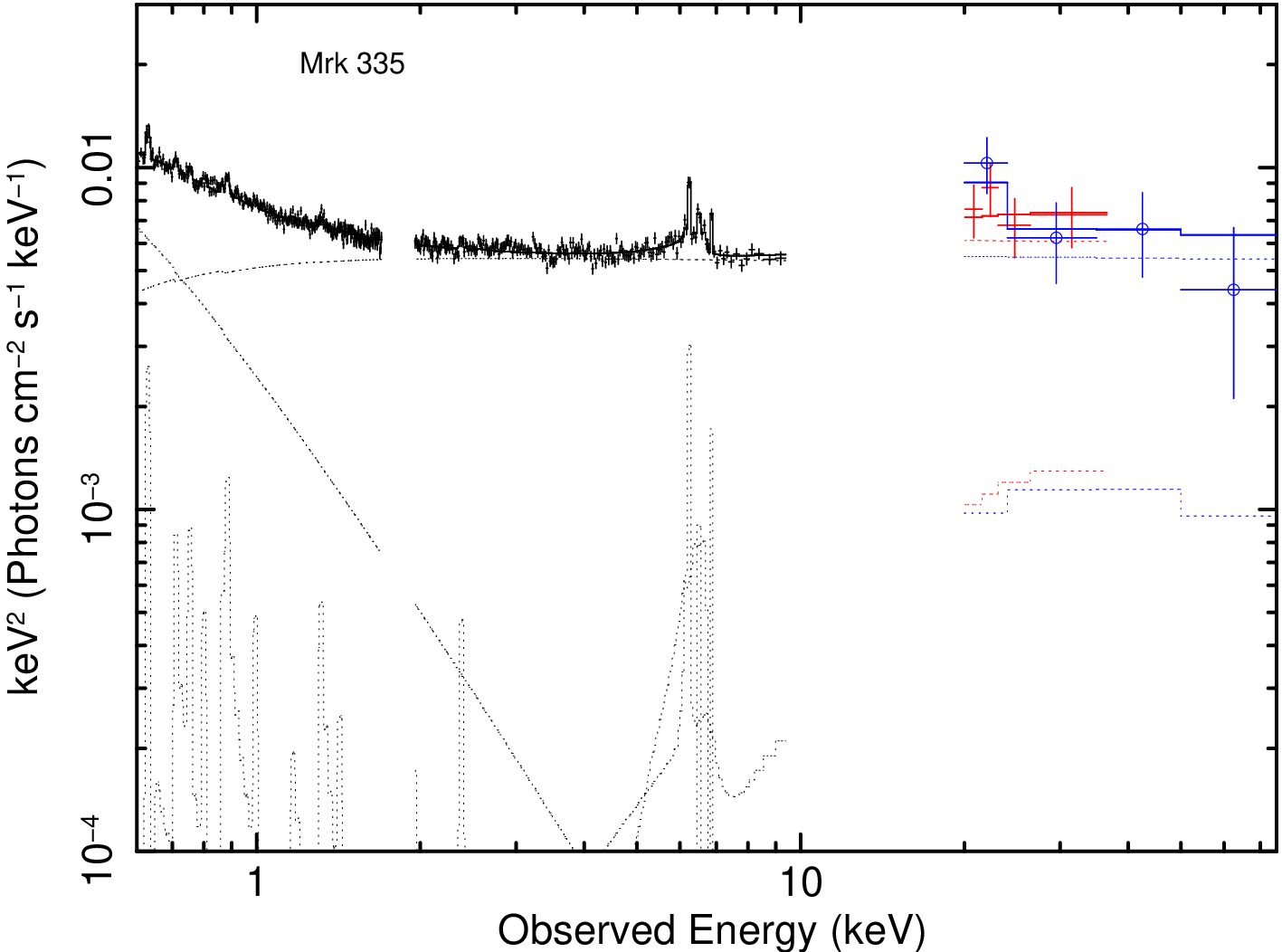}}
\rotatebox{-90}{\includegraphics[width=0.33\textwidth]{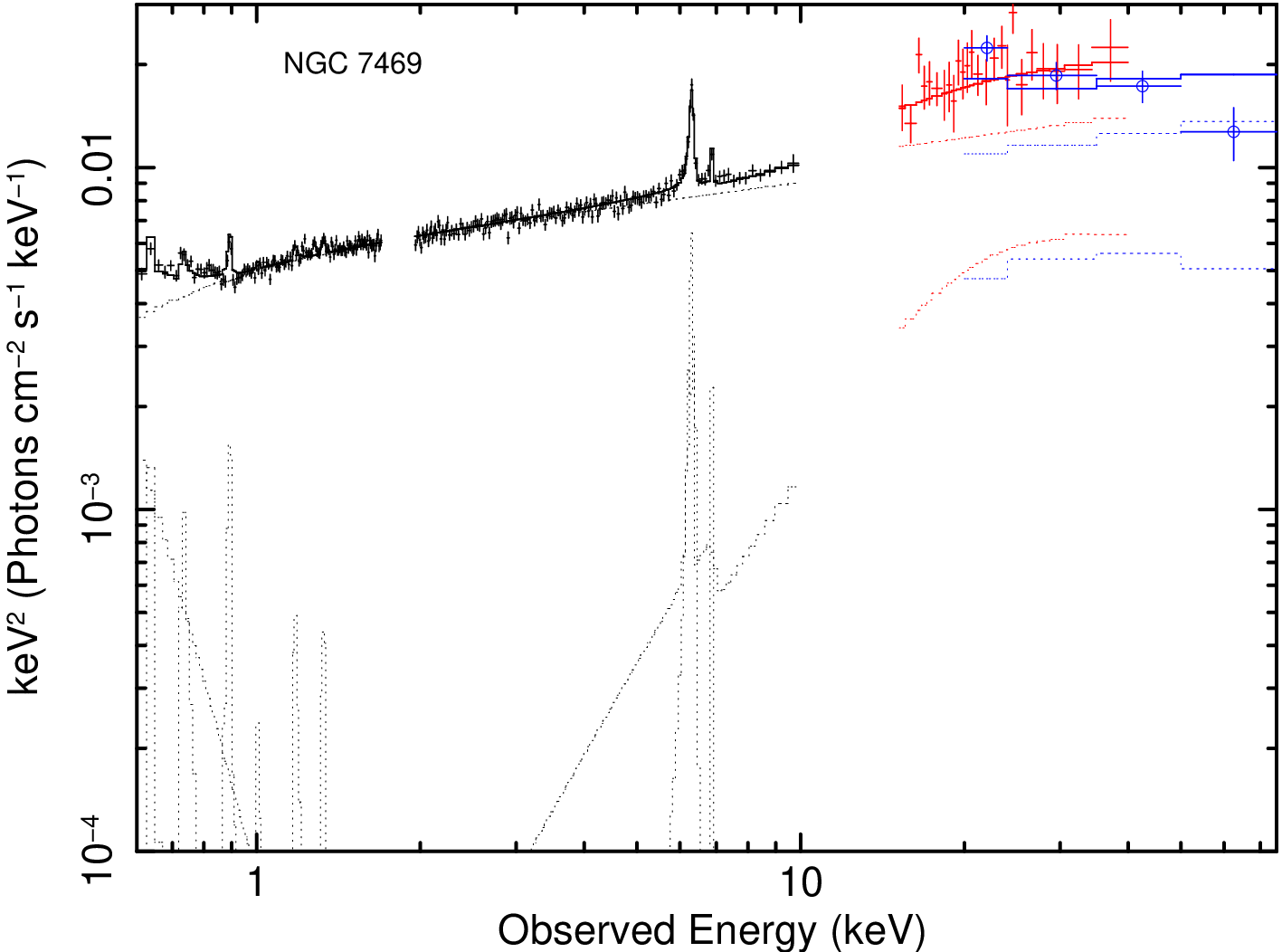}}
\rotatebox{-90}{\includegraphics[width=0.33\textwidth]{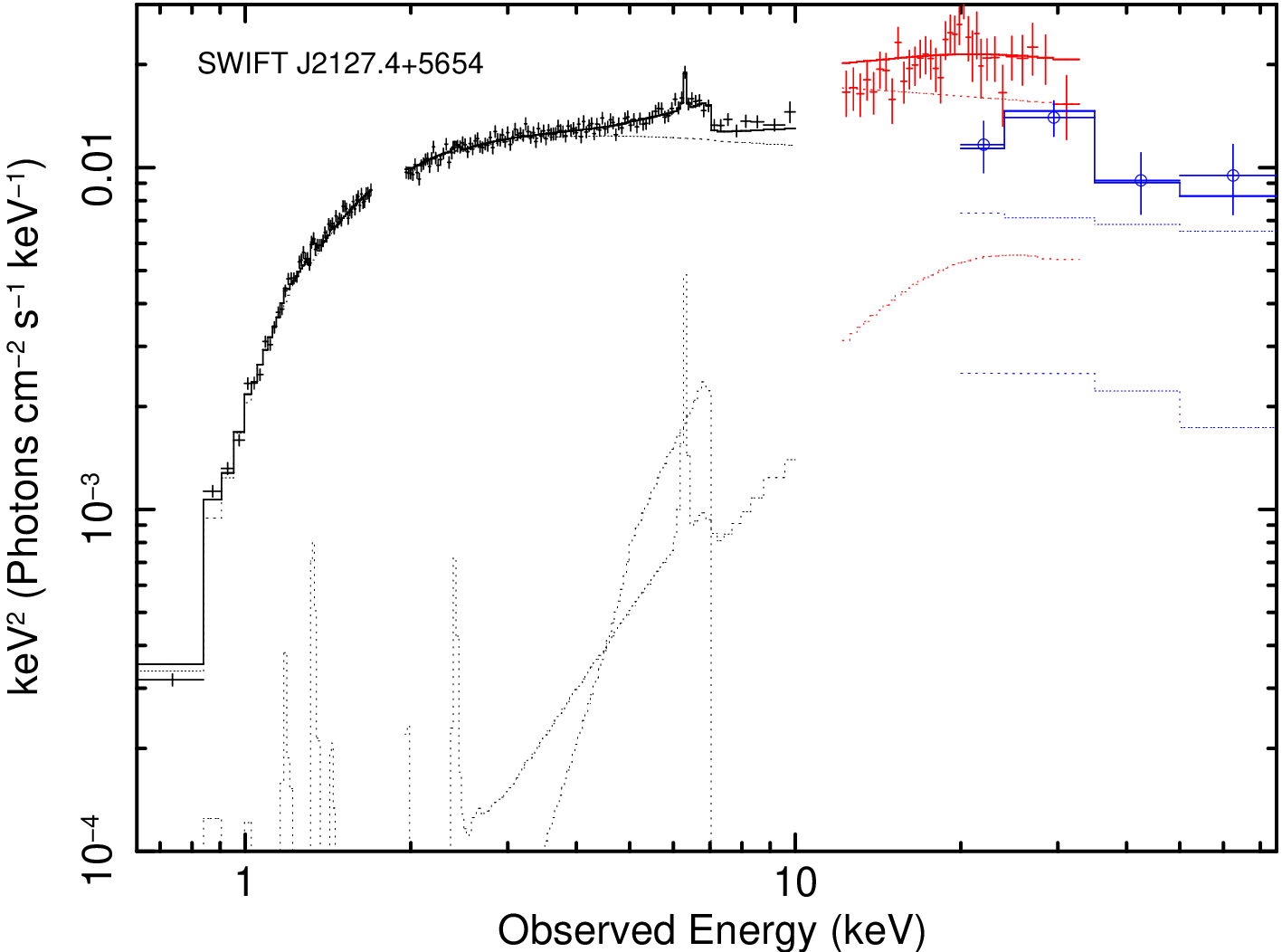}}
\caption{$\nu$F$\nu$ plots of Model D indicating the strength of any soft excess and relativistic line emission from the \textsc{kerrdisk} models. Consisting of \textsc{powerlaw} + \textsc{compTT} (where required) + unblurred \textsc{reflionx} + \textsc{kerrdisk} + narrow ionized lines as required. XIS data is in black, HXD in red and BAT data is represented by blue circles.}
\label{fig:ModelD_eeuf}
\end{figure*}

Model D provides good fits to all six objects, producing the best fitting physically motivated model for Mrk 335 and NGC 7469 prior to considering a blurred reflection component from the inner regions of the accretion disc (i.e. Models E \& F), see Table \ref{tab:ModelD} and Figure \ref{fig:ModelD_eeuf}. The presence of ionized emission lines is entirely consistent with Model C, adding weight to the probability that these lines are present in the spectra. Further to this, the inclinations of the accretion discs to the observer are all comparable to those obtained in Model C, as are the equivalent widths of the relativistic line profiles. Given that the spin of the central black hole is a free parameter and the assumptions within this model, the emissivity indicies are expected to vary from those obtained with the \textsc{laor} profile since the measured emissivity is degenerate to some extent with the inner radius of emission. Therefore as the spin parameter varies so does r$_{\rm ISCO}$, consequently affecting the measured $q$. 

The results obtained from the spin parameter $a$ for these six objects suggest that a maximally spinning central black hole can be ruled out at the 90\% confidence level, however for some objects such as Ark 120 and MCG-02-14-009 the spin is unconstrained and only an upper limit can be measured ($a<0.94$ and $a<0.88$, suggesting that emission does not occur within 2.02\,$r_{\rm g}$ and 2.45\,$r_{\rm g}$ respectively, consistent with the Nardini et al. 2010 analysis of Ark 120).

Here for Fairall 9 the best fitting model gives a measured spin value of $a=0.44^{+0.04}_{-0.11}$. A previous spin constraint for Fairall 9 by Schmoll et al. (2009) found $a=0.65^{+0.05}_{-0.05}$ using a blurred reflector model (\textsc{kerrconv}, Brenneman \& Reynolds (2006), convolved with \textsc{reflionx}). The measurement here also gives an intermediate value, however it is only consistent with their findings when they ignore the spectra below 2\,keV to ensure that the soft excess is not the component driving the main part of the fit. Schmoll et al. (2009) quote a worse spin constraint when these conditions are upheld $a=0.5^{+0.1}_{-0.3}$. Within error bars at the 90\% level these results are consistent with the value of the spin parameter found here. However, here we find the emissivity index is constrained to $q=2.7^{+0.7}_{-0.4}$ whereas Schmoll et al. only constrain this to $q>4.9$. 

Mrk 335 has been noted previously as an object with a broad relativistic Fe\,K line (Gondoin et al. 2002). Here we measure an intermediate spin $a=0.70^{+0.12}_{-0.01}$, ruling out maximally spinning and non-rotating black holes only at the 95\% confidence level. The emissivity index is measured at a moderate value of $q=2.6^{+0.5}_{-0.3}$ in agreement with previous studies of this object, for example by Longinotti et al. (2007). In agreement with Model C, the \textsc{kerrdisk} component also features a relatively broad equivalent width $EW=146^{+39}_{-39}$ with an improvement of $\triangle\chi^{2}\sim40$ (for four additional free parameters) over purely distant emission. 

In NGC 7469 we measure a broad Fe line featuring a low emissivity index of $q=1.7^{+0.4}_{-0.6}$, an inclination of $i^{\circ}=23^{+15}_{-7}$ and $EW=91^{+9}_{-8}$\,eV, providing good agreement with the previous measurements made within Model C. We obtain a black hole spin of $a=0.69^{+0.09}_{-0.09}$, ruling out a Schwarzschild black hole with 95\% confidence and a maximally rotating black hole at the 99\% confidence level. 

\begin{figure*}
\rotatebox{-90}{\includegraphics[width=5cm]{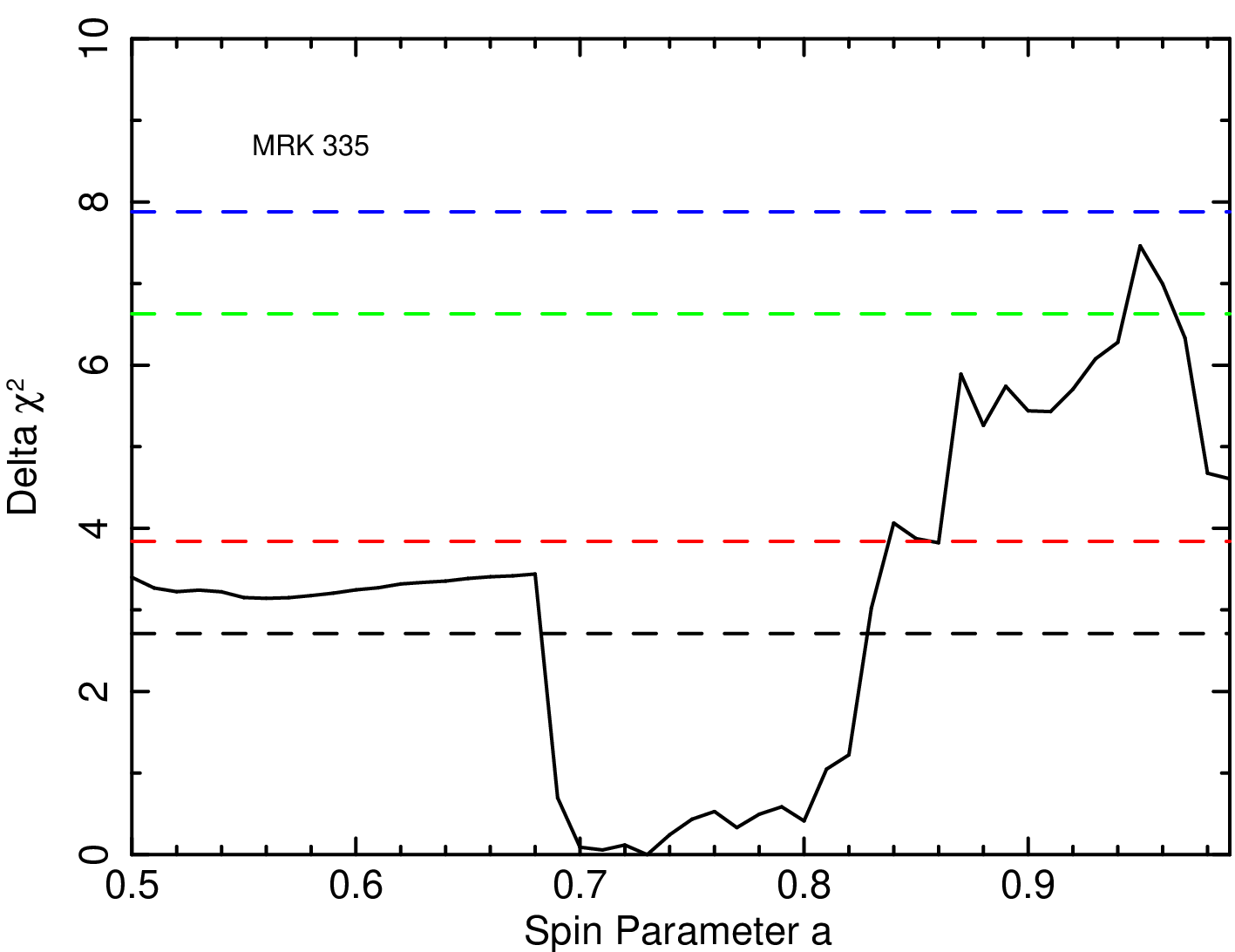}}
\rotatebox{-90}{\includegraphics[width=5cm]{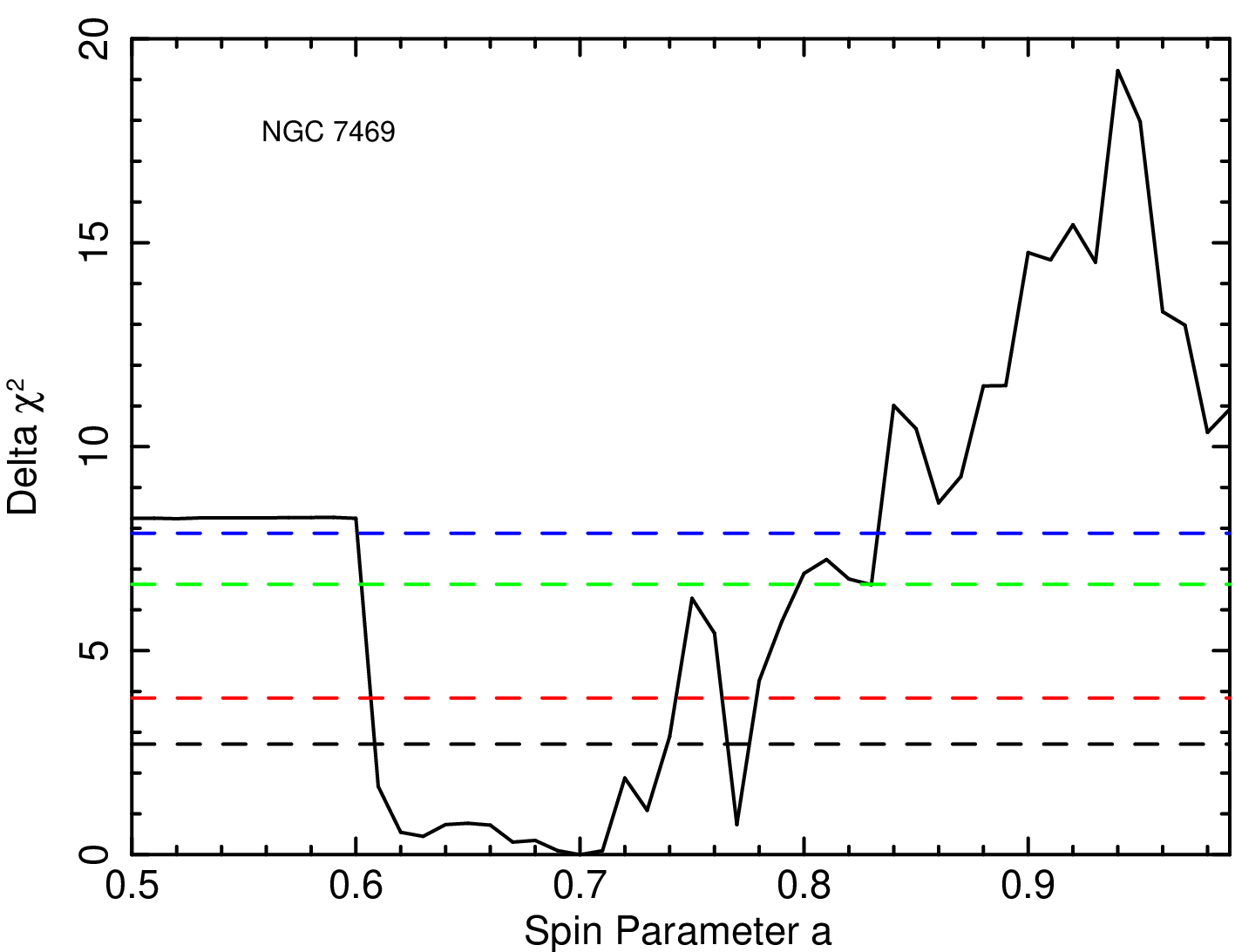}}
\rotatebox{-90}{\includegraphics[width=5cm]{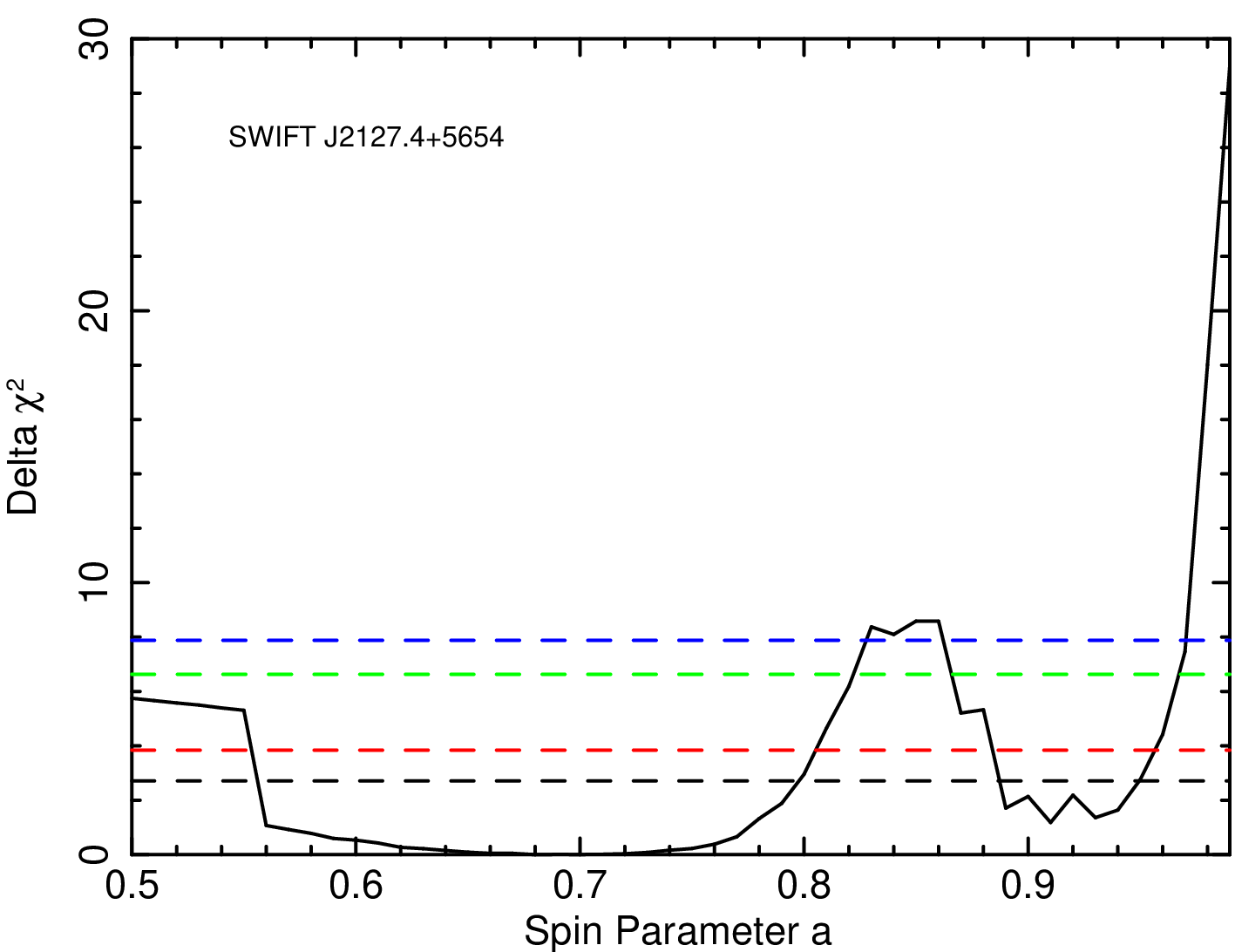}}
\caption{Confidence plots from Model D for the spin parameter $a$ for those objects in which it could be constrained. Dashed lines representing 90\% (black), 95\% (red), 99\% (green) and 99.5\% (blue) confidence levels. Note that the emissivity index q is a free parameter.}
\label{fig:spin_contours}
\end{figure*}

SWIFT J2127.4+5654 also shows a broad and statistically significant component, $EW=178^{+82}_{-69}$\,eV, improving the fit by $\triangle\chi^{2}\sim37$ (similarly to Model C) for four additional free parameters. The emissivity and the inclination of the accretion disc are typical of the objects in this sample. The spin parameter measured here is consistent with that measured by Miniutti et al. (2009), $0.70^{+0.10}_{-0.14}$ compared to $0.6^{+0.2}_{-0.2}$ found using a blurred reflection model. Here we reject a maximally rotating central black hole at greater than 99\% confidence and a non-rotating black hole with 95\% confidence. The spin constraints obtained with Model D for Mrk 335, NGC 7469 and SWIFT J2127.4+5654 can be seen in Figure \ref{fig:spin_contours}.

\begin{table*}
\caption{Fit parameters from Model E to {\sl Suzaku} XIS, HXD and BAT data from {\sl Swift}. (A) represents the unblurred \textsc{reflionx} and (B) represents the blurred \textsc{reflionx}.  The ionization parameter $\xi$ is given in units erg\,cm\,s$^{-1}$. $^{a}$ \textsc{powerlaw} normalization given in units $(10^{-2}\,{\rm ph\,keV^{-1}\,cm^{-2}\,s^{-1}})$. $^{b}$ \textsc{reflionx} normalization in units 10$^{-5}$. * denotes a parameter frozen at the best-fitting value from Model D. In some cases the spin parameter $a$ could not be constrained, denoted by -- .}
\begin{tabular}{l c c c c c c}
\hline
& Ark 120 & Fairall 9 & MCG-02-14-009 & Mrk 335 & NGC 7469 & SWIFT J2127.4+5654 \\
\hline
Soft Excess & \checkmark & \checkmark & X & \checkmark & \checkmark & X \\ 
\hline
$\Gamma$ & $2.12^{+0.01}_{-0.02}$ & $2.01^{+0.01}_{-0.01}$ & $1.90^{+0.03}_{-0.03}$ & $2.15^{+0.01}_{-0.02}$ & $1.80^{+0.01}_{-0.01}$ & $2.20^{+0.05}_{-0.05}$  \\ [0.8ex]
Norm $^{a}$ & $1.13^{+0.02}_{-0.02}$ & $0.78^{+0.02}_{-0.02}$ & $0.12^{+0.05}_{-0.03}$ & $0.57^{+0.02}_{-0.02}$ & $0.55^{+0.05}_{-0.07}$ & $1.65^{+0.05}_{-0.08}$ \\
\hline
\multicolumn{7}{c}{Kerrconv} \\
$q$ & $4.1^{+0.8}_{-0.8}$ & $>6.15$ & $<3.5$ & $6.6^{+2.0}_{-1.0}$ & $>3.0$ & $2.2^{+0.6}_{-0.4}$ \\ [0.8ex]
$a$ & $>0.97$ & $0.98^{+0.01}_{-0.01}$ & $<0.96$ & $0.87^{+0.05}_{-0.06}$ & $<0.97$ & -- \\ [0.8ex]
$i^{\circ}$ & $48^{+3}_{-6}$ & $70^{+5}_{-2}$ & $40^{+11}_{-14}$ & $53^{+6}_{-6}$ & $70^{+4}_{-3}$ & $43^{+18}_{-7}$ \\ [0.8ex]
\hline
Fe/Solar (A) & 1.4* & 1.9* & 0.4* & $2.0^{+0.8}_{-0.4}$ & 1.6* & $>0.2$ \\ [0.8ex]
$\xi$ (A) & $<13$ & $<11$ & $<543$ & $<11$ & $<11$ & $<26$ \\ [0.8ex]
Norm (A) $^{b}$ & $1.31^{+0.14}_{-0.15}$ & $1.13^{+0.08}_{-0.19}$ & $0.10^{+0.17}_{-0.10}$ & $0.47^{+0.05}_{-0.10}$ & $1.13^{+0.11}_{-0.21}$ & $0.67^{+0.60}_{-0.60}$ \\ [0.8ex]
Fe/Solar (B) & $1.5^{+0.3}_{-0.3}$ & $0.8^{+0.2}_{-0.3}$ & $>1.1$ & $1.0^{+0.1}_{-0.1}$ & $<0.4$ & $1.0^{+0.8}_{-0.3}$ \\ [0.8ex]
$\xi$ (B) & $56^{+12}_{-3}$ & $24^{+17}_{-9}$ & $<14$ & $207^{+5}_{-5}$ & $<24$ & $<14$ \\
Norm (B) $^{b}$ & $0.38^{+0.21}_{-0.13}$ & $0.24^{+0.04}_{-0.03}$ & $0.27^{+0.19}_{-0.20}$ & $0.06^{+0.01}_{-0.01}$ & $0.30^{+0.36}_{-0.12}$ & $3.68^{+1.50}_{-1.99}$ \\ [0.8ex]
\hline
Fe\,{\rm XXV} Line (keV) & $6.65^{+0.05}_{-0.07}$ & $6.74^{+0.04}_{-0.04}$ & -- & $6.67^{+0.03}_{-0.03}$ & -- & -- \\ [0.8ex]
$EW$ (eV) & $17^{+7}_{-7}$ & $16^{+6}_{-6}$ & -- & $34^{+8}_{-8}$ & -- & -- \\ [0.8ex]
Flux $(10^{-5}\,{\rm ph\,cm^{-2}\,s^{-1}})$ & $0.62^{+0.26}_{-0.27}$ & $0.44^{+0.17}_{-0.17}$ & -- & $0.55^{-0.13}_{-0.13}$ & -- & -- \\ [0.8ex]
$\triangle\chi^{2}$ & 8 & 10 & -- & 10 & -- & -- \\ 
\hline 
Fe\,{\rm XXVI} Line (keV) & $6.95^{+0.03}_{-0.03}$ & $6.88^{+0.02}_{-0.03}$ & $6.95^{+0.06}_{-0.07}$ & $6.97^{+0.07}_{-0.12}$ & -- & -- \\ [0.8ex]
$EW$ (eV) & $29^{+9}_{-9}$ & $27^{+7}_{-7}$ & $34^{+23}_{-23}$ & $16^{+9}_{-8}$ & -- & -- \\ [0.8ex]
Flux $(10^{-5}\,{\rm ph\,cm^{-2}\,s^{-1}})$ & $0.85^{+0.27}_{-0.27}$ & $0.63^{+0.17}_{-0.17}$ & $0.14^{+0.10}_{-0.10}$ & $0.23^{+0.13}_{-0.12}$ & -- & -- \\ 
$\triangle\chi^{2}$ & 16 & 25 & 4 & 10 & -- & -- \\
\hline
BAT const & $1.11^{+0.11}_{-0.11}$ & $0.87^{+0.10}_{-0.10}$ & -- & $1.06^{+0.23}_{-0.23}$ & $0.93^{+0.08}_{-0.08}$ & $0.68^{+0.10}_{-0.09}$ \\
\hline
$\chi^{2}_{\nu}$ & 770.2/647 & 879.6/832 & 612.0/538 & 811.6/722 & 835.4/808 & 830.4/865 \\
\hline
\end{tabular}
\label{tab:ModelE}
\end{table*}

\subsection{Model E -- Blurred reflector (1)}
In contrast to the previous models, Model E does not include a contribution from the Comptonization of soft photons (i.e. the \textsc{compTT} model). Alternatively, the excess seen at soft energies is modelled using a blurred reflector in addition to reflection consistent with distant emission. The ionized reflection model \textsc{reflionx} is convolved with a \textsc{kerrdisk} kernel, modelling a relativistically blurred reflection spectrum from the accretion disc originating from very near to the black hole instead of the \textsc{kerrdisk} model i.e. the \textsc{kerrconv} model (Brenneman \& Reynolds 2006). The reflection components consist of: (\textsc{reflionx} + \textsc{kerrconv}*\textsc{reflionx}). The assumptions in using this convolution model are consistent with those used in Model D, the inner radius of the accretion is assumed to extend down to r$_{\rm ISCO}$ and outer radius fixed at 400\,r$_{\rm ms}$ for a disc of uniform emissivity. Narrow emission lines from ionized matter are also included where required, while the narrow 6.4\,keV core is accounted for by the second unblurred \textsc{reflionx} component. 

Model E is a significantly worse fit in all 4 AGN with a soft excess (by an average of $\triangle\chi^{2}\sim23$ compared to Model D), requiring a significant amount of blurring to account for the excess at low energies (see Table \ref{tab:ModelE}). Given the amount of blurring required, parameters such as the emissivity index and the spin parameter approach more extreme values for all objects with a soft excess (e.g. $q>4$ and $a>0.9$). Whilst Model E is a good fit to the data for Mrk 335 and NGC 7469, Model D provides a better fit and without the use of such extreme parameters.

In addition, the best-fit obtained with Model E for Ark 120 requires an inclination of $i=48^{\circ}$, whilst not unreasonable, differs from the inclinations measured using Models C \& D. However freezing the inclination at $i=35^{\circ}$ worsens the fit further ($\chi^{2}_{\nu}=824.6/648$ compared to $\chi^{2}_{\nu}=770.2/647$). A similar dual-reflector fit as used here is considered by Nardini et al. (2010) in which the emissivity index is frozen at $q=5.0$ compared to $q=4.1^{+0.9}_{-0.9}$ measured here. Nardini et al. (2010) also note a relatively high inclination of $i^{\circ}=57^{+5}_{-12}$ obtained in their analysis is likely too large for an object such as Ark 120 although it possible for the accretion disc and obscuring material to be misaligned. One area of disagreement between the results obtained with Model E and the Nardini et al. (2010) fit is the determination of the spin parameter. Nardini et al. finding $0.24<a<0.93$ whereas here only a lower limit of $a>0.97$ is found, this may be due to the degeneracies between the spin and emissivity index which are both allowed to vary in Model E.

Similarly, NGC 7469 is best fit with a \textsc{kerrconv} inclination parameter of $i^{\circ}=70^{+4}_{-3}$ whereas Models C \& D suggest that the disc is inclined at $i\sim23^{\circ}$ to the observer, worsening the fit by $\triangle\chi^{2}\sim80$. These high inclinations are likely to be driven by the need to model the relatively smooth soft excess.

For the case of MCG-02-14-009, Model E provides an approximately equal quality of fit compared to Models C \& D. No significant amount of blurring is required to model the spectrum, obtaining parameters similar to those previously i.e. low emissivity and an unconstrained spin parameter. The narrow Fe\,{\rm XXVI} ionized emission line is also present, in line with findings from Models C \& D. The SWIFT J2127.4+5654 spectrum is fitted very well with Model E ($\chi^{2}_{\nu}=830.4/865$). Extreme parameters are not required to fit the data, which is in agreement with results from the previous disc line models. This may be due to the lack of any notable soft excess in these two objects. Since a significantly blurred spectrum simulates an excess at lower energies, a high emissivity and near maximally rotating black hole would provide an over-excess at soft energies, inappropriate for objects such as MCG-02-14-009 and SWIFT J2127.4+5654. Nonetheless Miniutti et al. (2009) find $q=5.3^{+1.7}_{-1.4}$ in SWIFT J2127.4+5654 compared to $q=2.2^{+0.6}_{-0.4}$ here, this may be due to the high level of Galactic absorption and the additional small amount of intrinsic absorption at the redshift of the source. An increase in the emissivity index could be compensated for by increased absorption which is found in the Miniutti et al. (2009) analysis.

\begin{table*}
\caption{Fit parameters from Model D to {\sl XMM-Newton} EPIC-pn data. * denotes a frozen parameter. In some cases the spin parameter $a$ could not be constrained, denoted by -- .$^{a}$ \textsc{powerlaw} normalization given in units $(10^{-2}\,{\rm ph\,keV^{-1}\,cm^{-2}\,s^{-1}})$. $^{b}$ Flux for \textsc{compTT} quoted over the 0.5-10.0\,keV range in units 10$^{-11}$erg\,cm$^{-2}$\,s$^{-1}$. $^{c}$ \textsc{reflionx} normalisation given in units $10^{-5}$.}
\begin{tabular}{l c c c c c }
\hline
& Ark 120 & Fairall 9 & MCG-02-14-009 & Mrk 335 & NGC 7469 \\
\hline
$\Gamma$ & $1.58^{+0.01}_{-0.01}$ & $1.89^{+0.01}_{-0.01}$ & $1.80^{+0.02}_{-0.02}$ & $1.64^{+0.06}_{-0.05}$ & $1.62^{+0.02}_{-0.04}$ \\ [0.8ex]
Norm$^{a}$ & $0.56^{+0.01}_{-0.02}$ & $0.35^{0.01}_{-0.01}$ & $0.16^{+0.01}_{-0.01}$ & $0.29^{+0.01}_{-0.01}$ & $0.07^{+0.01}_{-0.01}$ \\
\hline
Plasma Temperature (keV) & $<7.2$ & -- & -- & $9.0^{+55.3}_{-4.3}$ & $5.2^{+10.3}_{-0.1}$ \\ [0.8ex]
Plasma Optical Depth $\tau$ & $0.6^{+0.7}_{-0.1}$ & -- & -- & $<0.7$ & $<38.0$ \\ [0.8ex]
Flux$^{b}$ & $2.90^{+0.21}_{-0.21}$ & -- & -- & $2.15^{+0.18}_{-0.18}$ & $0.014^{+0.003}_{-0.002}$\\ [0.8ex]
$\triangle\chi^{2}$ & 2505 & -- & -- & 956 & 61 \\
\hline
\multicolumn{6}{c}{Kerrdisk Profile} \\
Line E (keV) & 6.4* & 6.4* & 6.4* & 6.4* & 6.4*  \\ [0.8ex]
Eqw (eV) & $114^{+10}_{-17}$ & $148^{+24}_{-29}$ & $149^{+76}_{-73}$ & $116^{+22}_{-22}$ & $57^{+16}_{-20}$  \\ [0.8ex]
$q$ & $1.8^{+0.3}_{-0.4}$ & $<1.8$ & $2.8^{+2.0}_{-0.8}$ & $1.9^{+0.2}_{-0.2}$ & $2.3^{+0.3}_{-0.4}$ \\ [0.8ex]
$a$ & $<0.87$ & $<0.96$ & -- & $<0.71$ & $<0.99$ \\ [0.8ex]
$i^{\circ}$ & $24^{+7}_{-2}$ & $<26$ & $37^{+12}_{-4}$ & $43^{+6}_{-15}$ & $24^{+3}_{-4}$ \\ [0.8ex]
$\triangle\chi^{2}$ & 97 & 17 & 10 & 62 & 12 \\
\hline
$\xi$ & $284^{+394}_{-53}$ & $<11$ & $<11$ & $233^{+24}_{-13}$ & $1526^{+302}_{-246}$ \\ [0.8ex]
Fe/Solar & $1.6^{+0.5}_{-0.3}$ & $<0.2$ & $0.4^{+0.1}_{-0.1}$ & $1.8^{+0.3}_{-0.2}$ & $0.5^{+0.1}_{-0.1}$ \\ [0.8ex]
Norm$^{c}$ & $<0.01$ & $0.56^{+0.07}_{-0.08}$ & $0.44^{+0.01}_{-0.01}$ & $0.01^{+0.01}_{-0.01}$ & $<0.01$ \\
\hline
Fe\,{\rm XXV} Line (keV) & $6.69^{+0.03}_{-0.03}$ & $6.78^{+0.06}_{-0.05}$ & -- & -- & -- \\ [0.8ex]
$EW$ (eV) & $21^{+6}_{-6}$ & $40^{+18}_{-18}$ & -- & -- & -- \\ [0.8ex]
Flux $(10^{-5}\,{\rm ph\,cm^{-2}\,s^{-1}})$ & $0.91^{+0.26}_{-0.27}$ & $0.55^{+0.25}_{-0.25}$ & -- & -- & -- \\ [0.8ex]
$\triangle\chi^{2}$ & 11 & 12 & -- & -- & -- \\
\hline 
Fe\,{\rm XXVI} Line (keV) & $7.01^{+0.03}_{-0.03}$ & $7.09^{+0.06}_{-0.06}$ & $7.00^{+0.05}_{-0.06}$ & $7.03^{+0.03}_{-0.03}$ & -- \\ [0.8ex]
$EW$ & $31^{+7}_{-7}$ & $49^{+22}_{-21}$ & $30^{+20}_{-20}$ & $34^{+10}_{-10}$ & -- \\ [0.8ex] 
Flux $(10^{-5}\,{\rm ph\,cm^{-2}\,s^{-1}})$ & $1.11^{+0.25}_{-0.25}$ & $0.58^{+0.26}_{-0.25}$ & $0.12^{+0.08}_{-0.08}$ & $0.54^{+0.16}_{-0.16}$ & -- \\ [0.8ex]
$\triangle\chi^{2}$ & 54 & 14 & 9 & 21 & -- \\
\hline
$\chi^{2}_{\nu}$ & 1757.3/1663 & 753.1/761 & 1147.8/1151 & 1464.4/1411 & 1049.4/885 \\
\hline
\end{tabular}
\label{tab:ModelDXMM}
\end{table*}

\subsection{Model F -- Blurred reflector (2)}
Model F starts with Model B as the base model but instead blurring the single \textsc{reflionx} reflection spectrum with the \textsc{kerrconv} convolution model to model any broad residuals present in the spectra. The Fe\,K line complex can still be modelled in this way since the \textsc{reflionx} model includes Fe\,K${\alpha}$ emission and blurring the spectrum emulates the resulting profile from a \textsc{kerrdisk} model. Since the narrow Fe\,K${\alpha}$ core included within \textsc{reflionx} is now relativistically blurred, a narrow Gaussian of fixed width 10\,eV was added to ensure that the narrow 6.4\,keV is still modelled. This may represent the case where the 6.4\,keV line is observed from Compton-thin matter, such as the BLR or NLR. Contrary to Model E, the soft excess (where present) is modelled using the \textsc{compTT} soft photon Comptonization model.

Any ionized emission due to Fe\,{\rm XXV} and Fe\,{\rm XXVI} required in Model D is also found to be required in this model. In general a good fit to all objects is obtained, with Model F clearly providing a better fit to the data compared to Model E for Fairall 9, NGC 7469 and SWIFT J2127.4+5654. The \textsc{kerrconv} parameters are also consistent within error bars with the emissivity index, inclination and spin parameter found using the \textsc{kerrdisk} line profile previously, yielding typically slightly lower emissivity indicies. 

Note that the best-fitting spin parameter value with Model F for Fairall 9 is $a=0.40^{+0.33}_{-0.40}$ (quoted at the 75\% confidence level and is unconstrained at the 90\% confidence level) in agreement with $a=0.44^{+0.04}_{-0.11}$ found in Model D (at the 90\% confidence level) and with $a=0.5^{+0.1}_{-0.3}$ found by Schmoll et al. (2009). 

Given the independent modelling of the soft excess within this model, it is interesting to note that $a=0.72^{+0.18}_{-0.17}$ is obtained for NGC 7469 at the 90\% confidence level. This value of the spin parameter is in agreement with that found in Model D, although providing a slightly worse constraint. 

\subsection{Comparison with {\sl XMM-Newton} results}
Analysis of the data for these objects obtained with {\sl XMM-Newton} (all objects other than SWIFT J2127.4+5654, see Table \ref{tab:observations}) yields no additional constraints upon the properties of the Fe\,K region, the reflection component is also difficult to constrain due to the lack of hard X-ray data. As with the {\sl Suzaku} data, the Fe\,{\rm XXVI} narrow ionized emission line is found throughout most objects and the Fe\,{\rm XXV} line is relatively uncommon, observed only in Ark 120 and Fairall 9. However, the results obtained with a \textsc{kerrdisk} relativistic line profile are consistent with those obtained with Model D in the {\sl Suzaku} data. Similarly, a good fit is obtained with this model for all of the objects in the sample, with maximal spin ruled out from the fits. The 0.5--10.0\,keV spectrum is generally well described by a combination of narrow distant emission and a small broad relativistic component, typically $EW<150$\,eV. 

In particular, the {\sl XMM} data for MCG-02-14-009 yields very similar results to the {\sl Suzaku} data. The main difference being that a statistically significant Fe\,{\rm XXVI} line is not present in the {\sl XMM-Newton} data. All measurements obtained with a \textsc{kerrdisk} profile are consistent at the 90\% confidence level with the {\sl Suzaku} data. Both data sets therefore reveal a relatively simple 0.5--10.0\,keV spectrum, there is no observed soft excess, it is well described by distant reflection in addition to a moderately broad relativistic line profile from neutral Fe occuring at $r_{\rm in}>2.5\,r_{\rm g}$ (obtained from the application of the \textsc{laor} model) with emissivity $q=2.8^{+2.0}_{-0.8}$, as seen in Table \ref{tab:ModelDXMM}.

These same observations of Ark 120 and Fairall 9 have previously been modelled with the blurred reflector approach, consisting of a combination of narrow distant emission and a smeared ionized reflector under the assumption of a Schwarzschild geometry, see Brenneman \& Reynolds (2009). Consistent results are found for Fairall 9 for which narrow ionized emission lines are also found, however the authors were unable to constrain the emissivity index, measured at $q<1.8$ here and the spin value requires that emission occurs at $r_{\rm in}>1.8\,r_{\rm g}$. The results found here for Ark 120 suggest consistent measurements for the emissivity index with Brenneman \& Reynolds (2009). In comparison with the {\sl Suzaku} data for this object, both approaches are consistent with Model D. Ionized emission lines are also found in Ark 120 throughout both analyses. The inner radius of emission is generally consistent, occurring at $r_{\rm in}<80\,r_{\rm g}$ from Brenneman \& Reynolds (2009) and $r_{\rm in}>2.5\,r_{\rm g}$ as measured here from the spin parameter within Model D (Table \ref{tab:ModelDXMM}). 

\section{Discussion}
\begin{figure*}
\rotatebox{-90}{\includegraphics[width=0.33\textwidth]{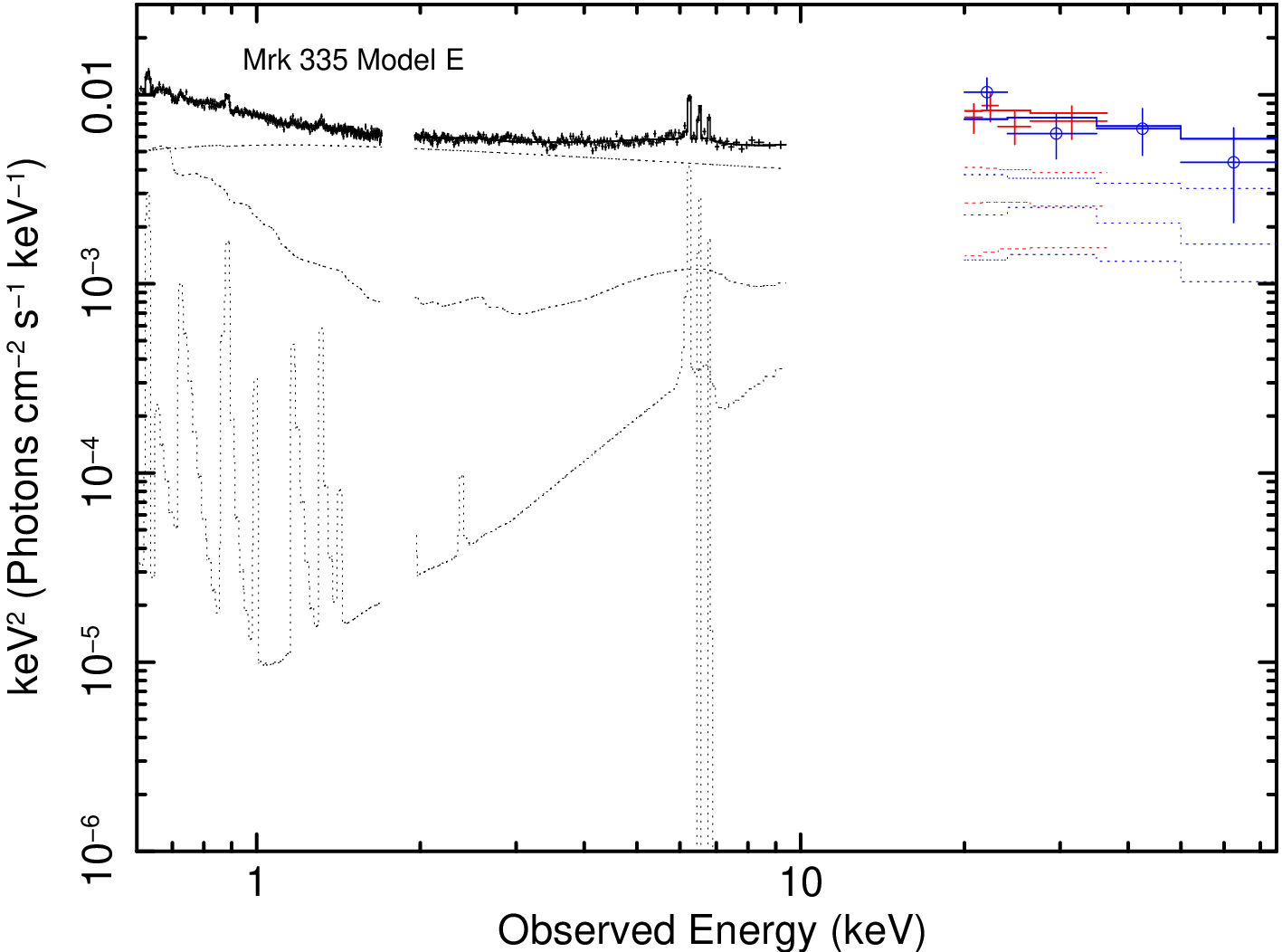}}
\rotatebox{-90}{\includegraphics[width=0.33\textwidth]{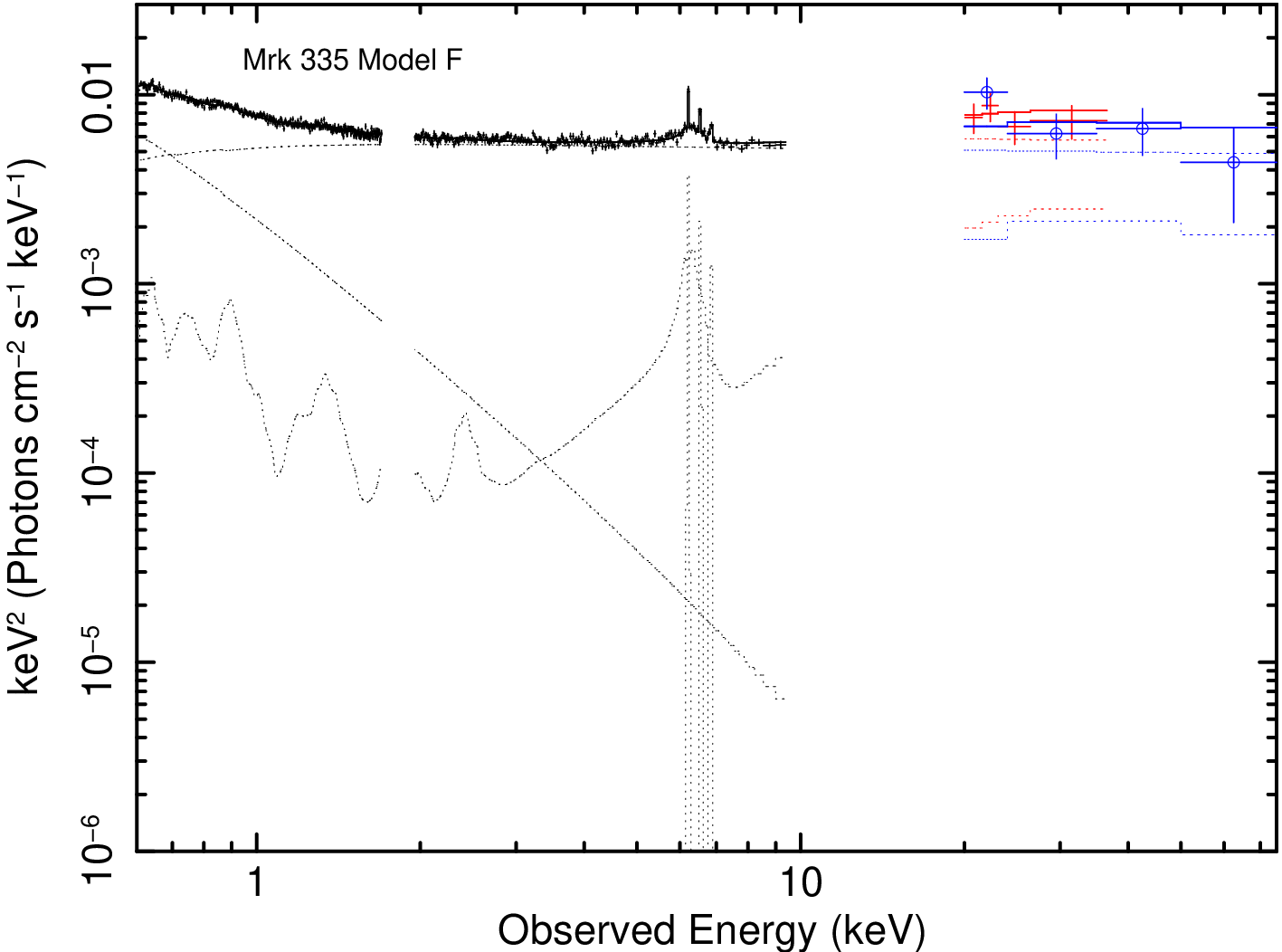}}
\caption{$\nu$F$\nu$ plots of Models E \& F for Mrk 335 showing to degree to which relativistic blurring of the reflection component is required according to differing interpretations of the soft excess. Note that the significant blurring required in Model E to account for the soft excess reduces the accuracy of the fit to the Fe\,K region features. XIS data is in black, HXD in red and BAT is represented by blue circles.}
\label{fig:ModelEF_comparison}
\end{figure*}

This small sample of six AGN includes typically bare Seyfert 1 galaxies featuring little or no intrinsic absorption. This property of these AGN is important since the spectra of these objects is simpler to model without complicating factors (such as absorption) allowing the observer to draw conclusions about the fundamentals of accretion disc properties and basic features of the Fe\,K region. These conclusions will therefore be less model dependent up on how the warm absorber is modelled (Turner \& Miller 2009). It is important to effectively model the spectra of these AGN and assess the likely origin of various components of the spectrum before we can proceed to draw conclusions about more complicated AGN. Modelling of the broad--band continuum is also essential prior to analysis of the Fe\,K region, this has been achieved using data from both HXD and BAT hard X-ray detectors onboard {\sl Suzaku} and {\sl Swift} respectively. Allowing the spectrum spanning 0.5--100.0\,keV to be modelled with better constraints upon distant reflection components modelled by \textsc{pexrav} and \textsc{reflionx}, whilst maintaining effective modelling of the soft excess. This approach furthers that taken by Nandra et al. (2007) in which EPIC-pn data from {\sl XMM-Newton} is analysed in the 2.5--10.0\,keV range.

The presence of narrow ionized emission lines due to Fe\,{\rm XXV} and Fe\,{\rm XXVI} assessed prior to modelling any broad residuals in the Fe K region has an important effect upon the parameters obtained with broad disc line profiles. In some spectra neglecting to model these lines (if present) can accentuate any apparent broad Fe\,K line, particularly those resulting from ionized species of Fe. Most commonly occurring in the objects analysed here is the 6.97\,keV line from H-like Fe which is observed in all objects, except NGC 7469 and SWIFT J2127.4+5654. This is a surprising result given the rarity of such lines found by Nandra et al. (2007) in which only 2/26 sources showed evidence for these lines (although more common in work by Bianchi et al. 2009). This may be due to higher quality (longer exposure) data obtained with {\sl Suzaku} allowing these lines to be more easily distinguished from broad residuals at these energies. Similarly, neglecting to include a narrow 6.97\,keV emission line where present forces the relativistic line profile to have the emissivity index increased to particularly large values ($q>6$) and the inclination is slightly increased to $\sim45^{\circ}-55^{\circ}$ whereby the blue wing of the profile is forced to model the narrow excess at ~$\sim$\,6.97\,keV. 

In agreement with Nandra et al. (2007), however, is the rarity of the 6.7\,keV line from He-like Fe which only features significantly here in Mrk 335. Models A \& B earlier suggested the possible presence of Fe\,{\rm XXV} in Fairall 9 and SWIFT J2127.4+5654, however it was found that excesses at these energies coincided with the peak in the blue-wing of the relativistic line profile. This suggests that high (i.e. calorimeter) resolution data is necessary to determine the presence of this particular emission line particularly when employing a relativistic line profile, such as that which would be obtained with {\sl Astro-H} (see Kelley et al. 2010).

The introduction of the \textsc{kerrdisk} model over a model considering emission purely from distant material (i.e. Model B) improves the fit to the six objects by an average of $\triangle\chi^{2}\sim20$. This implies that broad residuals are a statistically significant feature in all of these objects (although less so in MCG-02-14-009).

\subsection{Typical parameters of the accretion disc}
The average inclination of the accretion disc to the observer as inferred by the \textsc{kerrdisk} line profiles in Model D is $i=33^{\circ}\pm4^{\circ}$, this is consistent with Nandra et al. (2007) who find $i=38^{\circ}\pm6^{\circ}$. Also from Model D we find an average emissivity index of $q=2.3\pm0.2$, much lower than those used in previous work for some of these objects e.g. Fairall 9 (Schmoll et al. 2009) and SWIFT J2127.4+5654 (Miniutti et al. 2009), particularly when using blurred reflection models to model the whole continuum. The line profile produced within Model D also suggests a broad line profile with an average equivalent width $EW=119\pm19$\,eV consistent with an average $EW=91.3\pm12.8$ found by Nandra et al. (2007) using a Gaussian to model the broad residuals. Whilst here for Models C \& D the centroid rest energy of the line profile is not allowed to extend below 6.4\,keV in the rest frame, the basic parametrization within Model A suggests an average broad component line energy of $6.29\pm0.03$\,keV compared to $E_{K\alpha}=6.27\pm0.07$\,keV according to Nandra et al. (2007). 

The low average emissivity index of the objects in this small sample may be due to a number of factors: independent modelling of the soft excess through a Comptonization of soft photons; modelling of ionized emission lines where present in the data and the assumption that emission extends down to $r_{\rm ISCO}$ (the \textsc{kerrdisk} model). In accordance with this, within Models E \& F facilitating the use of a blurred reflector, the average emissivity index is also dependent upon the way in which the soft excess is modelled. Model F (in which the soft excess is modelled with a \textsc{compTT} component, as in Models A--D) suggests an average $q=2.0\pm0.1$ in general agreement with Model D. However Model E (in which purely the blurred reflection component is tasked with modelling the soft excess) suggests an average $q=5.0\pm0.7$ for those objects featuring a soft excess and an average $q=2.0\pm0.4$ for those without, namely MCG-02-14-009 and SWIFT J2127.4+5654. Thus the high emissivity in Model E for some objects is driven by the need to fit a featureless soft excess e.g. see Figure \ref{fig:ModelEF_comparison}. A blurred reflection model with high emissivity therefore appears to be ruled out given Model E produces a significantly worse fit in all 4 AGN with a soft excess.

Model C alternatively models reflection off a relativistic accretion disc under the assumption of a maximally rotating central black hole, without the inner radius of emission fixed at the {\rm ISCO}. This approach also suggests a similarly relatively small equivalent width of $EW=102\pm14$\,eV and an average emissivity $q=3.2\pm0.4$ (for those objects for which the emissivity could be constrained), higher than that derived from Model D and likely due to the fact that $r_{\rm in}$ is not fixed at $r_{\rm ISCO}$. The average inclination of the accretion disc is also very similar to Model D above with $i=34^{\circ}\pm3^{\circ}$. Given the assumptions made within this model, the inner radius of emission is found to originate at tens of $r_{\rm g}$ rather than $<6\,r_{\rm g}$ at an average $r_{\rm in}=39\pm8\,r_{\rm g}$. Indeed, in Ark 120, Fairall 9 and MCG-02-14-009 Model C and this interpretation of the accretion disc and SMBH provides a marginally better fit. However given the degeneracies between $q$, $a$ and $r_{\rm in}$ it would be impossible to reasonably constrain any of these values without assumptions similar to those mentioned above. 

\subsection{Model dependent Black Hole spin}
The results for determining the spin of the central black holes of the AGN in this sample suggest that the spin derived is very much dependent upon which interpretation of the Fe\,K line region is followed. Modelling the soft excess through an independent model such as {\sl compTT} tends to yield low to intermediate spin constraints for the objects in this sample, the exceptions being Ark 120 and MCG-02-14-009 in which only an upper limit could be placed. The employment of this interpretation also yields low to intermediate emissivity indicies for the accretion discs (i.e. $q\sim2$). Modelling the 0.5--100.0\,keV spectrum with a blurred reflection component and no other modelling of the soft excess (i.e. Model E) gives particularly high values of both emissivity index and spin parameter, typically $q\gtrsim4$ and $a\gtrsim0.9$, but these fits are statistically ruled out here.

This is, however, only in objects featuring a soft excess. In MCG-02-14-009, which has no obvious excess over a powerlaw at low energies, the derived parameters are very similar throughout Models D, E \& F. Similarly for SWIFT J2127.4+5654, the parameters obtained with these three models are all consistent and indeed the resulting spin parameter is also consistent with previous findings by Miniutti et al. (2009) although they find a much higher emissivity index is required ($q=5.3^{+1.7}_{-1.4}$ compared to $q=2.2^{+0.3}_{-0.9}$ from Model E). 

As discussed previously, the spin constraint obtained here for Fairall 9 agrees with that found by Schmoll et al. (2009), but only in the case where they ignore the spectrum below 2\,keV ($a=0.5^{+0.1}_{-0.3}$ compared to $a=0.44^{+0.04}_{-0.11}$ found here in Model D). According to Model D, an intermediate spin of the central black hole within NGC 7469 is found, $a=0.69^{+0.09}_{-0.09}$. This is also consistent with that found within Model F $a=0.72^{+0.18}_{-0.17}$, suggesting that the spin of this object is indeed $a\sim0.7$. 

These findings therefore suggest that the two differing interpretations give concurring spin constraints only when the blurred reflector is not required to model the soft excess, leading to the conclusion that the component responsible for the soft excess must be independently fitted before conclusive spin constraints can be made. Even so, the origin of the soft excess is not currently known, e.g. see Gierlinski \& Done (2004). The relatively constant temperature of the soft excess versus BH mass suggests that it may not arise from direct thermal emission from the accretion disc, since the accretion temperature properties should scale with $M_{\rm BH}^{-1/4}$ in a standard accretion disc. An atomic origin of the soft excess has been suggested, however no obvious spectral features are seen in high resolution data meaning that if atomic emission is responsible it must be significantly relativistically blurred such as here in Model E (Ross, Fabian \& Ballantyne 2002; Fiore, Matt \& Nicastro 1997). Alternatively, it is suggested that the soft excess could arise from relativistically blurred absorption (Gierlinski \& Done 2004) from a differentially rotating and outflowing disc wind (Murray \& Chiang 1997). The smeared absorption creates a smooth hole in the spectrum resulting in the apparent soft excess and a hardening at high energies. Whilst in this paper we do not aim to determine the origin of the soft excess, we find that a simple parametrization of the soft excess continuum through a model such as \textsc{compTT} provides a better fit to the spectra than an atomic origin from a highly blurred reflection component. 

\subsection{Reflection off a disk wind?}
One possible alternate origin of the observed broad component in AGN spectra is through reflection of primary continuum photons off a Compton thick disc wind, producing emission, absorption and broad features in the Fe\,K band (Sim et al. 2010). Fast outflows can add significant complexity to the Fe\,K region and can reproduce a wide range of spectral signatures owing to the many differing physical conditions. Sim et al. (2010) note that such outflows can, however, have a significant affect upon the soft X-ray spectrum resulting in features such as highly blueshifted absorption lines, although these absorption are weaker than the features seen in the Fe\,K band. The lines from a disc wind may not be observed in objects such as those in this sample if they are observed relatively face-on, however even if we are not looking directly through the line of sight to the wind, it may be possible to observe broadened emission features in reflection. For instance Sim et al. (2010) predict broadened Fe\,K emission down to $\sim5$\,keV in their disc wind spectra, with equivalent widths typically several tens of eV.

\section{Conclusions}
Resulting from the work above on this sample of six 'bare' Seyfert AGN, we conclude the following:
\begin{enumerate}
\item Narrow emission from distant material is very important when modelling the Fe\,K region and neglecting to include these components where present can have a significant affect upon the accretion disc parameters obtained with subsequent models. The narrow neutral 6.4\,keV core is ubiquitous amongst these six AGN.

\item Ionized emission at $\sim6.97$\,keV due to Fe\,{\rm XXVI} is more common in these spectra than previously thought (e.g. Bianchi et al. 2004 and Nandra et al. 2007) although Bianchi et al. (2009) find them more common. It is present here in 4/6 objects in the {\sl Suzaku} data and it is also present in the same objects during the 5 {\sl XMM-Newton} observations included here. 

\item Emission at $\sim6.70$\,keV due to Fe\,{\rm XXV} is still relatively uncommon amongst these AGN, present in only 1/6 {\sl Suzaku} and 2/5 {\sl XMM-Newton} observations. However, it is likely that this emission line may be more common amongst AGN since this is also the energy at which the blue-wing from a relativistically blurred neutral 6.4\,keV line occurs with typical model parameters (e.g.  $i=30^{\circ}-40^{\circ}$, $q=2-3$), making it difficult in most cases to distinguish between these two components in CCD resolution spectra.

\item The fit to all objects in the sample is significantly improved with the introduction of a relativistic line profile, such as that from a \textsc{kerrdisk} model. However this broad emission is weaker after effective modelling of narrow components (mean $EW=119\pm19$\,eV).

\item The average emissivity index of the accretion disc is a low to moderate $q=2.3\pm0.2$ when using \textsc{compTT} to model the soft excess. The emissivity index typically scales as $R^{-2.3}$, consistent with an $R^{-2}$ law for the illuminating continuum and therefore implying that strong GR effects (such as light bending) may not be required in these objects. The assumption that emission extends down to $r_{\rm ISCO}$ is still valid, however, since the flat emissivity of the disc indicates that emission is not centrally concentrated. This is consistent with emission occuring a typically tens of $r_{\rm g}$ rather than within $<6\,_{\rm g}$.

\item It is essential to effectively model the continuum and the soft excess prior to attempting to constrain parameters of the accretion disc and inner regions of AGN. Independent modelling of the soft excess with models such as \textsc{compTT} or a second soft \textsc{powerlaw}, rather than blurring reflection from inner parts of the accretion disc as a way to model the soft excess yields consistent results with respect to AGN with and without an excess at lower energies.

\item Purely using a blurred reflection component to model the soft excess results in typically higher values of spin and emissivity index. This would lead to a skewed distribution amongst AGN towards the higher values of these parameters for those objects with a soft excess i.e. $a>0.9$ and $q>4$. This would indicate that the regions of the accretion disc responsible for the soft excess (in an atomic origin of the soft excess) are not the same as those responsible for features in the Fe\,K region. In this small sample of six AGN, the fit is significantly worse using this interpretation for those objects with a soft excess, namely Ark 120, Fairall 9, Mrk 335 and NGC 7469. 

\item New constraints suggest further intermediate spin values for Mrk 335 and NGC 7469 of $a=0.70^{+0.12}_{-0.01}$ and $a=0.69^{+0.09}_{-0.09}$ respectively. The spin value for SWIFT J2127.4+5654 found here is consistent with that found by Miniutti et al. (2009), whereas the spin found here for Fairall 9 is consistent with the Schmoll et al. (2009) analysis above 2\,keV. Only upper bounds for the spin parameter can be placed for Ark 120 and MCG-02-14-009, requiring that emission does not occur within 2.02\,$r_{\rm g}$ and 2.45\,$r_{\rm g}$ respectively. 

\item Overall zero spin cannot be ruled out at a particularly high confidence level in all objects, as can be seen in Table \ref{tab:ModelD}. None of the objects in the sample show overwhelming evidence for a Kerr geometry and significantly centrally concentrated emission, maximal spin is ruled out in most cases (Models D \& F). 
\end{enumerate}

\section*{Acknowledgements}
This research has made use of data obtained from the {\sl Suzaku} satellite, a collaborative mission between the space agencies of Japan (JAXA) and the USA (NASA). Data obtained with {\sl XMM-Newton} has also been used within this paper, an ESA science mission with instruments and contributions directly funded by ESA Member States and NASA.

\end{document}